\documentclass[floatfix,superscriptaddress,noshowpacs]{revtex4}%
\usepackage{graphicx}
\usepackage{amsmath,amssymb,amsfonts}
\begin{document}

\title{Wigner-Mott quantum criticality: from 2D-MIT to $^3$He and Mott organics}

\author{V. Dobrosavljevi\'c}

\affiliation{Department of Physics and National High Magnetic Field Laboratory, Florida State University, Tallahassee, Florida 32306, USA}

\author{D. Tanaskovi\'c}

\affiliation{Scientific Computing Laboratory, Center for the Study of Complex Systems, Institute of Physics Belgrade, University of Belgrade, Pregrevica 118, 11080 Belgrade, Serbia}
\vspace{12pt}
\begin{abstract}
Experiments performed over the last twenty years have revealed striking similarities between several two-dimensional (2D) fermion systems, including diluted two-dimensional electron liquids in semiconductors, $^3$He monolayers, and layered organic charge-transfer salts.  These experimental results, together with recent theoretical advances, provide compelling evidence that strong electronic correlations - Wigner-Mott physics - dominate the universal features of the corresponding metal-insulator transitions (MIT). Here we review the recent theoretical work exploring quantum criticality of Mott and Wigner-Mott transitions, and argue that most puzzling features of the experiments find natural and physically transparent interpretation based on this perspective.     
\end{abstract}
\maketitle




\section{MIT in the strong correlation era: the mystery and the mystique}\vspace{-6pt}

The key physical difference between metals and insulators is obvious even at first glance. Its origin, however, has puzzled curious minds since the dark ages of Newton's Alchemy \cite{newton-alchemist}; centuries-long efforts have failed to unravel the mystery. Some basic understanding emerged in early 1900s with the advent of quantum mechanics, based on the nature of electronic spectra in periodic (crystalline) solids. The corresponding Band Theory of Solids (BTS) \cite{ashcroft} provided a reasonable description of many features of good metals such as silver or gold, but also of good insulators, such as germanium and silicon. 

From the perspective of the BTS, metals and insulators represent very different phases of matter. On the other hand, the rise of electronic technology, which rapidly accelerated in the second half of the 20th century, demanded the fabrication of materials with properties that can be conveniently tuned from the limit of good metals all the way to poor conductors and even insulators. This requirement is typically satisfied by chemically or otherwise introducing a modest number of electrical carriers into an insulating parent compound, thus reaching an unfamiliar and often puzzling regime. The need to understand the properties of this intermediate metal-insulator transition (MIT) region \cite{mott-book90} opened an important new Pandora's box, both from the technological and the basic science perspective.  

The fundamental question thus arises: what is the physical nature of the MIT phase transition \cite{dobrosavljevic2012conductor}, and what degrees of freedom play a dominant role in its vicinity? The answer is relatively simple in instances where the MIT is driven by an underlying thermodynamic phase transition to an ordered state. This happens, for example, with onset of antiferromagnetic (AFM) or charge-order (CO), which simply modify the periodic potential experienced by the mobile electrons, and the BTS picture suffices \cite{slater51}. Here, the MIT can simply be regarded as an experimental manifestation of the emerging order, reflecting the static symmetry change of the given material. 

The situation is much more interesting in instances where the MIT is not accompanied with some incipient order. Here, the physical reason for a sharp MIT must reside beyond the BTS description, involving either strong electron-electron interactions \cite{mott1949} or disorder \cite{anderson1958}. The interplay of both effects  proves difficult to theoretically analyze, although considerable effort has been invested, especially within Fermi-liquid (weakly interacting) approaches at weak disorder \cite{leeramakrishnan,fink-jetp83,fink-jetp84}. More recent work found evidence that disorder may even more significantly modify the properties of strongly correlated metals, sometimes leading to "non-Fermi liquid" metallic behavior \cite{RoP2005review}. In any case, the MIT problem in presence of both strong electron-electron interactions and disorder is still far from being fully understood, despite recent advances \cite{dobrosavljevic2012conductor}. 

On physical grounds, however, one may expect that in specific materials either the correlations or the disorder may dominate, so that a simpler theory may suffice. In this chapter we shall not discuss many of the interesting experimental systems where the MIT problem has been investigated so far; a comprehensive overview is given in the chapter by D. Popovi\'c. Instead, we shall focus on three specific classes of low-disorder materials showing remarkable similarities in their phenomenology. Our story begins with a short overview of the key experimental features observed in ultra-clean diluted two-dimensional electron gases (2DEG) in semiconductors, displaying the so-called ``Two-Dimensional Metal-Insulator Transition'' (2D-MIT) in zero magnetic field. Next, we review experiments where very similar behavior is observed for $^3$He monolayers, and also in quasi two-dimensional organic charge-transfer salts of the $\kappa$-family, both of which are believed to represent model systems displaying the Mott (interaction-driven) MIT. Finally, we provide a direct comparison between these experimental findings and recent theoretical results for interaction-driven (Mott and Wigner-Mott) MITs in absence of disorder. Our message is that the Mott picture represents the dominant physical mechanism in all these systems, and should be regarded as the proper starting point for the study of related materials with stronger disorder. 

\section{Phenomenology of 2D-MIT in the ultra-clean limit}

Two dimensional electron gas devices in semiconductors have been studied for almost fifty years, since the pivotal 2DEG discovery in 1966 \cite{fowler1966}. Because they served as basis for most modern electronics, 2DEG systems have been very carefully studied  and characterized \cite{AFS1982} from the materials point of view, allowing the fabrication of ultra-high mobility material by the 1990s. However, its use as an ideal model system for the study of MIT has been recognized relatively late, with the pioneering work of Sergey Kravchenko and collaborators \cite{kravchenko95}. The reason was the long-held belief that all electronic states remain localized at $T=0$, even in presence of  infinitesimally weak disorder, based on the influential "scaling theory of localization" \cite{gang4}. This result - the absence of a sharp MIT in two dimensions - was firmly established only for models of noninteracting electrons with disorder, and although even early results on interacting models \cite{fink-jetp83,fink-jetp84} suggested otherwise, Kravchenko's initial work was initially met with extraordinary skepticism and disbelief \cite{maslov2001}.

Soon enough, however, clear evidence of a sharp 2D-MIT in high mobility silicon was confirmed on IBM samples (the original material Kravchenko used came from Russia), with almost identical results \cite{popovic97}. This marked the beginning of a new era; the late 1990s witnessed a veritable avalanche of experimental and theoretical work on the subject, giving rise to much controversy and debate. This article will not discuss all the interesting results obtained in different regimes, many of which are reviewed in the chapter by D. Popovi\'c in this volume. Our focus here will be on documenting evidence that strong electronic correlations - and not disorder - represent the main driving force for 2D-MIT in high mobility samples. In the following, we follow the historical development of the field, with emphasis on those experimental signatures that emerge as robust and rather universal features of interaction-driven MITs in a variety of systems.  
\begin{figure}[b]
\begin{center}
\includegraphics[width=0.41\columnwidth]{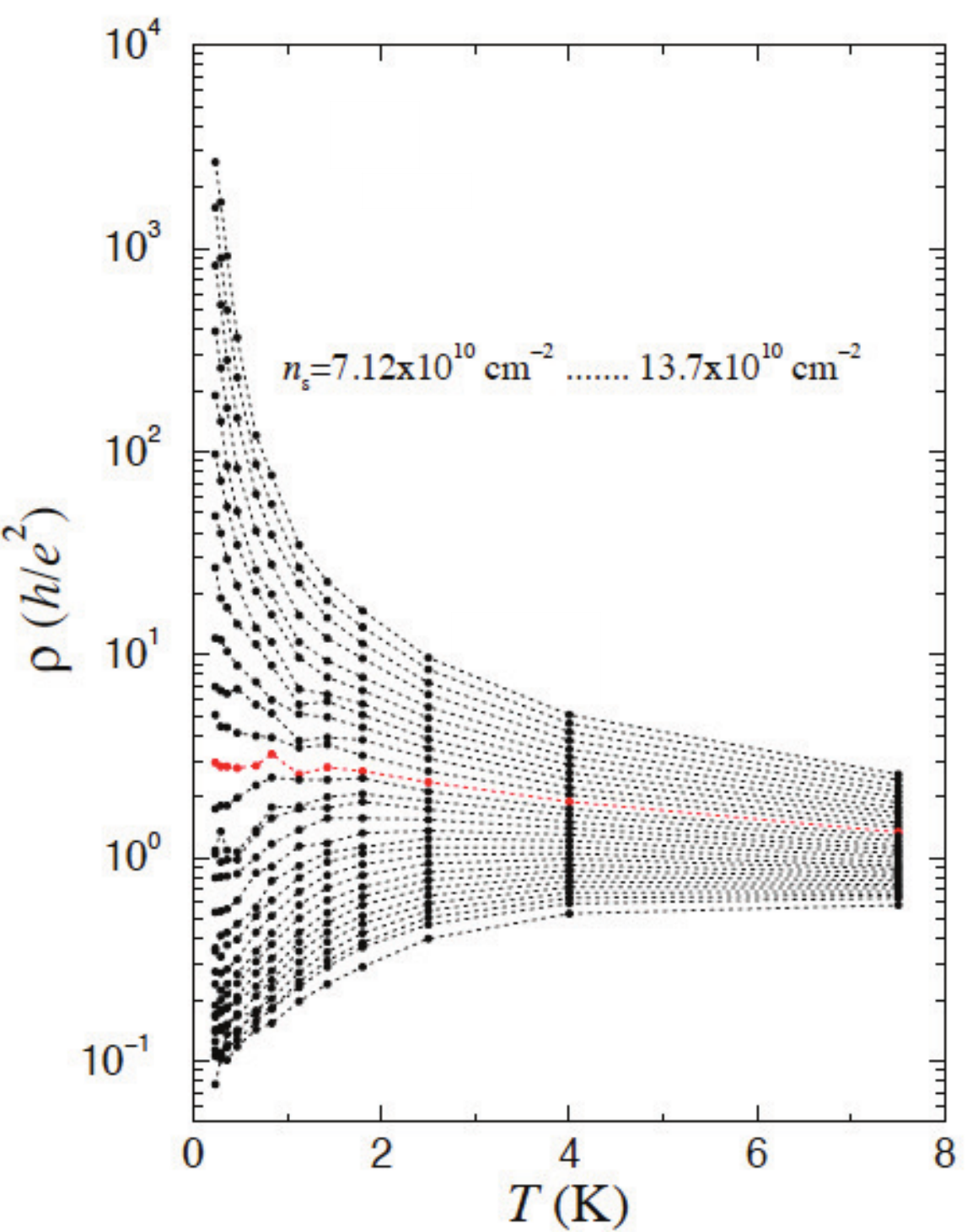}
\includegraphics[width=0.43\columnwidth]{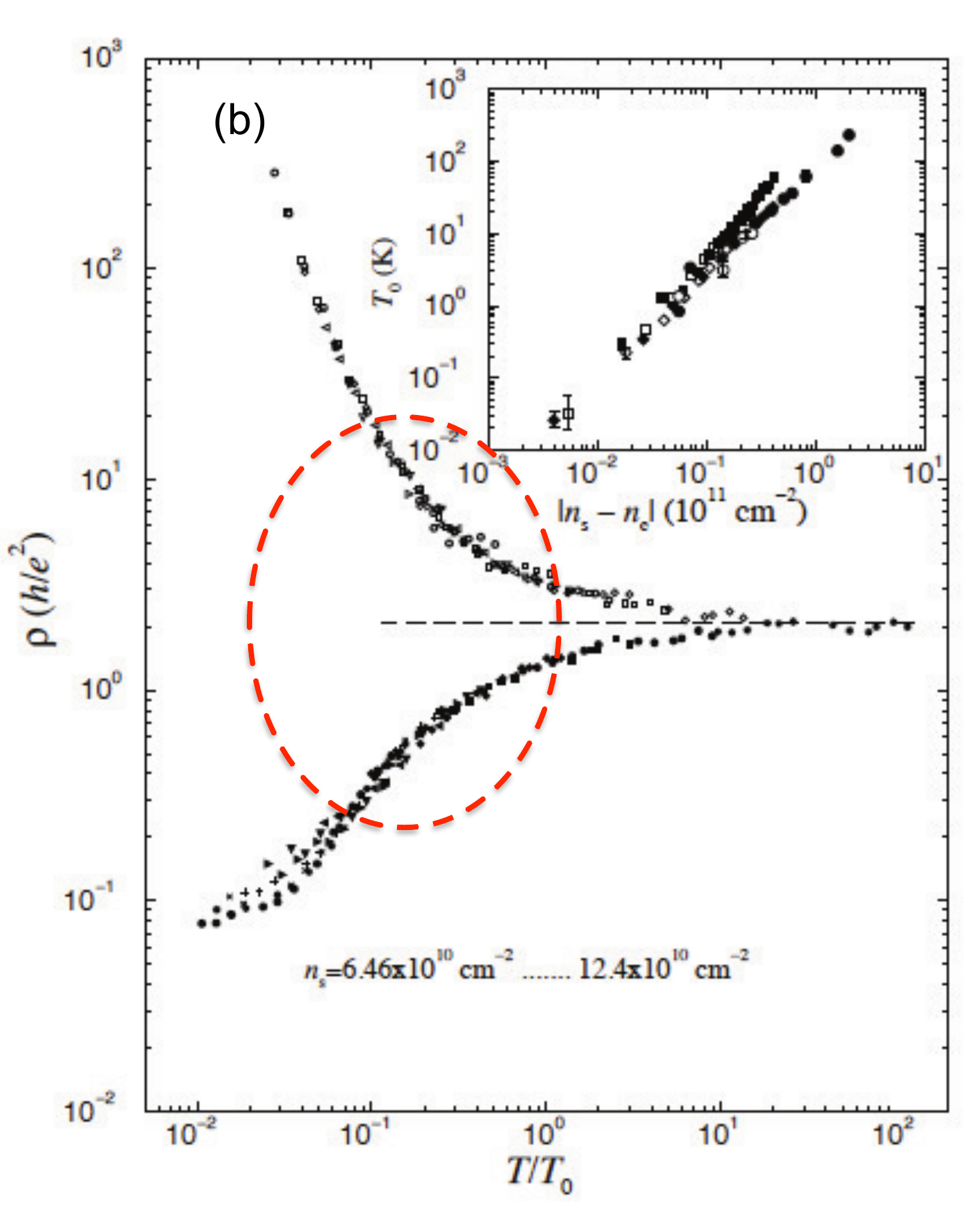}
\end{center}
\caption{(a) Resistivity curves (left panel) for a high mobility 2DEG in silicon. (b) Scaling behavior is found over a comparable temperature range. The remarkable ``mirror symmetry'' of the scaling curves (indicated by red dashed line) seems to hold over more then an order of magnitude for the resistivity ratio (from \cite{kravchenko95}).} \label{2deg curves}
\end{figure}

\subsection{Finite temperature transport}

2DEG devices provide a very convenient setup for the study of MITs, because one can readily control the carrier density in a single device. Thus one easily obtains an entire family of resistivity curves, covering a broad interval of densities and temperatures. Typical results for high mobility 2DEG in silicon are shown in Fig. \ref{2deg curves}(a), showing a dramatic metal-insulator crossover as the density is reduced below $n_{c}\sim10^{11}$cm$^{-2}$.

One can observe a rather weak temperature dependence of the resistivity around the critical density (shown as a red curve in Fig. \ref{2deg curves}(a)), with much stronger metallic (insulating) behavior at higher (lower) densities. Note that the system has ``made up its mind'' whether to be a metal or an insulator even at, relatively speaking, surprisingly high temperatures $T\sim T_{F}\approx10K$. This behavior should be contrasted to that typical of good metals such as silver or gold, where at accessible temperatures $T \ll T_F \sim 10^4 K$. Physically, this means that for 2D-MIT the physical processes relevant for localization already kick in at relatively high temperatures. Here, the system can no longer be regarded as a degenerate electron liquid, with dilute elementary excitations. In contrast, simple estimates show \cite{abrahams01} that around the critical density, the characteristic Coulomb energy (temperature) $T_{Coul} \sim 10 T_F \sim 100K$. Clearly, the transition occurs in the regime where Coulomb interactions dominate over all other energy scales in the problem. We should also mention that in the Kelvin range, there essentially are no phonons in silicon \cite{AFS1982}.
\begin{figure}[h]
\vspace{-6pt}
\begin{center}
\includegraphics[width=0.8\columnwidth]{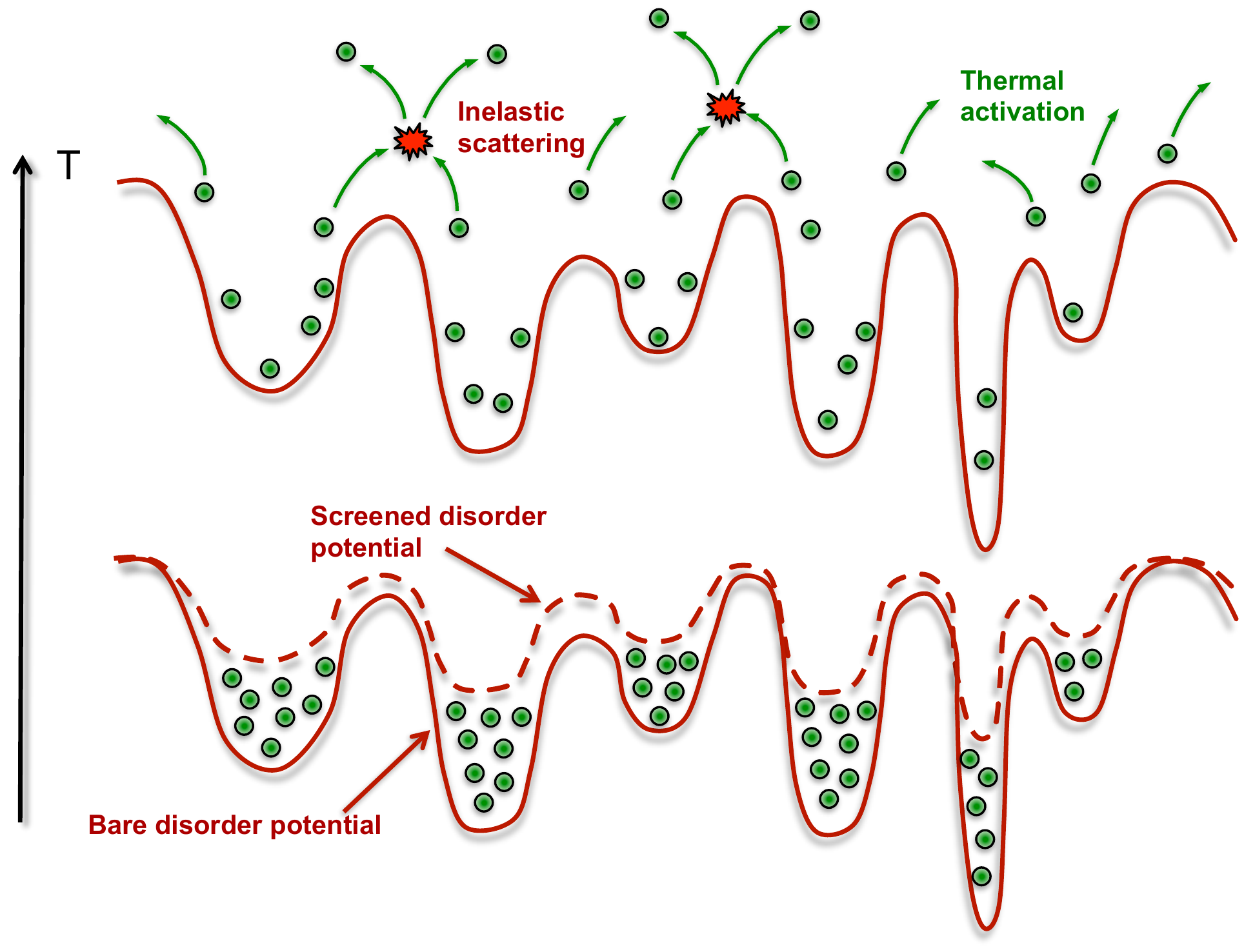}
\end{center}
\caption{In the disordered Fermi Liquid (FL) picture, the leading low-temperature
dependence of transport reflect elastic scattering off a renormalized,
but temperature-dependent random potential (dashed line). At low temperatures
(bottom), the potential wells ``fill-up'' with electrons; in presence
of repulsive (Coulomb) interactions, the screened (renormalized) potential
has reduced amplitude (dashed line), leading to effectively weaker
disorder. As the temperature increases (top), electrons thermally
activate (shown by arrows) out of the potential wells, reducing the
screening effect. This physical mechanism, which operates both
in the ballistic and in the diffusive regime  \protect\cite{Zala}, is at the origin of
all ``quantum corrections'' found within the FL picture \cite{leeramakrishnan}.
It is dominant, provided that inelastic electron-electron scattering
can be ignored. While this approximation is well justified in good
metals, inelastic scattering (star symbol) is considerably enhanced
in presence of strong correlation effects, often leading to disorder-driven
non-Fermi liquid behavior \protect\cite{RoP2005review} and electronic
Griffiths phases \protect\cite{andrade09prl}.}
\label{unscreening}
\end{figure}

We conclude that in the relevant temperature regime transport should be dominated by inelastic electron-electron scattering, and not the impurity-induced (Anderson) localization of quasiparticles, as described by disordered Fermi liquid theories of Finkel'stein and followers \cite{fink-jetp83,fink-jetp84,punnoose02}. As illustrated schematically in Fig. \ref{unscreening}, transport in this incoherent regime has very different character than in the coherent "diffusive" regime found in good metals with disorder. Incoherent transport dominated by inelastic electron-electron scattering is typical of systems featuring strong electronic correlations, such as heavy fermion compounds \cite{stewartrev1} or transition-metal oxides \cite{goddenough63book} close to the Mott (interaction-driven) MIT \cite{mott-book90}. Understanding this physical regime requires not only a different set theoretical tools, but also an entirely different conceptual picture of electron dynamics. As we shall explain below, modern Dynamical Mean-Field Theory (DMFT) methods \cite{dmft96} make it possible to qualitatively and even quantitatively explain most universal features within this incoherent transport regime dominated by strong correlation effects. 

\subsection{Scaling phenomenology and its interpretation}

There is no doubt that dramatic changes of transport arise in a narrow density range around $n \sim n_c$, but this alone is not enough to mark the existence of a second order (continuous) phase transition. In particular, all known classical \cite{goldenfeldbook} and even quantum \cite{sachdevbook} critical points also display the characteristic scaling phenomenology. In the case of MITs, one expects \cite{dobrosavljevic2012conductor} that the resistivity assumes the following scaling form, as a function of the reduced density $\delta_n = (n - n_c)/ n_c$ and temperature
\begin{equation}
\rho^*(\delta_n, T) = F(T/T_o(\delta_n )),
\end{equation}
where $\rho^*(T) =  \rho (\delta_n ,T) / \rho (\delta_n =0,T)$ is the reduced resistivity, and $T_o (\delta_n ) \sim \delta_n^{\nu z}$ is the corresponding crossover temperature. 
While similar families of resistivity curves were reported earlier, a big breakthrough for 2D-MIT occurred when Kravchenko demonstrated \cite{kravchenko95} precisely such scaling behavior, providing strong evidence of quantum critical behavior expected at a sharply defined $T=0$ MIT point. It should be stressed, though, that such scaling behavior must arise at any quantum phase transition (QPT), but this  alone does  not provide direct insight into its physical content or mechanism. 

\begin{figure}[h]
\begin{center}
\includegraphics[width=0.5\columnwidth]{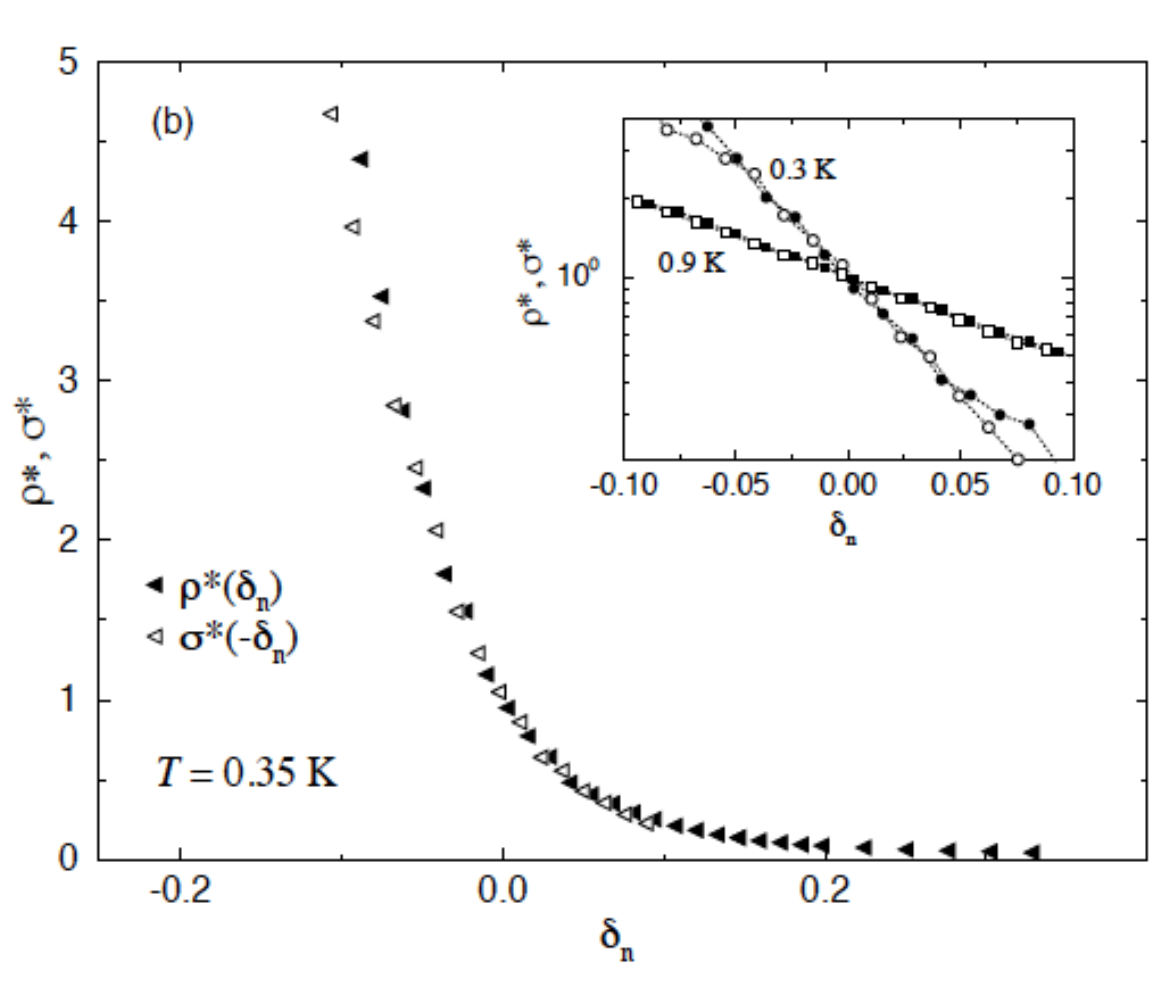}
\end{center}
\vspace{-6pt}
\caption{Experimental data displaying "mirror symmetry" (from \cite{simonian97}). The inset shows the same data on a semi-logarithmic scale, emphasizing linear density dependence of $\ln \rho$ consistent with Eq.~(\ref{stretched}), i.e. the "stretched exponential"  phenomenology \cite{gang4me}.}
\label{duality}
\end{figure}

On the other hand, soon after the discovery of 2D-MIT, another interesting feature was noticed \cite{simonian97}, which is not necessarily expected to hold for any general QPT. By carefully examining the form of the relevant scaling function (see Fig. \ref{2deg curves}(b)), one observes a remarkable "mirror symmetry" of the two branches, such that 
\begin{equation}
\rho^* (\delta_n, T) \approx \sigma^* (-\delta_n, T) = [\rho^* (-\delta_n, T)]^{-1},
\end{equation}
which was found to hold within the entire quantum critical (QC) region, e. g. at $T > T_o (\delta_n )$. In practical terms, this means that in the corresponding QC regime the resistivity assumes a "stretched-exponential" form
\begin{equation}
 \rho^* (\delta_n, T) \sim \exp\{ A \delta_n /T^x\}, \label{stretched}
\end{equation}
where $A$ is a constant, and $x = 1/ \nu z$. Such exponentially strong temperature dependence is not surprising on the insulating side, where it obtains from activated or hopping transport \cite{efros-book}. In contrast, such "inverse activation" behavior is highly unusual on the metallic side, where it reflects the dramatic drop of resistivity at low temperatures, at densities just above the transition.

A plausible physical interpretation was quickly proposed \cite{gang4me} soon after the experimental discovery, based on general scaling considerations. It emphasized that such "stretched exponential" form for the relevant scaling function should be interpreted as a signature of "strong coupling" behavior. Physically, it indicates that the critical region is more akin to the insulating than to the metallic phase. This observation, while having purely phenomenological character, emphasized the key question that one should focus on: the clue to the nature of 2D-MIT should be contained in understanding the physical nature of the insulating state. Is the insulating behavior essentially the result of impurities trapping the mobile electrons, or is the electron-electron repulsion preventing them to move around? If the correlation effects - and not disorder - represent the dominant physical mechanism for localization in other systems, then all the scaling features observed in 2DEG should again be found. Remarkably, this is exactly what is observed (see below) in very recent experiments on organic Mott systems \cite{kanoda2015nphys}. 
\vspace{-6pt}

\subsection{Resistivity maxima in the metallic phase}

\begin{figure}[h]
\begin{center}
\includegraphics[width=0.44\columnwidth]{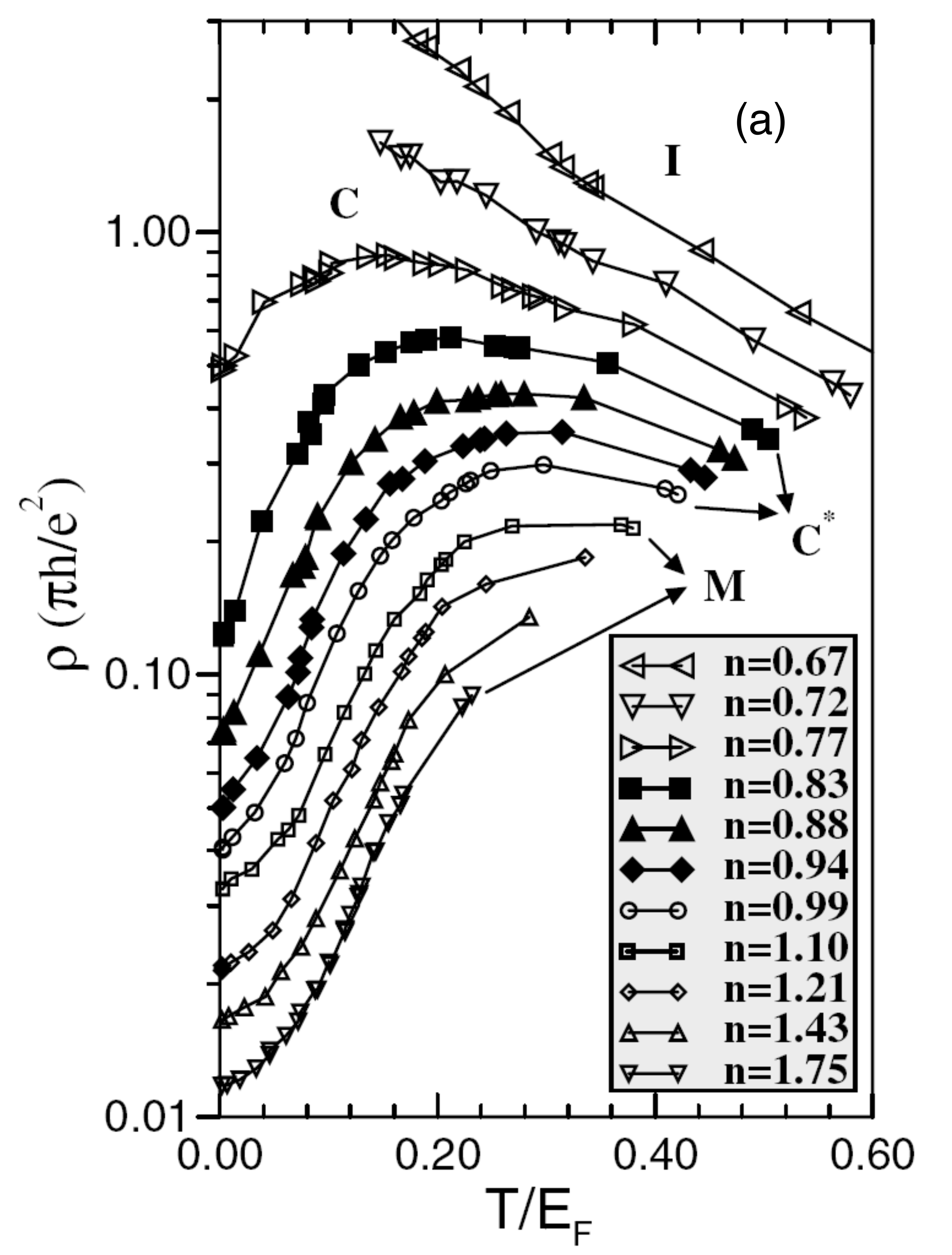}
\includegraphics[width=0.435\columnwidth]{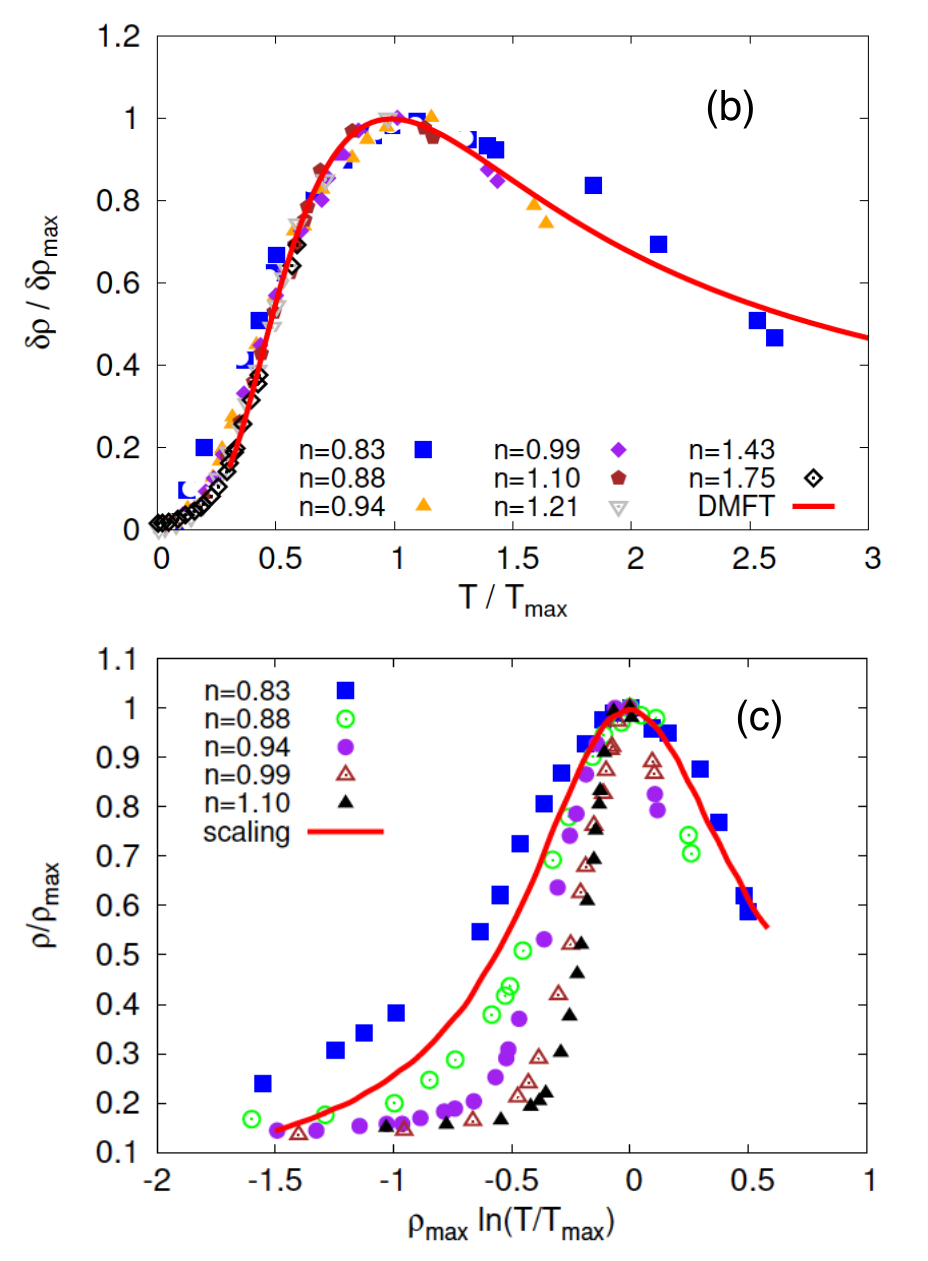}
\end{center}
\vspace{-0.2cm}
\caption{(a) Resistivity as a function of temperature from the experiments on 2DEG in silicon \cite{pudalov98} (from \cite{punnoose02}). Only data points indicated by full black symbols have been estimated to lie in the diffusive (coherent) regime, while all metallic curves have an almost-identical shape. (b) Same data scaled according to the procedure based on the Wigner-Mott (disorder-free) picture (from \cite{radonjic12prb}). (c) The scaling procedure based on the disorder screening scenario \cite{punnoose02} does not produce a convincing collapse \cite{radonjic12prb}. In both scaling plots, the full line is the result of the respective theoretical calculations.}
\label{resmaxima}
\end{figure}

The "quantum critical" resistivity scaling has been observed close to the critical density, but other interesting features are also found, further out on the metallic side. Here, especially in the cleanest samples, one observes increasingly pronounced resistivity maxima at temperatures $T=T_{max} (\delta )$, which decrease as the transition is approached. Remarkably, the family of curves displaying such maxima persists even relatively deep inside the metallic phase, far outside the critical region. However, even a quick glance at the experimental data (Fig. \ref{resmaxima}) reveals that all such curves have essentially the same qualitative shape, and only differ by the density-dependent values of $T_{max}$ and $\rho_{max} = \rho (T_{max} )$.

While this phenomenology is clearly seen, its origin is all but obvious, since resistivity maxima can arise in many unrelated situations. One possibility is the competition of two different disorder-screening mechanisms, as envisioned within  the Finkel'stein picture of a disordered Fermi liquid \cite{punnoose02}. However, this mechanism applies (as the authors themselves pointed out) only within the disorder-dominated diffusive regime, which is confined to low temperatures and sufficiently strong impurity scattering. According to reasonable estimates \cite{punnoose02} for the experiment of Fig. \ref{resmaxima}, this picture should only apply to the three curves (full black symbols) close to the critical density, and completely different physics should kick-in deeper in the metallic regime (open symbols). This seems unreasonable, since the experimental curves show very similar form at all metallic densities. The key test of these ideas, therefore, should be to examine different materials with strong correlations and weaker disorder, where most of the relevant temperature range is well outside the diffusive regime. If the resistivity maxima, with similar features, persist in such systems, their origin must be of completely different nature that the disorder screening mechanism described by the disordered Fermi liquid picture. 

\begin{figure}[b]
\begin{center}
\includegraphics[width=0.4\columnwidth]{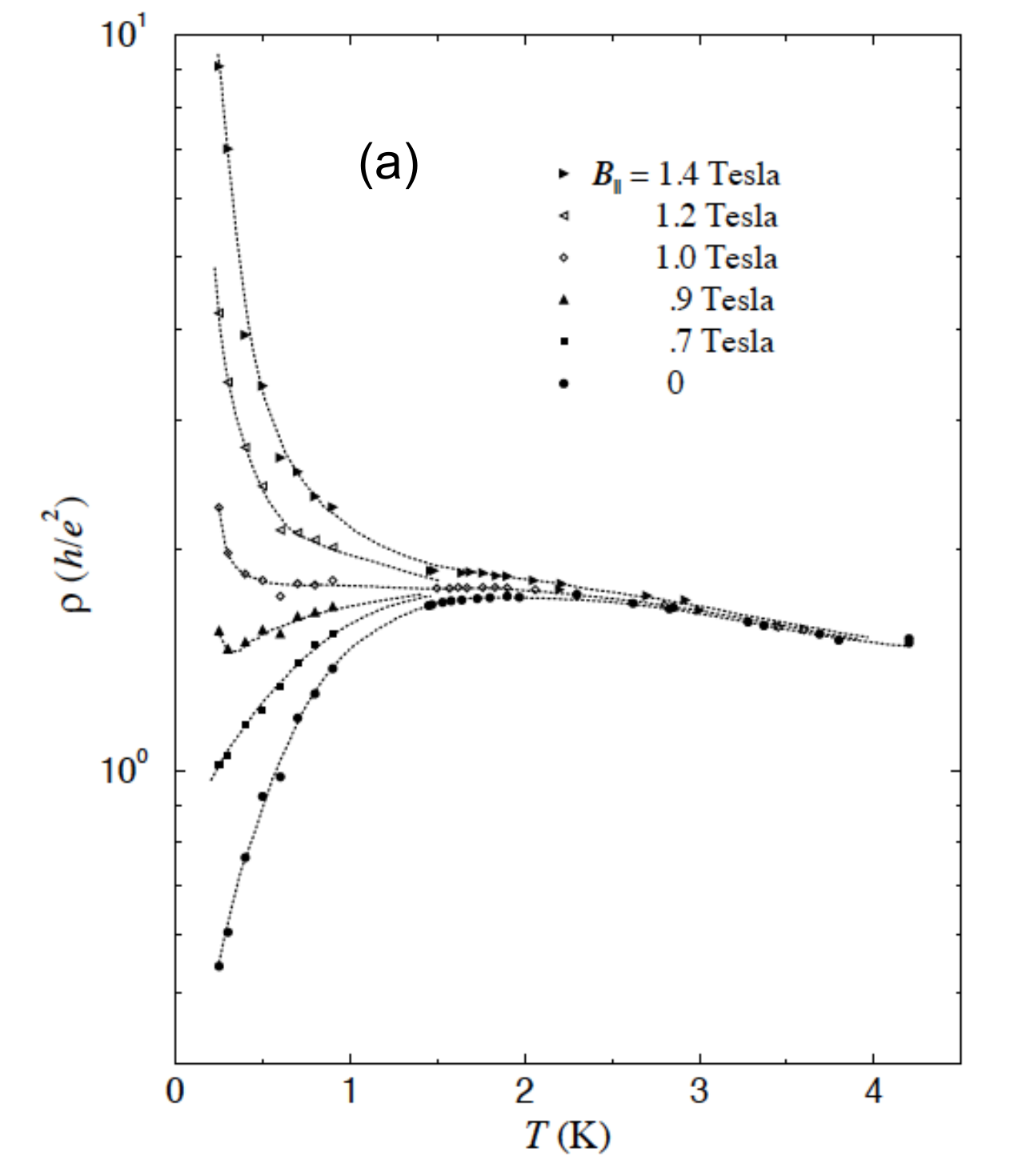}
\includegraphics[width=0.55\columnwidth]{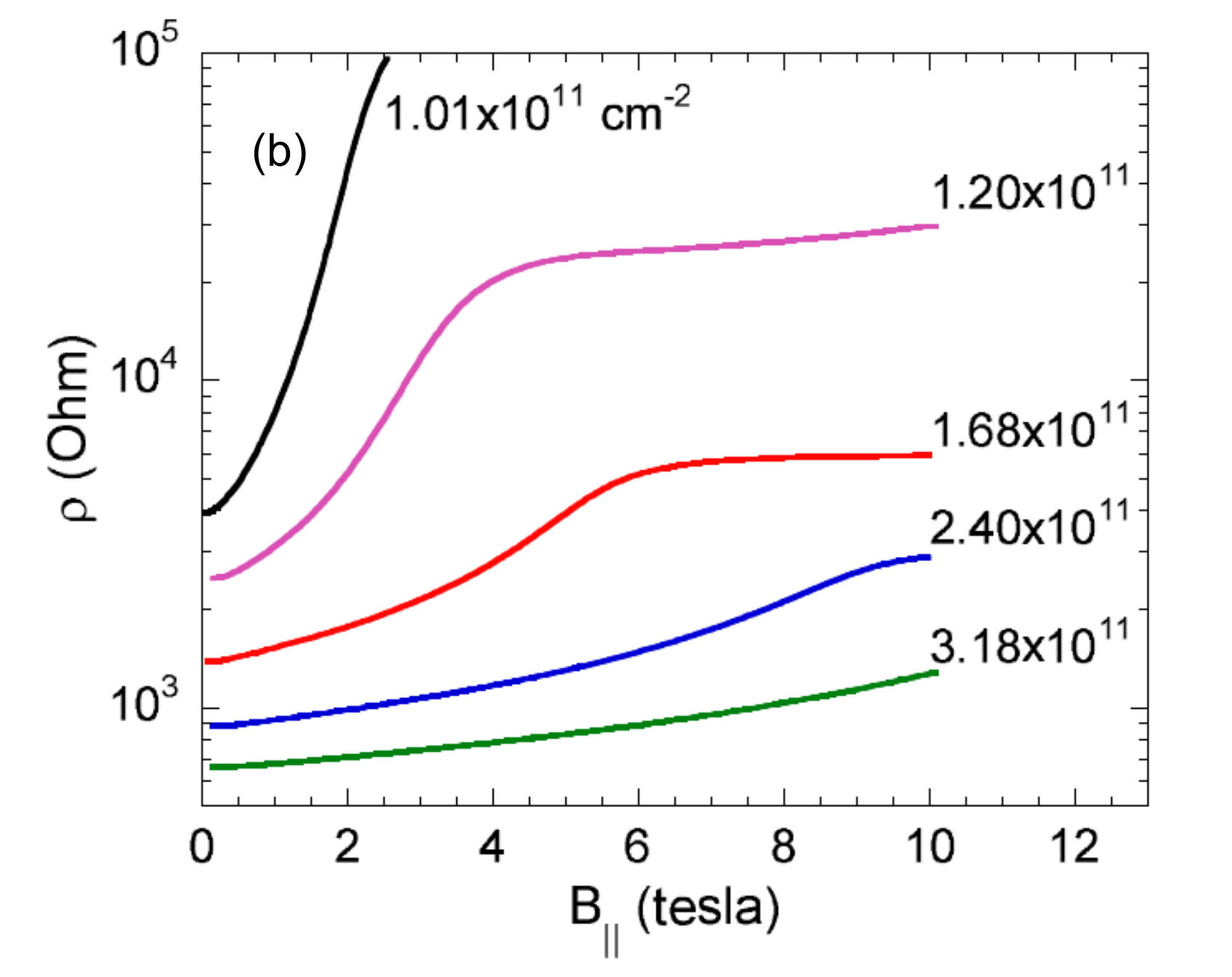}
\end{center}
\caption{(a) Resistivity versus temperature for five different fixed magnetic fields applied parallel to the plane of a low-disordered silicon MOSFET. Here $n =$ 8.83x10$^{10}$cm$^{-2}$ (from \cite{simonian1997}). (b) Low-temperature magnetoresistance as a function of parallel magnetic field, at different electron densities above $n_c$ (from \cite{shashkin2001}).}   
\label{magnetoressitance}
\end{figure}

A related issue is the origin of the pronounced resistivity drop at $ T < T_{max}$, which can be as large as one order of magnitude. Here again very different theoretical interpretations have been proposed, which may or may not apply in a broad class of systems. One issue is the role of impurities \cite{punnoose02}, which may be addressed by experimentally examining materials with weak or negligible amounts of disorder. Another curious viewpoint is that of Kivelson and Spivak \cite{spivak05}, who proposed that the microscopic phase coexistence (bubble and/or stripe phases) between the Wigner crystal and a Fermi liquid may explain this behavior, even in absence of disorder. This possibility may be relevant in some regimes, but certainly not in lattice systems such as heavy fermion compounds, or systems displaying the (interaction-induced) Mott transition, in absence of any significant disorder.

\subsection{Effect of parallel magnetic fields}

One of the most important clues about 2D-MIT have been obtained soon following its discovery, from studies of magneto-transport. While much of the traditional work on 2DEG systems focuses on the role of perpendicular magnetic fields and the associated Quantum Hall regime, for 2D-MIT the application of magnetic fields parallel to the 2D layer shed important new light. It was found \cite{simonian1997} that the application of parallel magnetic fields can dramatically suppress (Fig. \ref{magnetoressitance}) the resistivity maxima and the pronounced low-temperature resistivity drop on the metallic side of the transition. This finding is significant, because  parallel fields couple only to the electron spin (Zeeman splitting), and not to the orbital motion of the electron. It points to the important role of spin degrees of freedom, in stabilizing the metallic phase. 
\begin{figure}
\begin{center}
\includegraphics[width=0.7\columnwidth]{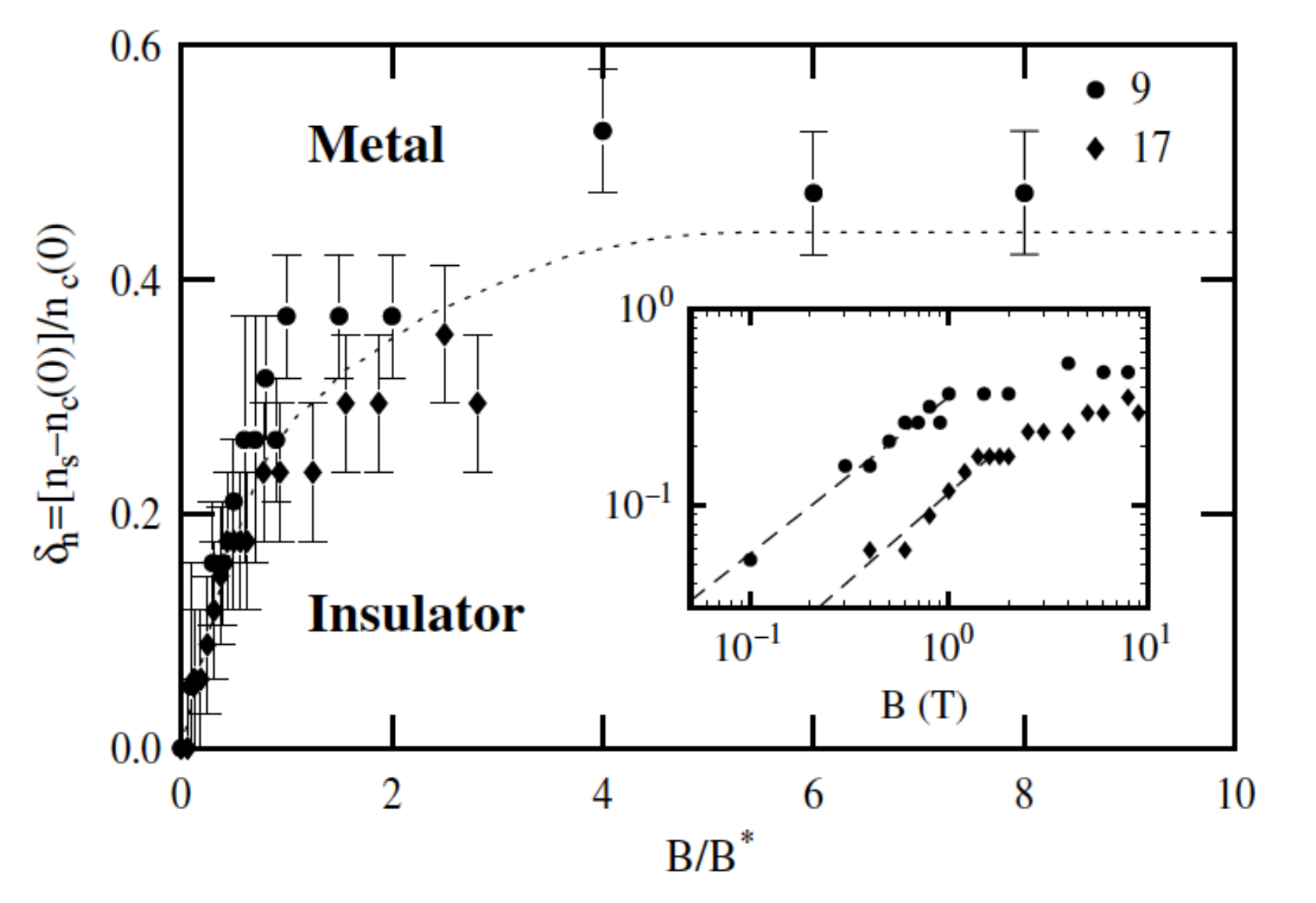}
\end{center}
\caption{Phase diagram for high mobility 2DEG in silicon (from \cite{eng2001}), as a function of electron density and parallel magnetic field. The field-driven MIT is found only in the narrow density range $0 < \delta_n < 0.4$, close to the $B=0$ critical density.}   
\label{magnetic phase diagram}
\end{figure}

An interesting fundamental issue relates to the precise role of the Zeeman splitting on the ground state of 2DEG. Is an infinitesimally small value of the parallel field sufficient to suppress the true metal at $T=0$, or is there a field-driven MIT at some finite value of the parallel field? This important question was initially a subject of much controversy, but careful experimental investigation at ultra-low temperatures have established \cite{shashkin2001,eng2001} the existence of a field-driven MIT (Fig. \ref{magnetic phase diagram}) at finite parallel field, but only within a density range close to the $B=0$ MIT. Interestingly, this field-driven transition appears to belong to a different universality class  \cite{eng2001})  than the zero-field transition (see also the chapter by D. Popovi\'c), somewhat similarly to what happens in bulk doped semiconductors \cite{sarachik95}. 

\subsection{Thermodynamic response}

The early evidence for the existence of 2D-MIT relied on scaling features of transport in zero magnetic field \cite{abrahams01}, but despite the obvious beauty and elegance of the data, some people \cite{maslov2001} remained skeptical and unconvinced. Complementary insight was
 therefore of crucial importance, and in the last fifteen years this sparked a flurry of experimental work focusing on thermodynamics response. Detailed account of this large and impressive body of work has been given elsewhere \cite{kravchenko2004}; it is also covered in the chapter by Shashkin and Kravchenko. Several outstanding features have been established by these experiments, as follows.

\begin{figure}[h]
\begin{center}
\includegraphics[width=0.525\columnwidth]{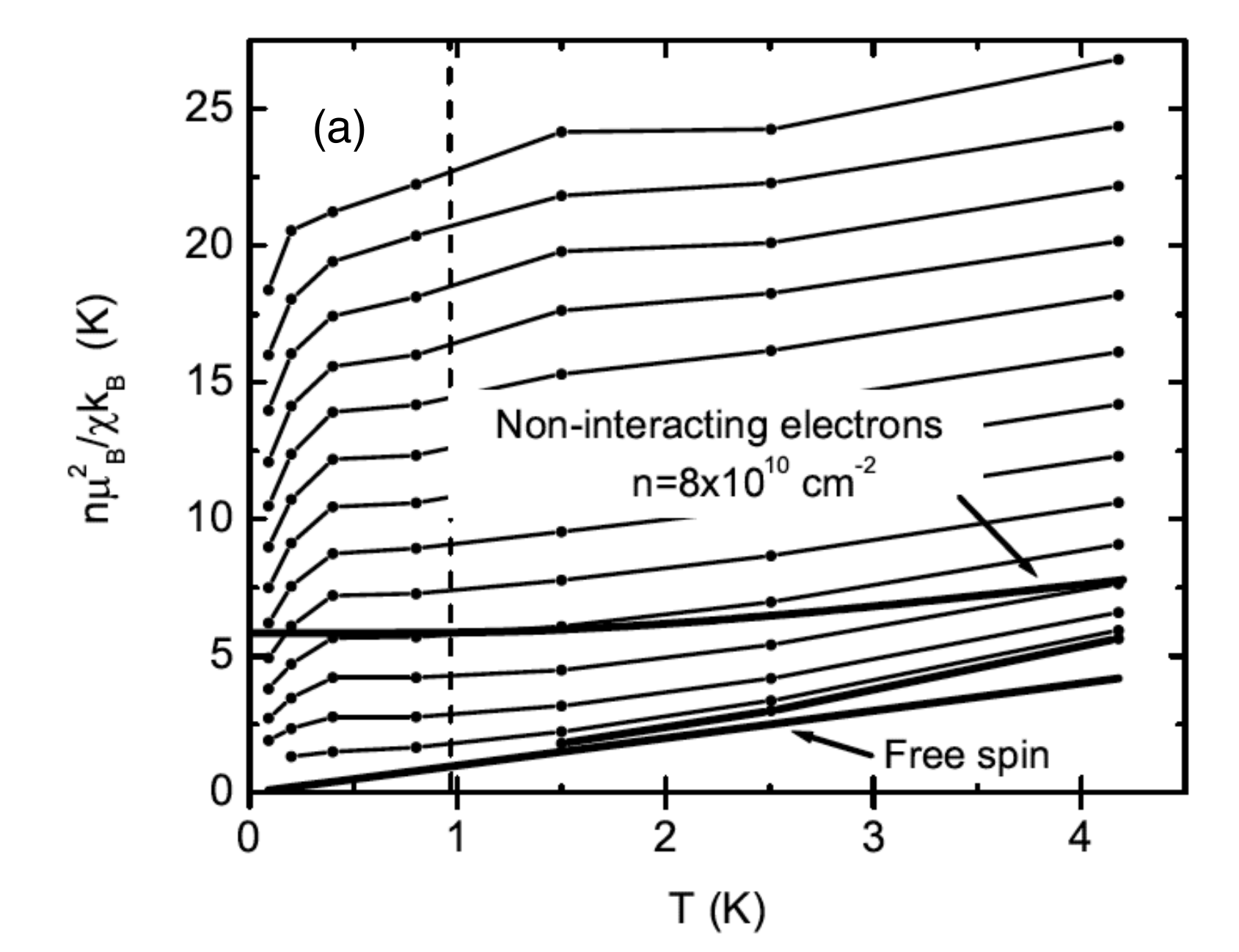}\includegraphics[width=0.5\columnwidth]{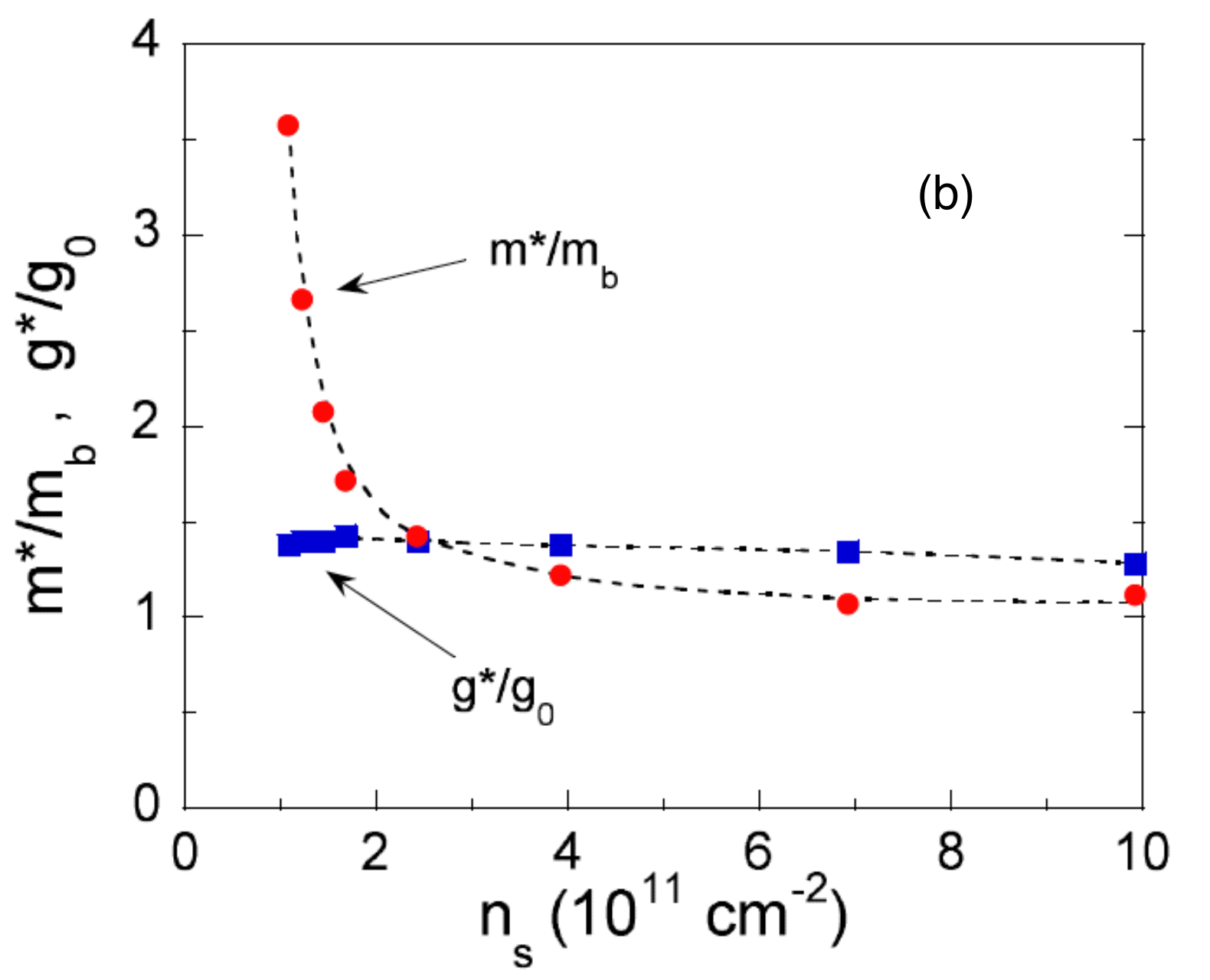}
\end{center}
\vspace{-12pt}
\caption{(a) Inverse spin susceptibility normalized per carrier, as obtained from magneto-capacitance experiments 
(from \cite{reznikov2003}). Experimental curves from bottom to top correspond to densities $0.8  - 6$x$10^{11}$cm$^{-2}$ in  4x$10^{10}$cm$^{-2}$ steps. The thick straight line depicts the Curie law (slope given by the Bohr magneton), and the dashed line marks $T = (g \mu_B /k_B ) \times 0.7$ T. (b) Renormalization of the effective mass  (dots)  and the g-factor  (squares),  as a function of electron density (adapted from \cite{shashkin2002}).}
\label{thermodynamic}
\end{figure}
%

\begin{enumerate}

\item The spin susceptibility is strongly enhanced at low temperatures, and it seems to evolve \cite{reznikov2003}  from a Pauli-like form at higher densities (typical for metals), to a Curie-like form (expected for localized magnetic moments) as the transition is approached. This result indicates that most electrons convert into spin-1/2 localized magnetic moments within the insulating state. 

\item Various complementary experimental methods have firmly established  \cite{kravchenko2004}  pronounced interaction-induced enhancement of the (quasiparticle) effective mass $m^*$, which appears to linearly diverge at $n = n_c$. In contrast, the corresponding g-factor does not display any significant renormalization. This important observation rules out incipient ferromagnetism as a possible origin of enhanced spin susceptibility.  Instead, it points to local magnetic moment formation within the insulating phase. These earlier findings were recently confirmed by spectacular thermo-power measurements, where values of $m^* /m$ as large as 25 were reported \cite{kravchenko2012}. \vspace{6pt}

\item The vast majority of the relevant experiments in this category have been performed within the ballistic regime. This is significant, because the predictions of the disordered Fermi-liquid picture of Finkel'stein \cite{punnoose02} apply only within the diffusive regime, and thus cannot provide any insight into the origin of these thermodynamic anomalies.\vspace{6pt}

\item All these thermodynamic signatures are best seen in the cleanest samples, and they display  surprisingly weak dependence on the sample mobility (level of disorder). This again points to strong correlations as the dominant mechanism in this regime.   

\end{enumerate}

\vspace{-18pt}

\section{Comparison to conventional Mott systems}

Experiments on 2DEG systems have suggested that 2D-MIT should best be viewed as an interaction driven MIT, where the insulator consists of localized magnetic moments. To put these ideas in perspective, it is useful to compare these findings to the properties of other 2D systems known to display the Mott (interaction-driven) MIT \cite{mott1949,mott-book90}.

\begin{figure}[h]
%
%
\begin{center} 
\includegraphics[width=0.7\columnwidth]{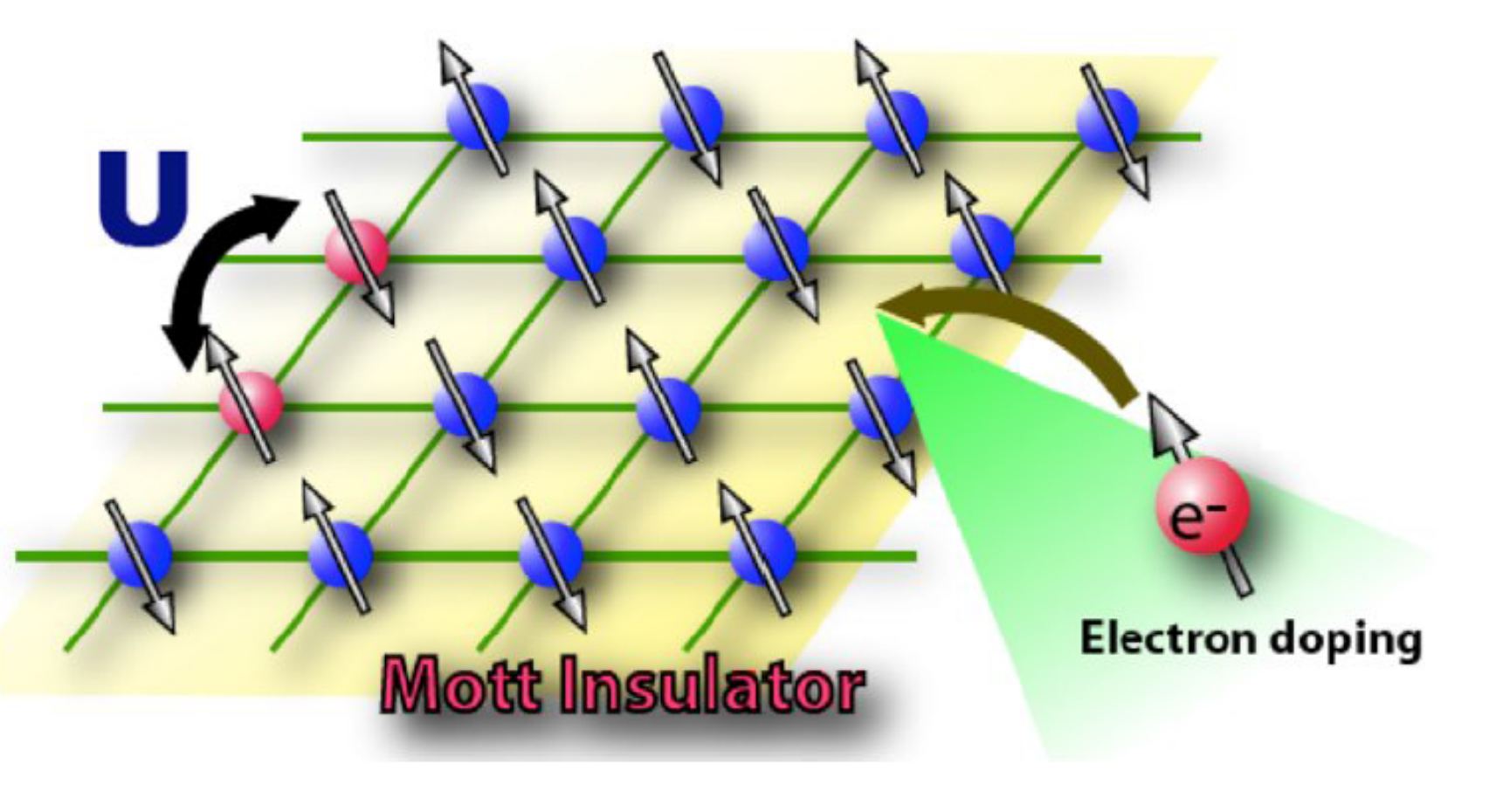}
\end{center}
%
%
\caption{In a Mott insulator, each electron resides in a bound electronic state, forming a spin-1/2 local magnetic moment. The lowest energy charge excitation costs an energy  comparable to the on-site Coulomb repulsion $U$.}
\label{mott insulator}
\end{figure}

Mott insulators often display some form of magnetic order at low temperatures, but in contrast to the band picture \cite{slater34rmp}, their insulating nature is not tied down to magnetic order. Indeed, a substantial gap in charge transport is typically seen even at temperatures much higher than the magnetic ordering temperature. Such a quantum phase transition \cite{dobrosavljevic2012conductor} between a paramagnetic FL metal and a paramagnetic Mott insulator - the "pure" Mott transition - is our main focus here. In particular, we shall focus on two specific examples of the broad family of Mott materials, where it proved possible to experimentally study several features with striking similarity to the behavior found in 2DEG displaying 2D-MIT. 

\subsection{Mott transition in $^3$He monolayers on graphite}

The $^3$He liquid consists of charge-neutral spin-1/2 atoms with Fermi statistics; its normal phase at ambient pressure is often regarded as a model system for a strongly correlated Fermi liquid \cite{vollhardt84rmp}. This includes significant enhancements of the specific heat and the spin susceptibility, reflecting the emergence of heavy quasiparticles \cite{pinesnozieres}, as first predicted by Landau \cite{landauFL1}. $^3$He will solidify under increasing hydrostatic pressure, due to strong inter-atomic repulsion at short distances. Each spin-1/2 $He^3$ atom then sits on a different lattice site, thus forming a Mott insulator.

The associated solidification transition has long been viewed  as a realization of the Mott MIT. Many of its low temperature properties, on the fluid side, can be described \cite{vollhardt84rmp} by the variational solution of the Hubbard model  based on the Gutzwiller approximation  \cite{brinkmann70prb}, predicting the divergence of the effective mass as
\begin{equation}
m^* /m \sim (P_c - P)^{-1}.
\end{equation}
However, similarly as in most other freezing transitions, bulk $^3$He displays a rather robust first-order solidification phase transition, associated with the emergence of  ferromagnetic (FM) order within the crystalline phase. Consequently, one is able to observe only a restricted range of $m^*$ values, making it difficult to compare to theory. Despite these limitations, many qualitative and even quantitative features can be understood from this perspective, including the findings that application of strong magnetic fields may trigger the destruction of the FL state \cite{vollhardt84rmp} and the associated field-induced solidification of $^3$He. 

More recent work by Saunders and collaborators \cite{casey03prl} has focused on $^3$He fluid monolayers placed on graphite, where again one can study the approach to the Mott transition by controlling hydrostatic pressure. Most remarkably, the first order jump at solidification is then suppressed, presumably because the solid $^3$He  monolayer on graphite displays no magnetic order down to $T=0$. Instead, nonmagnetic spin-liquid behavior is found within the solid phase, due to important ring-exchange processes in this 2D triangular lattice \cite{misguich98prl}. The corresponding Mott transition is found to be of second order, with rather spectacular agreement with Brinkmann-Rice (BR) theory for the $m^*$ divergence. According to FL theory \cite{pinesnozieres}, the Sommerfeld coefficient $\gamma = C/T \sim m^*$ depends only on the effective mass, while the spin susceptibility $\chi \sim m^* /(1+F_o^a) = m^* g^*$ also involves the (renormalized) g-factor. This is significant, because the behavior of the g-factor distinguishes the Brinkmann-Rice (Mott) transition from a FM instability, although $\chi$ diverges in both instances. Namely, $g^*$ is expected to diverge at the FM transition \cite{pinesnozieres}, while it should remain constant at the BR point \cite{vollhardt84rmp}.

To address this important issue, it suffices to determine both $m^*$ and $g^*$, or equivalently to measure both $\gamma$ and $\chi$. Precisely such an analysis was carried out for $^3$He monolayers \cite{casey03prl}, where $m^*$ enhancement as large as 15 was reported, while $g^*$ was found to remain constant in the critical region, and assume a value close to the BR prediction. This finding is strikingly similar to what is observed in analogous studies \cite{kravchenko2004} for ultra-clean 2DEG systems close to 2D-MIT; it suggests that both phenomena may have a common physical origin: the behavior of a strongly correlated Fermi liquid in the proximity of the Mott insulating phase.   

Transport properties are more difficult to probe in $^3$He, because its charge neutrality precludes direct coupling of electric fields to density currents. One cannot, therefore, easily compare its transport properties to that of 2DEG systems, in order to further test the Mott picture. To do this, we next turn to other classes of "canonical" Mott systems where recent work has provided important insights. Over the last thirty years, many studies of the Mott transition have focused on various transition metal oxides \cite{goddenough63book}, especially after the discovery of high temperature superconductivity in late 1980s. These materials, however, are notorious in displaying all kinds of problems in material preparation and sample growth, and often contain substantial amount of impurities and defects. Another class of systems where better sample quality is somewhat easier to obtain are the so-called organic Mott systems, which we discuss next.

\begin{figure}[h]
\vspace{-6pt}
\begin{center}
\includegraphics[width=0.45\columnwidth]{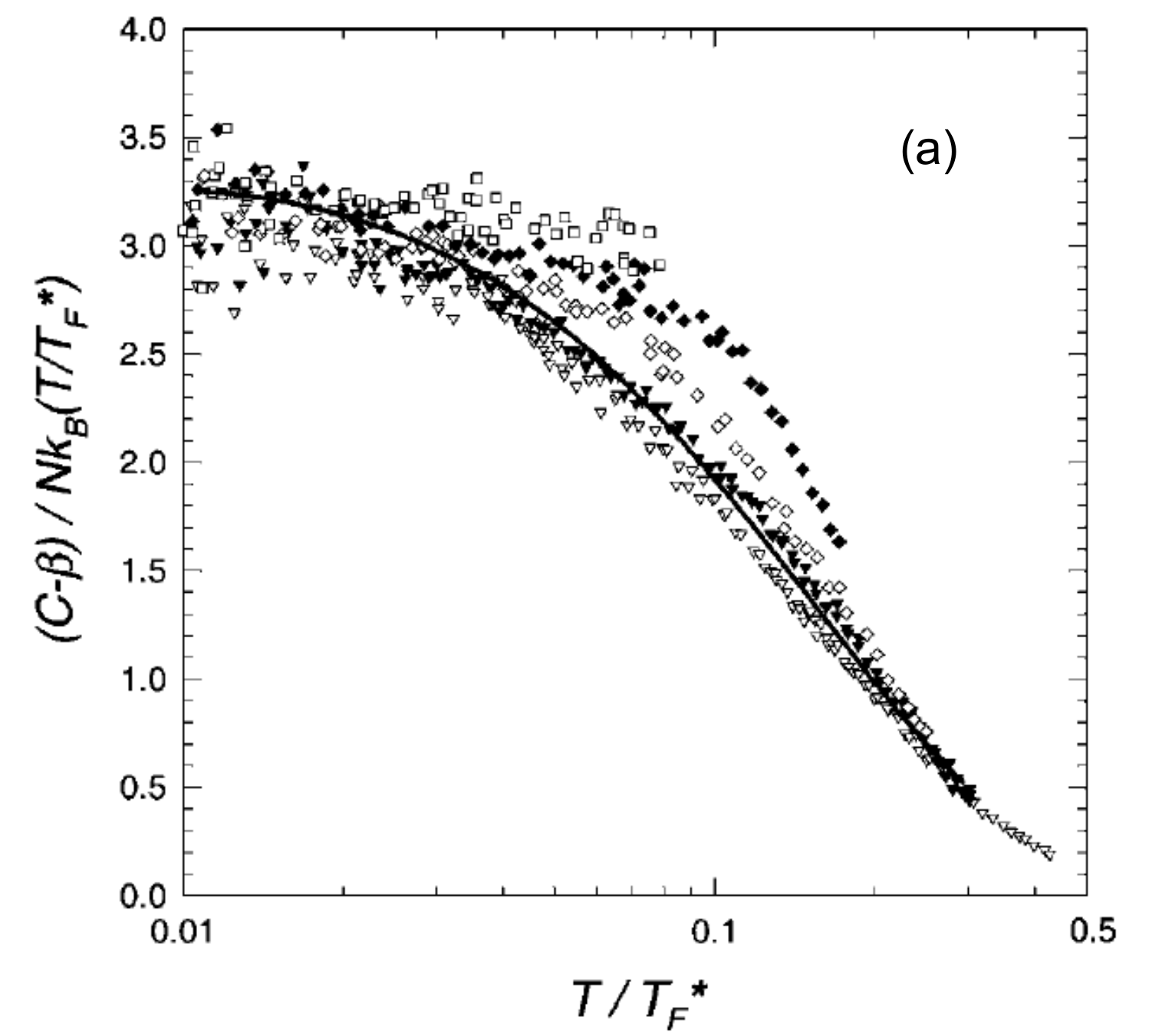}
\includegraphics[width=0.44\columnwidth]{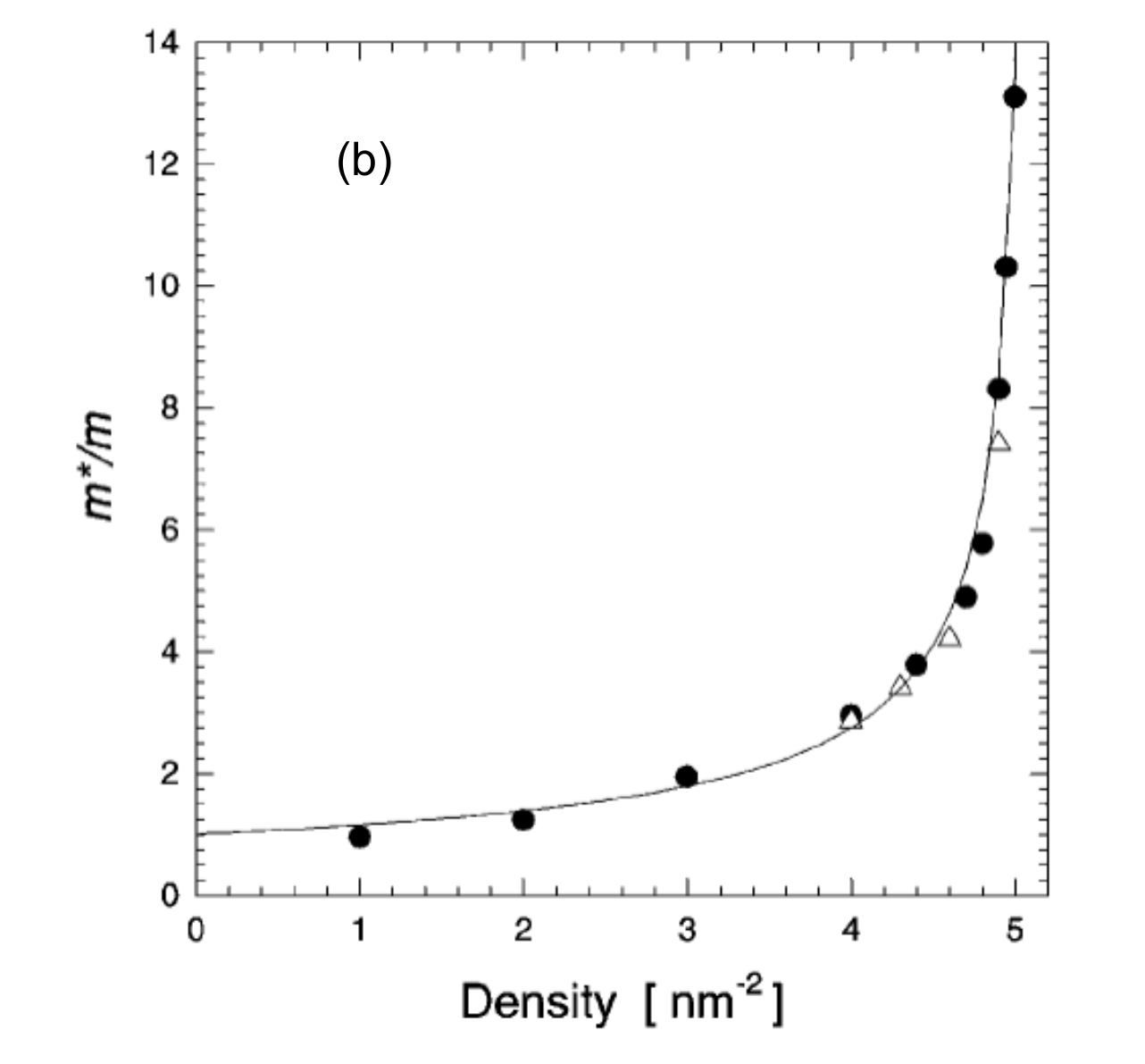}
\end{center}
\vspace{-6pt}
\caption{Approaching the Mott transition in $^3$He monolayers on graphite (from \cite{casey03prl}). (a) Reduced fluid heat capacity as a function of reduced temperature, showing emergence of Fermi liquid at low temperatures. Here $T^*_F =  T_F /(m^* /m)$ is the quasiparticle Fermi temperature. (b) Effective mass ratio as a function of $^3$He fluid density inferred from heat capacity (full circle) and magnetization (open triangles), showing apparent divergence. The fact that both quantities display the same critical behavior indicates that $g^* \sim const.$} 
\label{3He}
\end{figure}

\subsection{Mott organics}

The organic Mott systems have emerged, over the thirty years, as a very popular set of materials for the study of the Mott MIT. These are organic charge-transfer salts consisting of a quasi two-dimensional lattices of rather large organic molecules, with typically a single molecular orbital close to the Fermi energy. Since the inter-molecular overlap of such orbitals is typically modest, these crystals usually have very narrow electronic bands with substantial intra-orbital (on-site) Coulomb repulsion, hence are well described by two-dimensional single band Hubbard-type models \cite{mckenzie2011review}. 

\begin{figure}[h]
\begin{center}
\includegraphics[width=0.6\columnwidth]{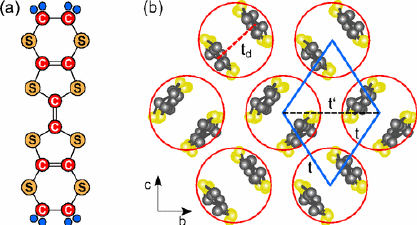}
\end{center}
\caption{Sketch of the BEDT-TTF molecule (from \cite{elsasser2012}). (b) For $\kappa$-(BEDT-TTF)$_2 X$, the molecules are arranged in dimers, which constitute an anisotropic triangular lattice within the conduction layer.} 
\label{mott organics crystal}
\end{figure}

There exist two general families of organic Mott crystals, the $\kappa$-family corresponding to a half-filled Hubbard model, and the $\theta$-family, corresponding to quarter filling. The former class is particularly useful for the study of the bandwidth-driven Mott transition at half filling, since the electronic bandwidth can be conveniently tuned via hydrostatic pressure. A notable feature of these materials is a substantial amount of magnetic frustration, due to nearly isotropic triangular lattices formed by the organic molecules within each 2D layer. As a result, several members of this family (such as $\kappa {\mathrm{- (BEDT-TTF)_2Cu_2(CN)_3}}$) display no magnetic order down to the lowest accessible temperatures, while others  (such as 
$\kappa {\mathrm { - (BEDT-TTF)_2Cu[N(CN)_2]Cl}}$) 
feature antiferromagnetic (AFM) order in the insulating phase. Another important property of these materials is the excellent quality of crystals, with very little disorder, making the dominance of strong correlation effects obvious  in all phenomena observed. 

\begin{figure}[t]
\begin{center}
\includegraphics[width=0.46\columnwidth]{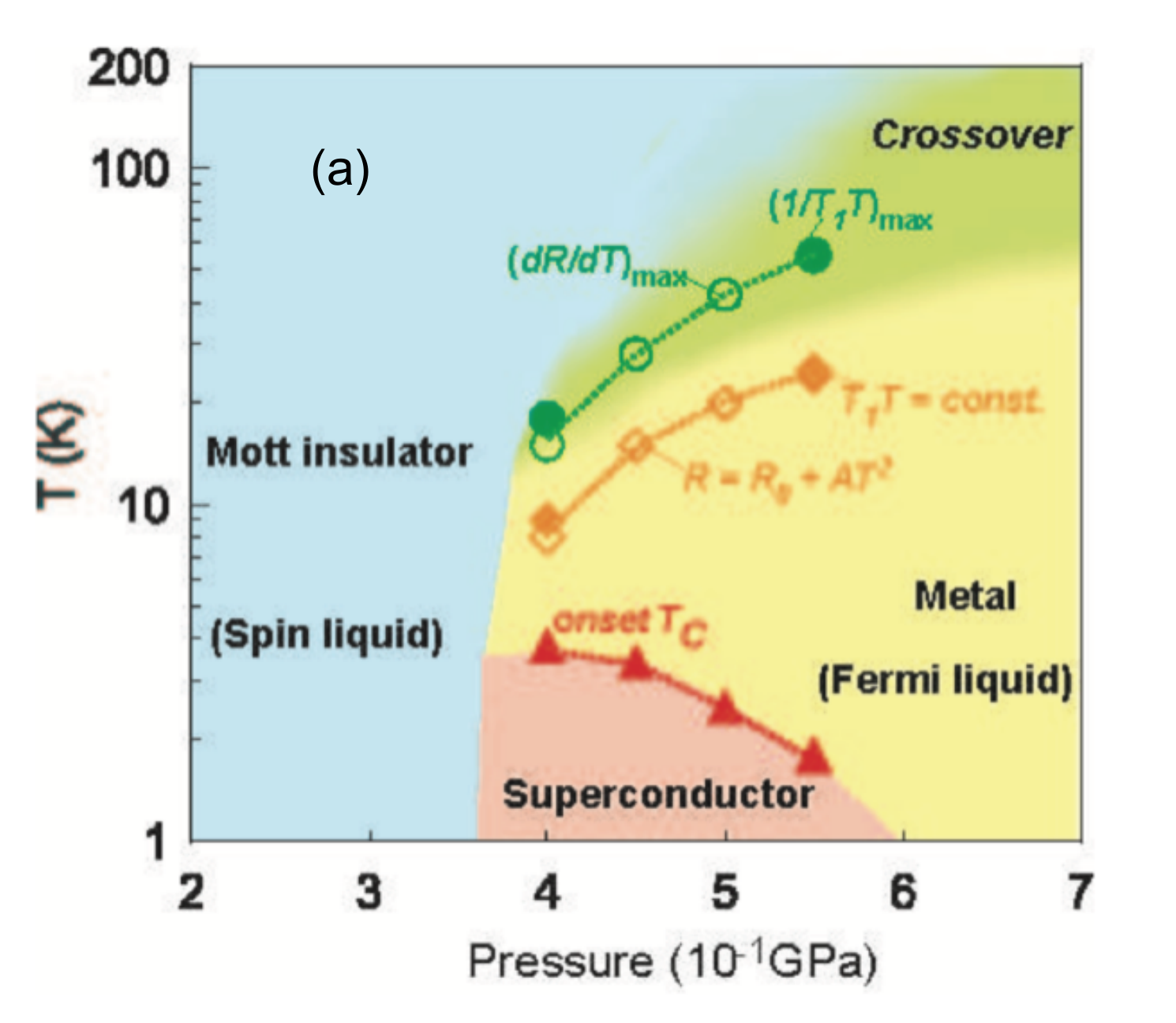}
\includegraphics[width=0.51\columnwidth]{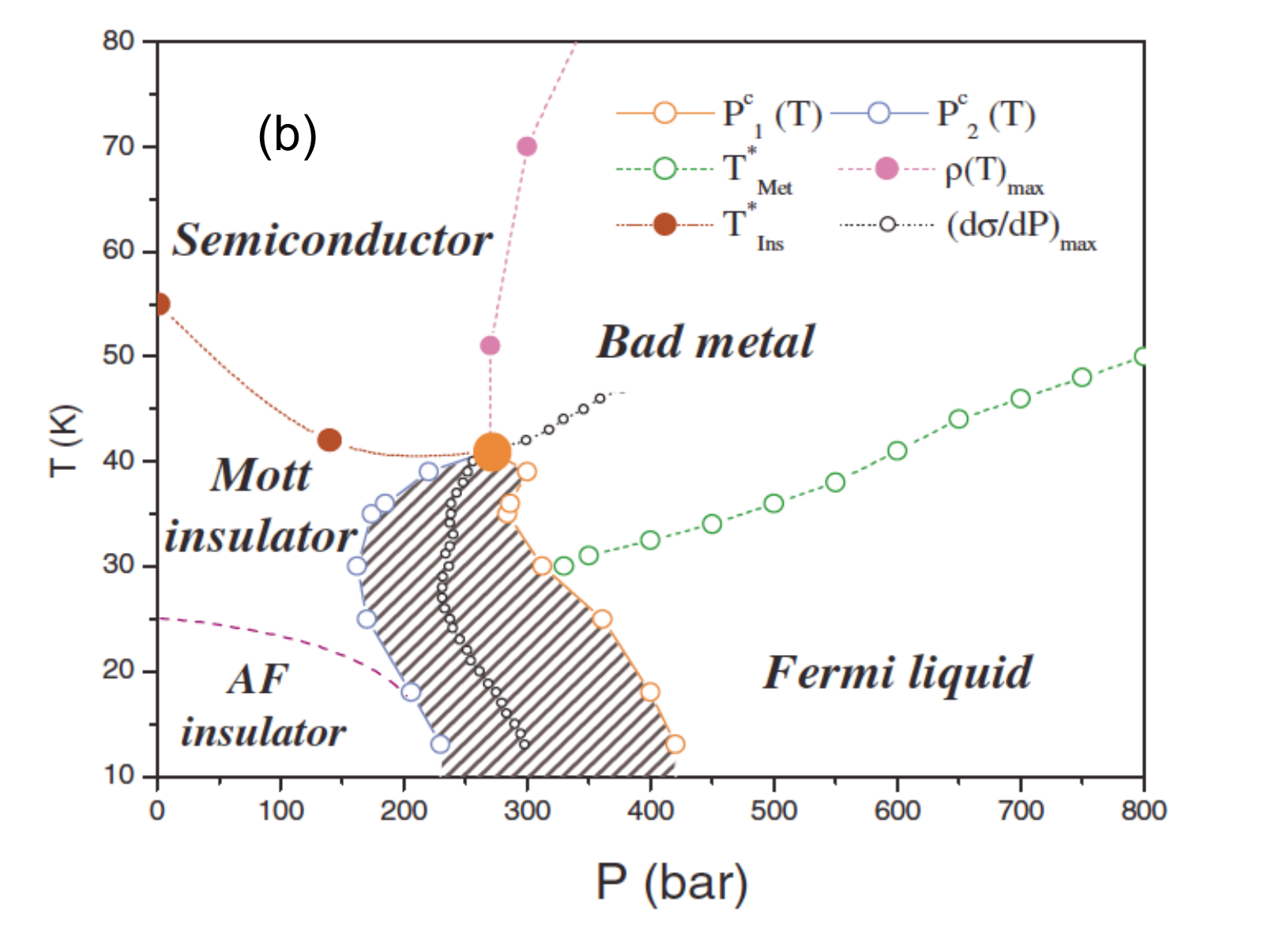}
\end{center}
\caption{Phase diagrams of (a) $\kappa - ({\mathrm {BEDT-TTF}})_2{\mathrm{Cu}}_2({\mathrm{CN}})_3$ displaying no magnetic order (from \cite{kanoda2005prl}), and (b) $κ\kappa - ({\mathrm{BEDT-TTF}})_2{\mathrm{Cu}}[{\mathrm{N}}({\mathrm{CN}})_2]{\mathrm{Cl}}$ featuring AFM order on the insulating side (from \cite{limelette03prl}).} 
\label{phase_diagrams}
\end{figure}

\begin{figure}[b]
\begin{center}
\includegraphics[width=0.49\columnwidth]{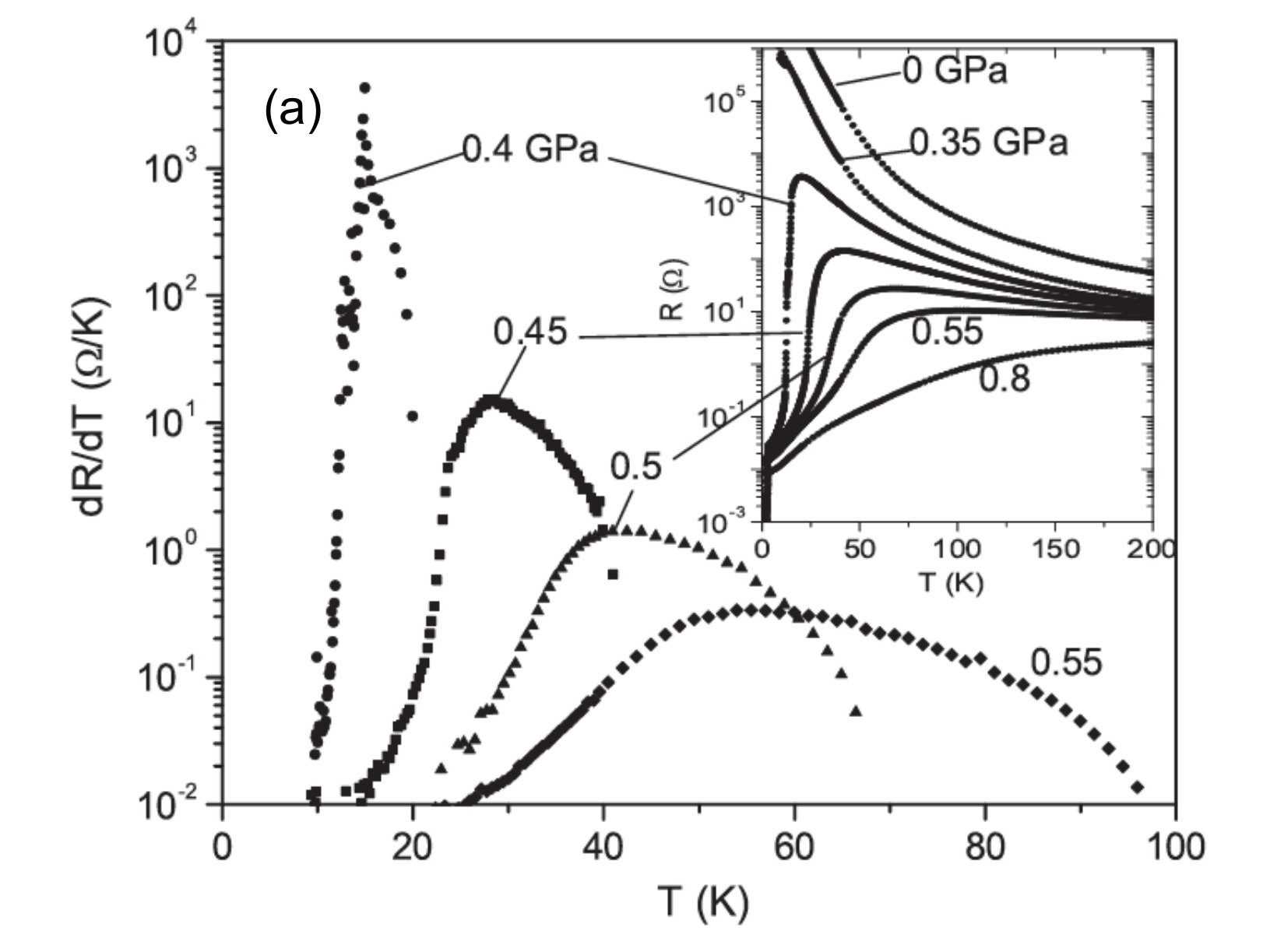}
\includegraphics[width=0.475\columnwidth]{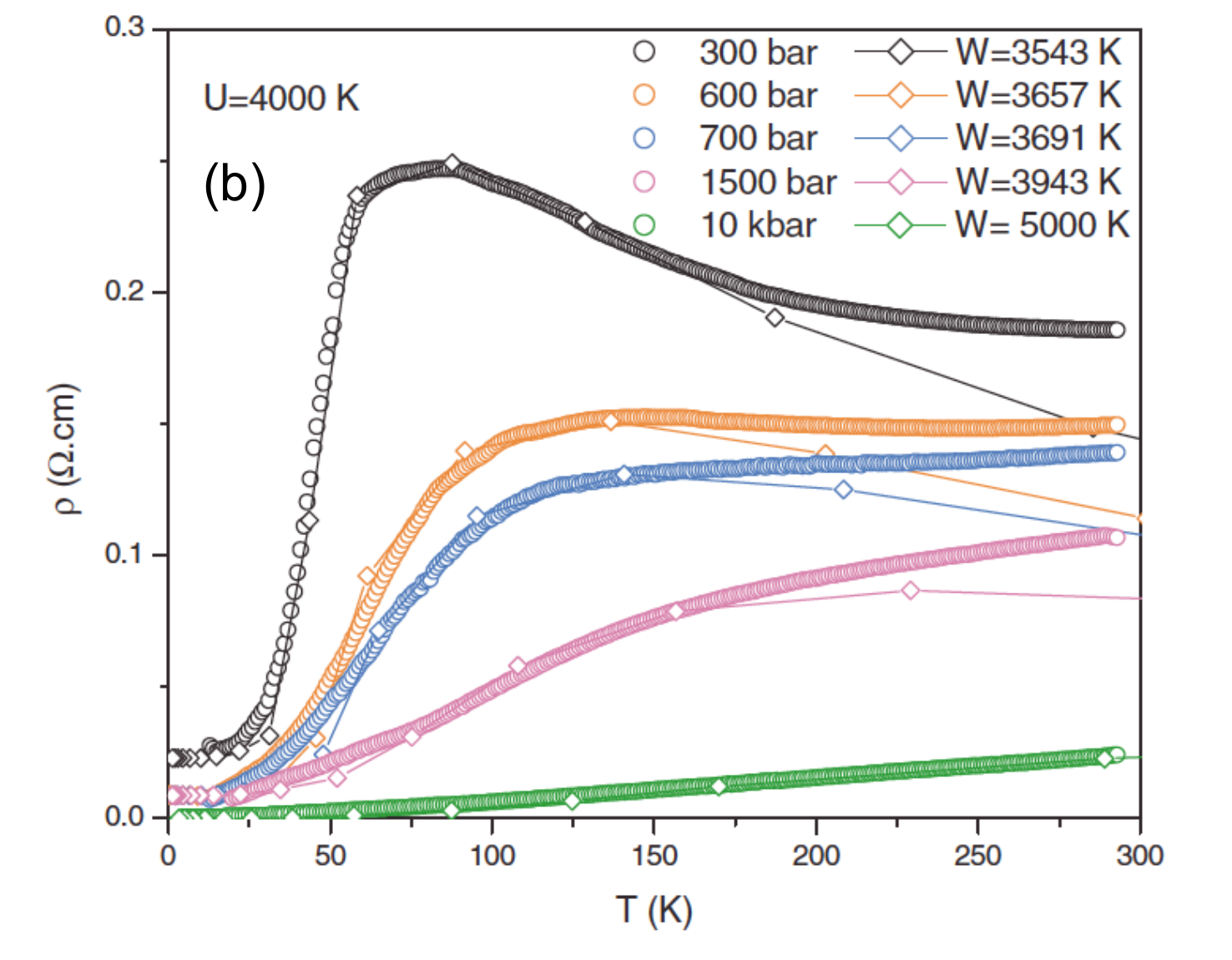}
\end{center}
\caption{(a) Transport in Mott organic materials; (a)  $\kappa - ({\mathrm{BEDT-TTF}})_2{\mathrm{Cu}}_2({\mathrm{CN}})_3$ (from \cite{kanoda2005prl}), and (b) $κ\kappa - ({\mathrm{BEDT-TTF}})_2{\mathrm{Cu}}[{\mathrm{N}}({\mathrm{CN}})_2]{\mathrm{Cl}}$. The lines are obtained from DMFT theory (from \cite{limelette03prl}).} 
\label{mott organics tranport}
\end{figure}

On the metallic side, all these materials display well characterized Fermi liquid behavior, with substantial effective mass enhancements, similarly to other Mott systems. In contrast to the conventional quantum critical scenario, however, here the phase diagram features a first-order MIT and the associated coexistence region, terminating at the critical end point $T = T_c$. Still, as emphasized in recent theoretical \cite{terletska11prl}  and experimental  \cite{kanoda2015nphys} work, the temperature range associated with this Mott coexistence region is typically very small ($T_c \sim 20-40$ K), in comparison to the Coulomb interaction or the (bare) Fermi temperature ($T_F \sim 2000$ K). Note that the comparable temperature range in 2DEG systems would be extremely small viz.  $T < 10^{-2}T_F \sim$ 0.1K, while most experimental results (especially those obtained in the ballistic regime) correspond to much higher temperatures.

We should emphasize that different materials in this class display various magnetic or even superconducting orders within the low temperature regime $T < T_c$ (Fig. \ref{phase_diagrams}), which strongly depend on material details, such as the precise amount of magnetic frustration. In contrast, the behavior above the coexistence dome is found to be remarkably universal, featuring very similar behavior in all compounds. Here, one finds a smooth crossover between a FL metal and a Mott insulator, displaying a  family of resistivity curves with a general form very similar to those found in 2DEG systems close to 2D-MIT, in the temperature range comparable to a fraction of $T_F$. In addition to a "fan-like" shape at high temperatures, we note the existence of resistivity maxima on the metallic side, which become more pronounced closer to the transition, precisely as in many 2DEG examples. 

\begin{figure}
\begin{center}
\includegraphics[width=1\columnwidth]{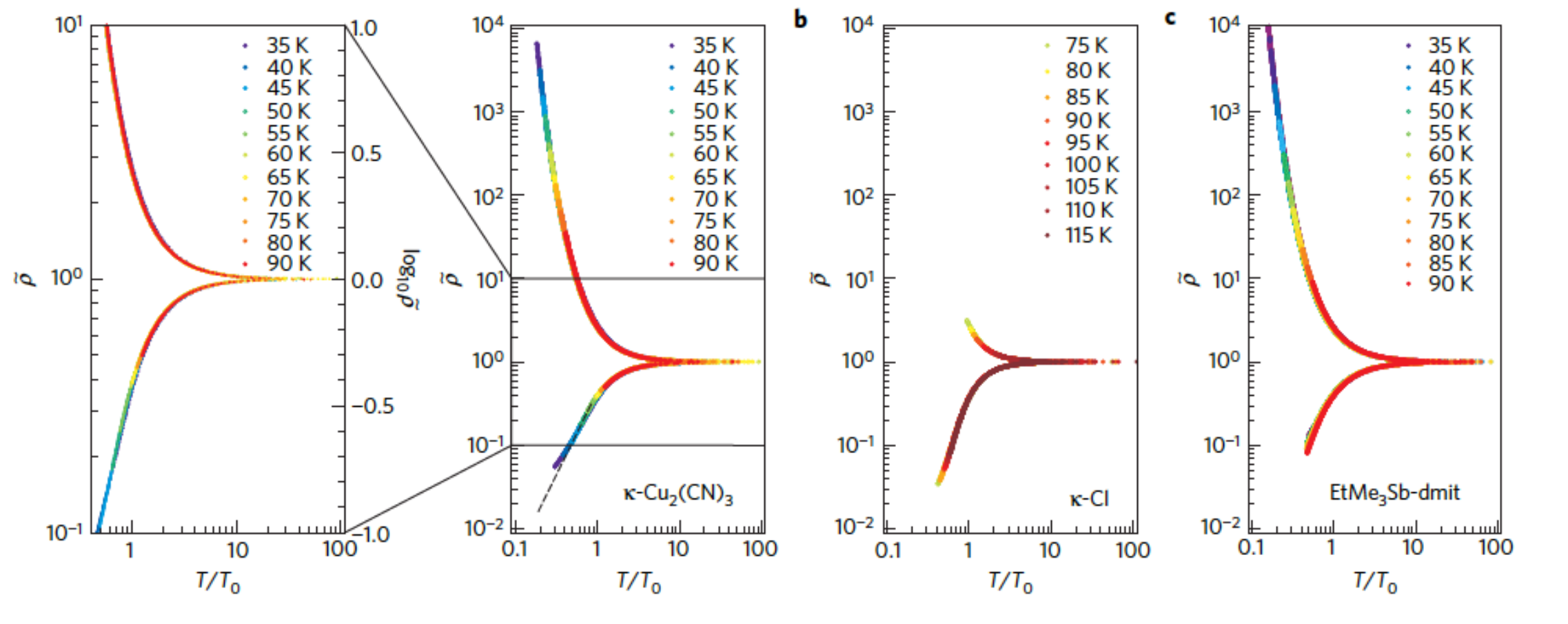}
\end{center}
\caption{Quantum critical scaling of the resistivity curves for three different organic Mott materials, demonstrating  universal behavior insensitive to low-temperature magnetic order (from \cite{kanoda2015nphys}).} 
\label{qc_scaling}
\end{figure}

In addition, very recent experimental work \cite{kanoda2015nphys}  performed a careful scaling analysis of the resistivity data in the high temperature region ($T > T_c$), providing clear and convincing evidence of Quantum Critical (QC) behavior associated with the Mott point. The analysis was performed on three different materials (including two compounds discussed above), featuring very different magnet ground states (AFM order vs. spin liquid behavior) on the insulating side. Nevertheless, the scaling analysis succeeded in extracting the universal aspects of transport in this QC regime, finding impressive universality (Fig. \ref{qc_scaling}) both in the form of the corresponding scaling function, and in the values of the critical exponents, with excellent agreement with earlier DMFT predictions \cite{terletska11prl}.

Most remarkably, the experimentally determined scaling functions display precisely the same "mirror symmetry" as earlier found on 2DEG systems \cite{simonian97}, with exactly the "stretched exponential" form proposed by the phenomenological scaling theory \cite{gang4me}. Given the fact that 2DEG systems and these organic Mott crystals have completely different microscopic character, these experimental findings provide direct and clear evidence on a surprisingly robust universality of the quantum critical behavior in interaction-driven MITs. If one and the same basic physical processes - Mott localization - indeed dominate all these phenomena, then one can hope that even a theory based on simple models may account for the main trends. 

\section{Theory of interaction-driven MITs}

Interaction driven MIT in absence of static symmetry breaking - the Mott transition - has first been proposed \cite{mott1949} as a possible mechanism in doped semiconductors. Soon enough, it became apparent that it is the right physical picture for many systems at the brink of magnetism, for example various transition-metal oxides (TMO) \cite{goddenough63book,mott-book90}. Such systems are typically characterized by a narrow valence band close to half-filling, with substantial on-site (intra-orbital) Coulomb interaction $U$. The simplest generic model describing this situation is the single-band Hubbard model given by the Hamiltonian

\begin{equation}
H=-t\sum_{\langle ij\rangle\sigma}c_{i\sigma}^{\dagger}c_{j\sigma}+U\sum_{i}n_{i\uparrow}n_{i\downarrow}.\label{EHM}
\end{equation}
Here,  $t$ is the hopping amplitude, $c_{i\sigma}^{\dagger}$
($c_{i\sigma}$) are the creation (annihilation) operators, $n_{i}=n_{i\uparrow}+n_{i\downarrow}$
is the occupation number operator on site $i$, and $U$ is the on-site Coulomb interaction. 

Theoretical investigations of the Hubbard model go back to the pioneering investigations in early 1960s \cite{hubbard3} when it already became clear that, even in absence of any magnetic order, a Hubbard-Mott gap will open for sufficiently large $U/t$. Approaching the transition from the metallic side was predicted  \cite{brinkmann70prb} to result in substantial effective mass enhancement as a precursor of the Mott transition, as confirmed in numerous TMO and other materials. A reliable description of the transition region, however, remained elusive for many years. Still, the existence a a broad class of Mott insulators, containing localized magnetic moments, has long been regarded as well established \cite{goddenough63book}.

\subsection{The DMFT approach}

Modern theories of Mott systems were sparked by the discovery of high temperature superconductors in late 1980s, which triggered considerable renewed interest in the physics of TMOs and related materials. A veritable zoo of physical ideas and theoretical approaches were put forward, but few survived the test of time. Among those, Dynamical Mean-Field Theory (DMFT) \cite{dmft96} proved to be the most useful, reliable, and flexible tool, applicable both to model Hamiltonians but also in first-principles theories \cite{LDA-DMFT06rmp} of correlated matter. 

\begin{figure}[h]
\begin{centering}
\includegraphics[width=0.6\columnwidth]{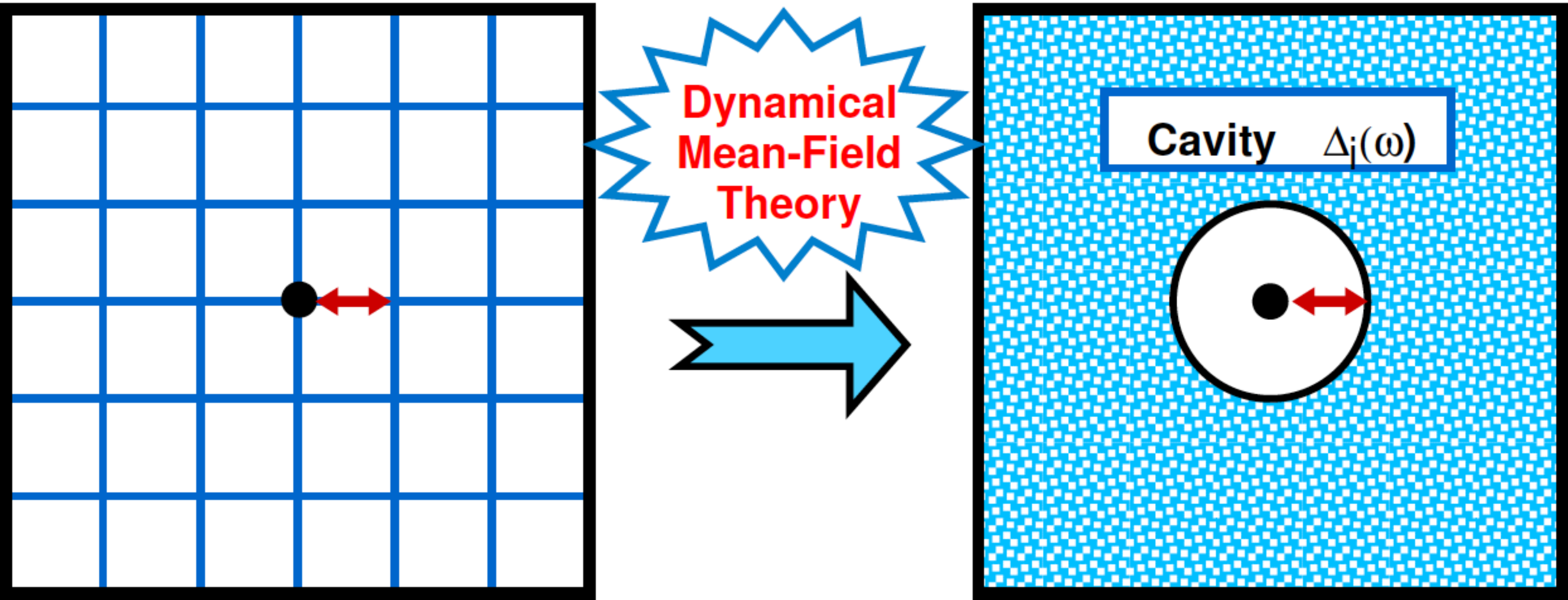}
\par\end{centering}
\caption{In dynamical mean-field theory, the environment of a given site is
represented by an effective medium, represented by its ``cavity
spectral function'' $\Delta_{i}(\omega)$. }

\end{figure}

In its simplest ("single-site") implementation, DMFT provides a local approximation for many-body corrections in strongly interacting systems. The central quantity it self-consistently calculates is the single-particle self-energy $\Sigma_i (\omega ,T)$, which is obtained within a local approximation. Its real part describes the interaction (or temperature) dependence of the electronic spectra, allowing substantial narrowing of the QP band width (i.e. the effective mass $m^*$ enhancement), and the eventual opening of the Mott gap. Its imaginary part, on the other hand, describes electron-electron (inelastic) scattering processes, leading to the ultimate destruction of the coherent QPs at sufficiently large $T$ or $U$. 

Most importantly, DMFT does not suffer from limitations of standard Fermi liquid approaches, which are largely restricted to situations with dilute QPs (i.e. the lowest temperatures). In fact, DMFT is most reliable at high temperatures, within the incoherent regime (with large inelastic scattering), where its local approximation becomes essentially exact. Indeed, systematic (nonlocal) corrections to single-site DMFT have established \cite{tanaskovic11prb} that there exists, in general, a well defined crossover temperature scale $T^* (U)$, above which the nonlocal corrections are negligibly small. This scale is typically much smaller than the basic energy scale of the problem (e. g. the bandwidth or the interaction $U$), so that DMFT often proves surprisingly accurate over much of the experimentally relevant temperature range. 

Another important aspect of the DMFT approach deserves special emphasis. In contrast to conventional (Slater-like) theories focusing on effects of various incipient orders, DMFT focuses on dynamical correlation effects unrelated to orders. From the technical perspective, its local character suppresses the spatial correlations associated with magnetic, charge, or structural correlations. Physically, such an approximation becomes accurate in presence of sufficiently strong frustration effects due to competing interactions or competing orders, a situation which is typical for most correlated electronic systems. Frustration effects generally lead to a dramatic proliferation of low-lying excited states, which further invalidates the conventional Fermi liquid picture of dilute elementary excitations; this situation demands an accurate description of incoherent transport regimes - precisely what is provided by DMFT.  

Historically, since its discovery more then twenty years ago, DMFT has been applied to many models and various physical systems. Recent developments have also included the application to inhomogeneous electronic systems in presence of disorder, and were able to incorporate the relevant physical processes such as Anderson localization and even the glassy freezing of electrons (the description of quantum Coulomb glasses). These interesting developments have been reviewed elsewhere \cite{RoP2005review, dobrosavljevic2012conductor} but will not be discussed here. In the following, we briefly review the key physical results of DMFT when applied to the simplest model of a Mott transition at half filling, and the associated Wigner-Mott transition away from half filling. 

\subsection{The Mott transition}

Many insulating materials have an odd number of electrons per unit
cell, thus band theory would predict them to be metals - in contrast
to experiments. Such compounds (e.g. transition metal oxides) often
have antiferromagnetic ground states, leading Slater to propose that
spin density wave formation \protect\cite{slater51} is likely at
the origin of the insulating behavior. This mechanism does not require
any substantial modification of the band theory picture, since the
insulating state is viewed as a consequence of a band gap opening
at the Fermi surface.

According to Slater \protect\cite{slater51}, such insulating behavior
should disappear above the Neel temperature, which is typically in
the $10^{2}$ K range. Most remarkably, in most antiferromagnetic
oxides, clear signatures of insulating behavior persist at temperatures
well above any magnetic ordering, essentially ruling out Slater's
weak coupling picture.

\begin{figure}[b]
\begin{center}
\includegraphics[width=0.6\columnwidth]{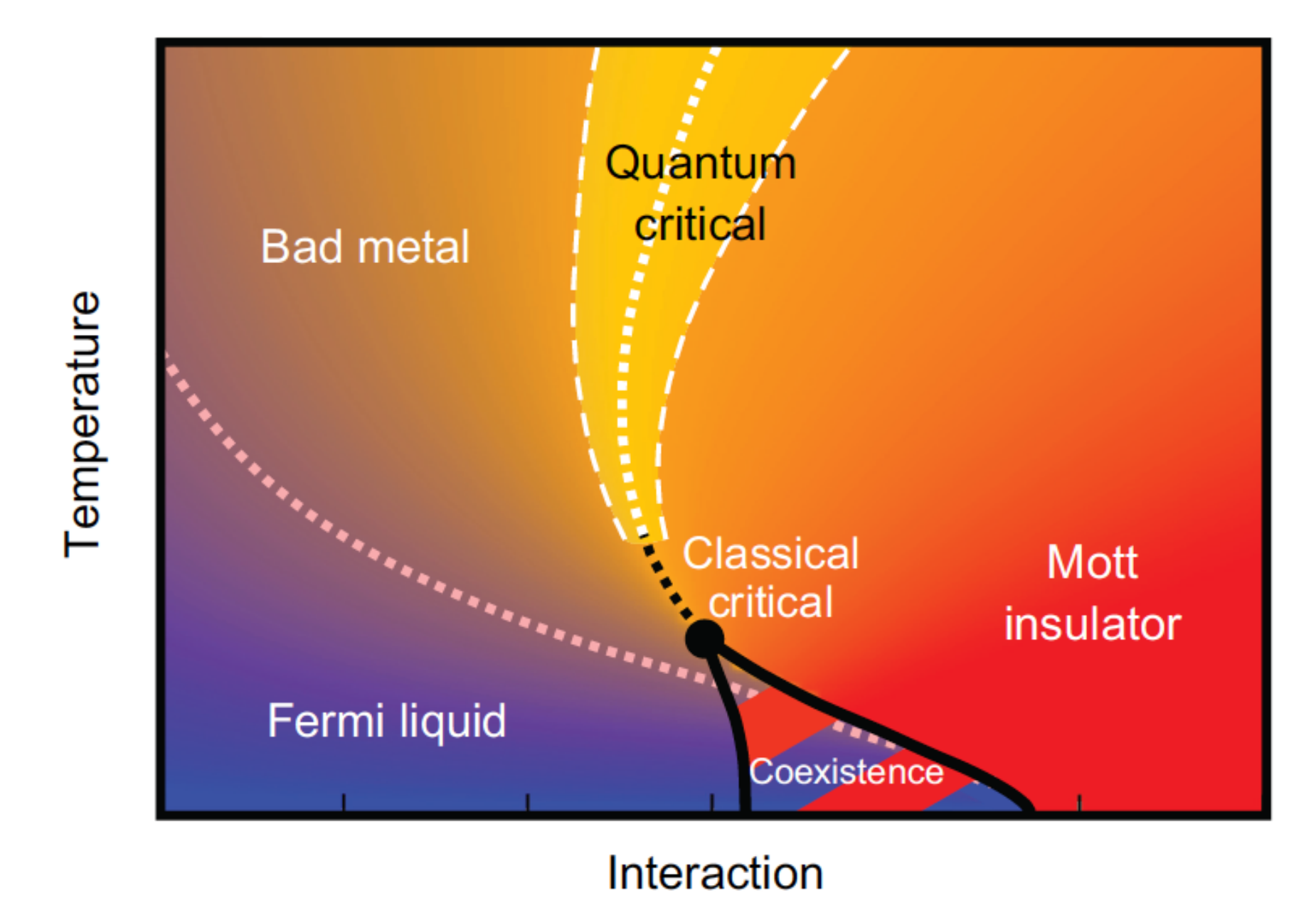}
\end{center}
\caption{Phase diagram of the half-filled Hubbard model calculated
from DMFT theory (from \cite{vucicevic13prb}). At very low temperatures  $T < T_{c} \le 0.03 T_F$ the Fermi liquid and the Mott
insulating phases are separated by a first order transition line,
and the associated coexistence region. Very recent work \protect \cite{terletska11prl, vucicevic13prb, vucicevic15prl} has established that in a very broad intermediate temperature region $T_{c} < T < T_F$, one finds characteristic metal-insulator crossover behavior showing all features expected of quantum criticality. }
\label{hubbard} 
\end{figure}

What goes on in such cases was first clarified in early works by Mott
\cite{mott1949} and Hubbard \protect\cite{hubbard3}, tracing the
insulating behavior to strong Coulomb repulsion between electrons
occupying the same orbital. When the lattice has integer filling per unit cell, then electrons
can be mobile only if they have enough kinetic energy ($E_{K}\sim t$)
to overcome the Coulomb energy $U$. In the narrow band limit of $t\ll U$,
the electrons do not have enough kinetic energy, and a gap opens in
the single-particle excitation spectrum, leading to Mott insulating
behavior. This gap $E_{g}\approx U-B$ (here $B\approx$ $2zt$ is
the electronic bandwidth; $z$ being the lattice coordination number)
is the energy an electron has to pay to overcome the Coulomb repulsion
and leave the lattice site. 

In the ground state, each lattice site
is singly occupied, and the electron occupying it behaves as a spin
1/2 local magnetic moment. These local moments typically interact
through magnetic superexchange interactions \cite{anderson59superexchange}of the order $J\sim t^{2}/U$, leading to magnetic ordering at temperatures
of order $T_J$. The insulating behavior, however, is not
caused by magnetic ordering, and typically persist all the way to temperatures
$T\sim E_{g}\gg T_{J}$. In oxides, $E_{g}\sim10^{3}-10^{4}$ K is typically on the atomic (eV) scale, while magnetic ordering
emerges at temperatures roughly an order of magnitude lower $T_{J}\sim100-300$ K.

\subsection{Correlated metallic state}

An important step in elucidating the approach to the Mott transition
from the metallic side was provided by the pioneering work of Brinkmann
and Rice \protect\cite{brinkmann70prb}. This work, which was motivated
by experiments on the normal phase of $^3$ He \cite{vollhardt84rmp}, predicted a strong effective mass
enhancement close to the Mott transition. In the original formulation,
as well as in its subsequent elaborations (slave boson mean-field
theory \protect\cite{kotliarruckenstein}, DMFT 
\protect\cite{dmft96}), the effective mass is predicted to continuously
diverge as the Mott transition is approached form the metallic side\begin{equation}
\frac{m^{\ast}}{m}\sim(U_{c}-U)^{-1}.\end{equation}

A corresponding coherence (effective Fermi) temperature \begin{equation}
T^{\ast}\sim T_{F}/m^{\ast}\end{equation}
is predicted above which the quasiparticles are destroyed by thermal
fluctuations. As a result, one predicts (e. g. within DMFT) a large resistivity increase
around the coherence temperature, and a crossover to insulating (activated)
behavior at higher temperatures. Because the low temperature Fermi
liquid is a spin singlet state, a modest magnetic field of the order\begin{equation}
B^{\ast}\sim T^{\ast}\sim(m^{\ast})^{-1}\end{equation}
 is expected to also destabilize such a correlated metal and lead to large
and positive magnetoresistance. Note that both $T^*$ and $B^*$ are predicted to continuously vanish as the Mott transition is approached from the metallic side, due to the corresponding $m^*$ divergence. Precisely such behavior is found in all the three classes of physical systems we discussed above. 

\subsection{Effective mass enhancement}

How should we physically interpret the large effective mass enhancement
which is seen in all these systems? What determines its magnitude
if it does not actually diverge at the transition? An answer to this
important question can be given using a simple thermodynamic argument,
which does not rely on any particular microscopic theory or a specific
model. In the following, we present this simple argument for the case
of a clean Fermi liquid, although its physical context is, of course,
much more general.

In any clean Fermi liquid \protect\cite{pinesnozieres} the low temperature
specific heat assumes the leading form\begin{equation}
C(T)=\gamma T+\cdots,\end{equation}
 where the Sommerfeld coefficient \begin{equation}
\gamma\sim m^{\ast}.\end{equation}
 In the strongly correlated limit ($m^{\ast}/m\gg1$) this behavior
is expected only at $T\lesssim T^{\ast}\sim(m^{\ast})^{-1}$, while
the specific heat should drop to much smaller values at higher temperatures
where the quasiparticles are destroyed. Such behavior is indeed observed
in many systems showing appreciable mass enhancements.

On the other hand, from general thermodynamic principles, we can express
the entropy as \begin{equation}
S(T)=\int_{0}^{T}dT\frac{C(T)}{T}.\end{equation}
 Using the above expressions for the specific heat, we can estimate
the entropy around the coherence temperature\begin{equation}
S(T^{\ast})\approx\gamma T^{\ast}\sim O(1).\end{equation}
 The leading effective mass dependence of the Sommerfeld coefficient
$\gamma$ and that of the coherence temperature $T^{\ast}$ cancel
out!

Let us now explore the consequences of the assumed (or approximate)
effective mass divergence at the Mott transition. As $m^{\ast}\longrightarrow\infty$,
the coherence temperature $T^{\ast}\longrightarrow0+$, resulting
in large residual entropy\begin{equation}
S(T\longrightarrow0+)\sim O(1).\end{equation}
 We conclude that the effective mass divergence indicates the approach to
a phase with finite residual entropy!

Does not this result violate the Third Law of Thermodynamics?! And
how can it be related to the physical picture of the Mott transition?
The answer is, in fact, very simple. Within the Mott insulating phase
the Coulomb repulsion confines the electrons to individual lattice
sites, turning them into spin 1/2 localized magnetic moments. To the
extent that we can ignore the exchange interactions between these
spins, the Mott insulator can be viewed as a collection of free spins
with large residual entropy $S(0+)=R\ln2$. This is precisely what
happens within the Brinkmann-Rice picture; similar results are obtained
from DMFT, a result
that proves exact in the limit of large lattice coordination  \protect\cite{dmft96}. 

\begin{figure}[t]
\begin{centering}
\includegraphics[width=0.6\columnwidth]{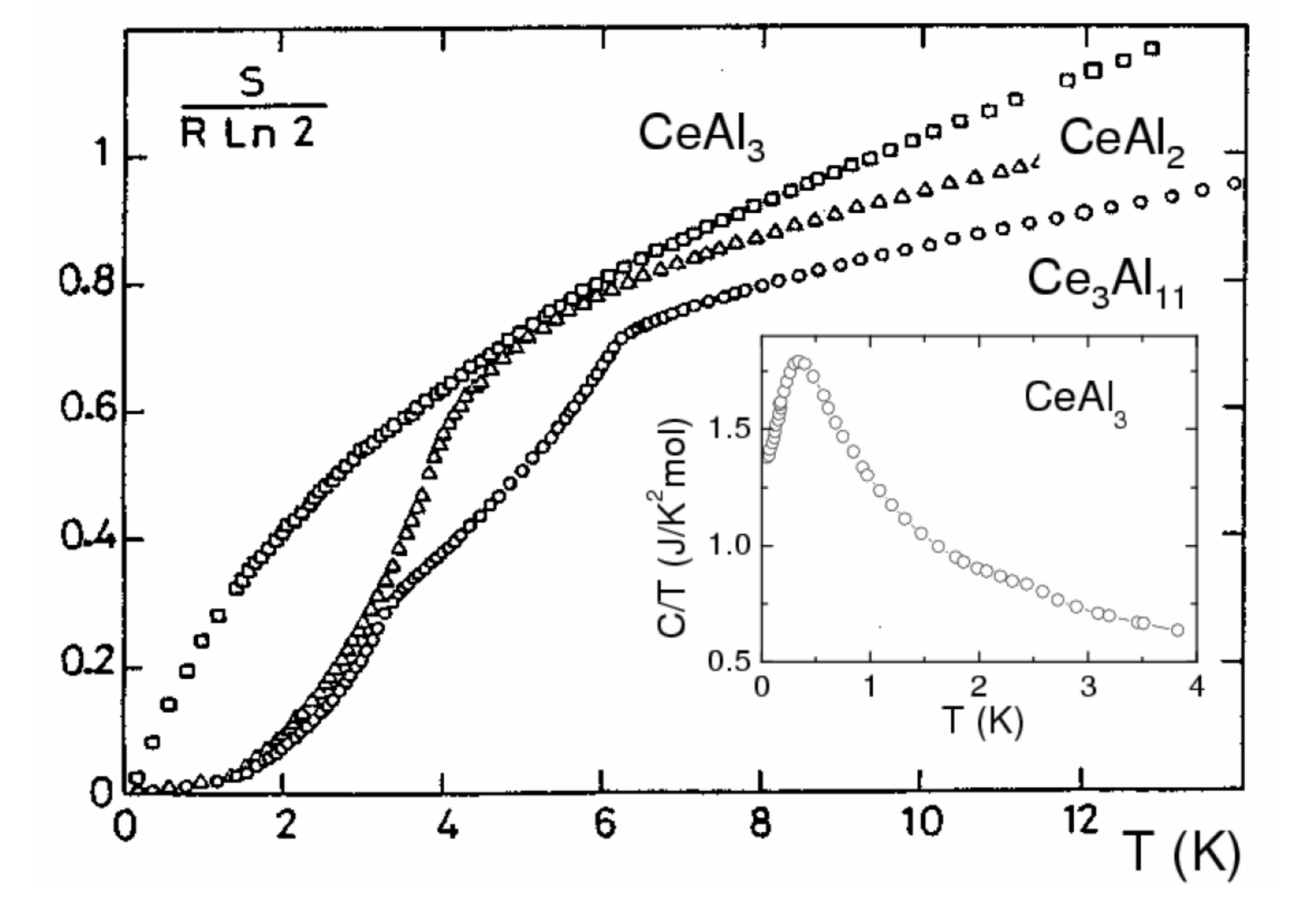}
\par\end{centering}
\caption{Temperature dependence of entropy extracted from specific heat (inset)
experiments on several heavy-fermion materials (\cite{flouquet1982,Flouquet2005}). 
Essentially the entire doublet entropy $S=R\ln2$ is recovered
by the time the temperature has reached $T^{\ast}\approx10K$, consistent
with a large mass enhancement $m^{\ast}\sim1/T^{\ast}$. }
\centering{}\label{flouquet05lanl}
\end{figure}

In reality, the exchange interactions between localized spins always
exist, and they generally lift the ground state degeneracy, restoring
the Third Law. This happens below a low temperature scale $T_{J}$,
which measures the effective dispersion of inter-site magnetic correlations
\cite{moeller99prb,park08prl} emerging from such exchange interactions.
In practice, this magnetic correlation temperature $T_{J}$ can be very low,
either due to effects of geometric frustration, or additional ring-exchange
processes which lead to competing magnetic interactions. Such a situation is found both in organic Mott systems, and for 2D Wigner crystals, where the insulating state corresponds to a geometrically frustrated triangular lattice. In addition, numerical calculations by Ceperley and others \cite{candido-2004-70} have established that for 2D Wigner crystals significant ring-exchange processes indeed provide additional strong frustration effects, further weakening the inter-site spin correlation effects. 

We conclude that the effective mass enhancement, whenever observed
in experiment, indicates the approach to a phase where large amounts
of entropy persist down to very low temperatures. Such situations
very naturally occur in the vicinity of the Mott transition, since
the formation of local magnetic moments on the insulating side gives
rise to large amounts of spin entropy being released at very modest
temperatures. A similar situation is routinely found  \protect\cite{stewartrev1,hewson,Flouquet2005}
 in the so-called ``heavy fermion" compounds (e.g. rare-earth inter-metallics) 
featuring huge effective mass enhancements.
Here, local magnetic moments coexist with conduction electrons giving
rise to the Kondo effect, which sets the scale for the Fermi liquid
coherence temperature $T^{\ast}\sim1/m^{\ast}$, above which the entire
free spin entropy $S(T^{\ast})\sim R\ln2$ is recovered (Fig. \ref{flouquet05lanl}).
This entropic argument is, in turn, often used to experimentally prove the
very existence of localized magnetic moments within a metallic host.

We should mention that other mechanisms of effective mass enhancement
have also been considered. General arguments \cite{millis} indicate
that $m^{*}$ can diverge when approaching a quantum critical point
corresponding to some (magnetically or charge) long-range ordered
state. This effect is, however, expected only below an appropriate
upper critical dimension \cite{sachdevbook}, reflecting an anomalous
dimension of the incipient ordered state. In addition, this is a mechanism
dominated by long wavelength order-parameter fluctuations, and is thus
expected to contribute only a small amount of entropy per degree of
freedom, in contrast to local moment formation. 

It is interesting to mention that weak-coupling approaches, such as the popular ``on-shell'' interpretation of the Random-Phase Approximation (RPA)  \cite{quinn75prl}, often result in inaccurate or even misleading predictions \cite{dassarma05prb} for the effective
mass enhancement behavior. Indeed, more
accurate modern theories such as DMFT can be used \cite{dobrosavljevic2012conductor} to benchmark these and other weak coupling theories, and to reveal the origin of the pathologies resulting from their inappropriate  applications to strong coupling situations. 

\subsection{Resistivity maxima}

Within the strongly correlated regime close to the Mott transition, transport is carried by heavy quasiparticles. As temperature increases, inelastic electron-electron scattering kicks-in, leading to a rapid increase of resistivity with temperature. Within Fermi liquid theory, this gives rise to the temperature dependence of the form $\rho (T) \approx AT^2$, with $A \sim (m^*)^2$ according to the Kadowaki-Woods law \cite{hewson}. Such behavior is indeed observed in most known correlated system, in agreement with DMFT predictions \cite{powell09nphys}. This argument makes it clear that increasingly strong temperature dependence emerges in the correlated regime, but it does not tell us what happens above the corresponding QP coherence temperature $T^* \sim 1/m^*$.

\begin{figure}[h]
\vspace{-6pt}
\begin{centering}
\includegraphics[width=0.48\columnwidth]{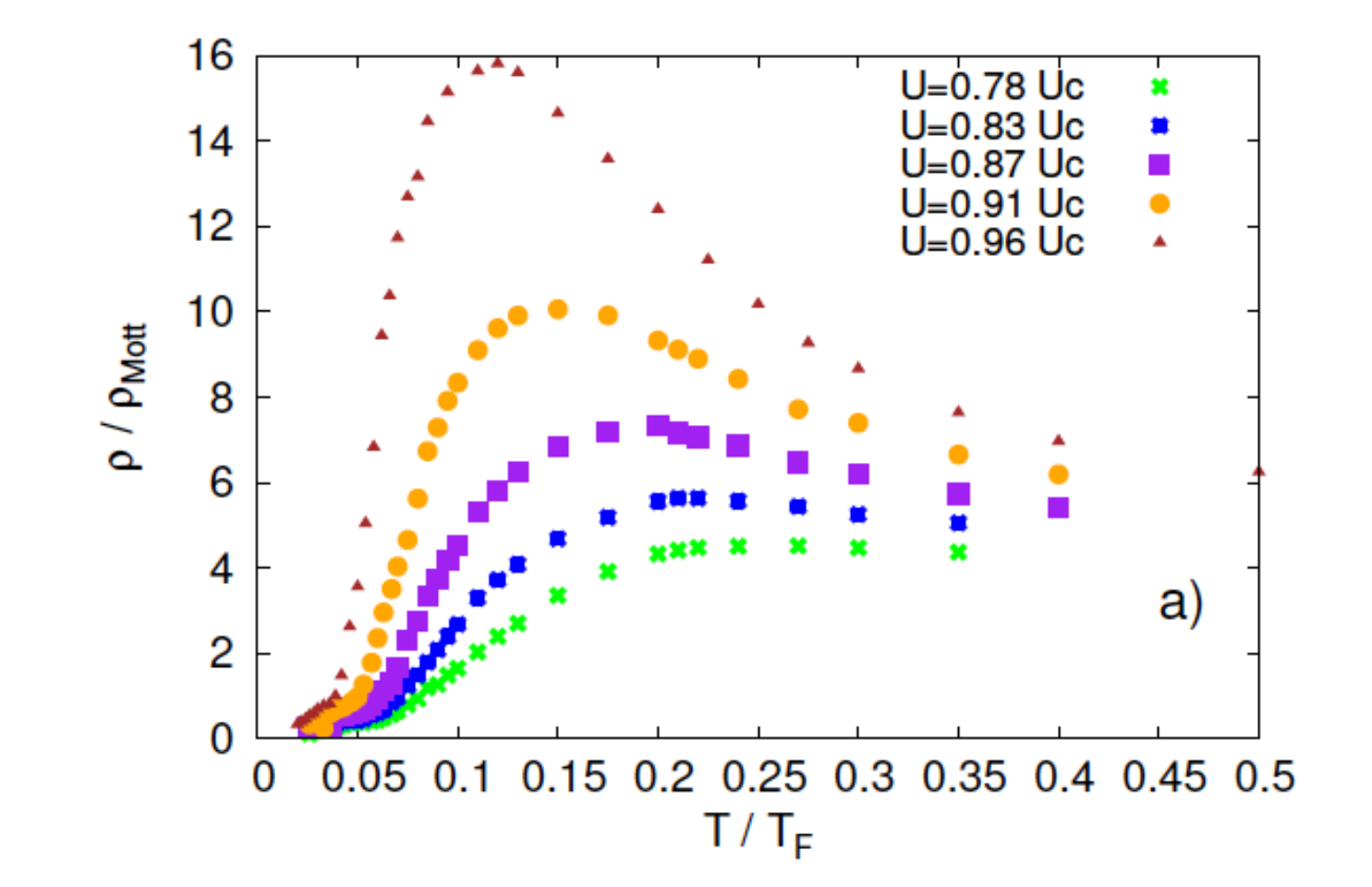}
\includegraphics[width=0.46\columnwidth]{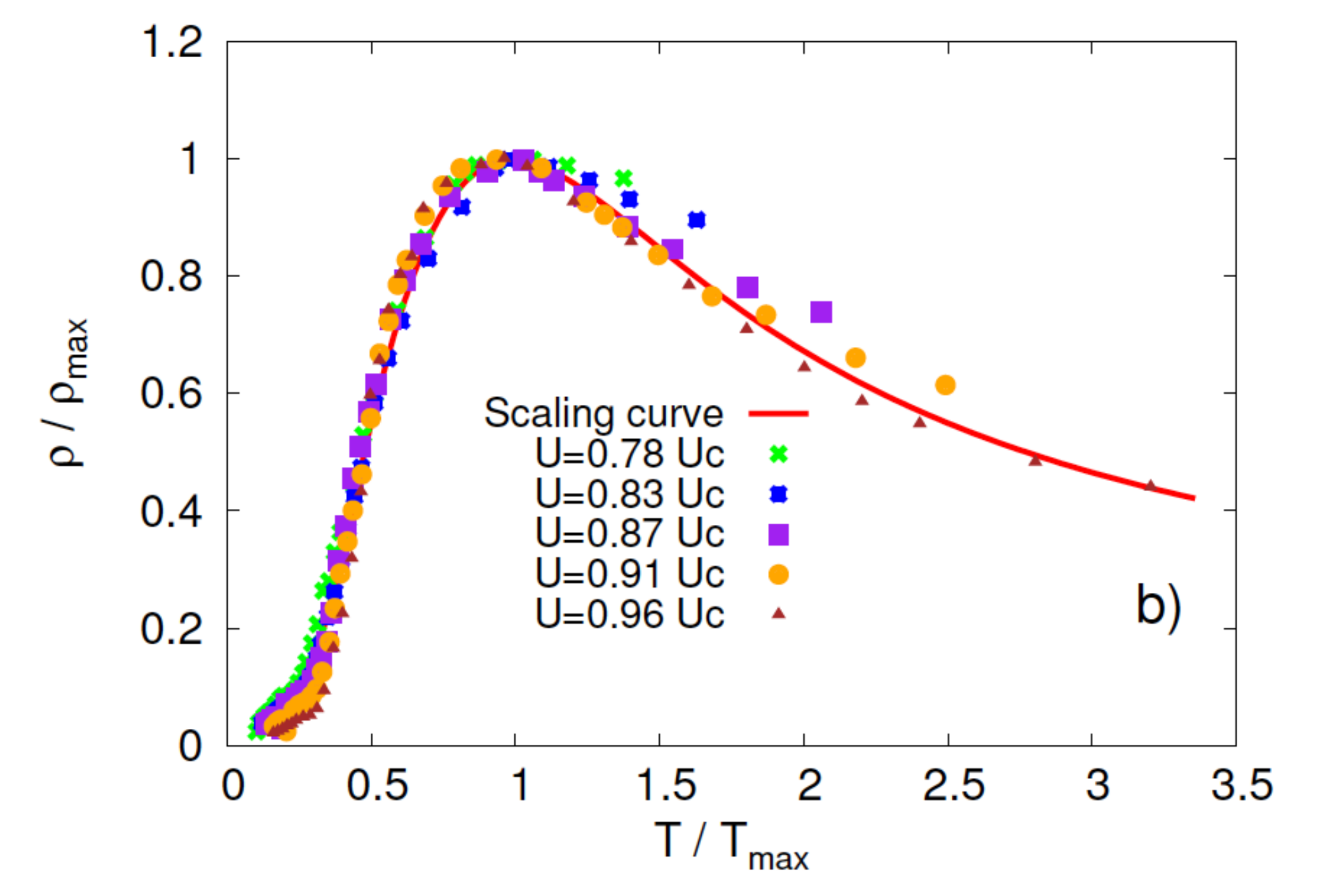}
\par\end{centering}
\vspace{-6pt}
\caption{ (a) Resistivity maxima in the strongly correlated metallic regime of a half-filled Hubbard model within DMFT theory. (b) Scaling collapse of the resistivity maxima. (from \cite{radonjic12prb})}
\centering{}\label{maxima}
\end{figure}

In this regime inaccessible to conventional FL theories, DMFT provided an illuminating answer \cite{radonjic12prb}. It described the thermal destruction of QPs, and the consequent opening of a Mott pseudogap at $T > T^*$, leading to resistivity maxima. While much of the earlier DMFT work focused on the low $T$ regime, more recent advances \cite{ctqmc2011rmp} in Quantum Monte Carlo methods  (needed to solve the DMFT equations) allowed a careful and precise characterization of this transport regime. One finds a family of curves displaying pronounced resistivity maxima with increasing height, at temperatures that decrease close to the transition. 

These curves all assume essentially the same functional form (Fig. \ref{maxima}), and therefore can all be collapsed on a single scaling function, by rescaling the temperature with that of the resistivity maximum $T_{max}$, and the resistivity with its maximum value $\rho_{max}$. From these numerical results one can extract a universal scaling function describing the entire family of curves (as shown by the thick red line Fig. \ref{maxima}).  Direct comparison can now be made with data obtain from experiments on 2DEG in silicon close to 2D-MIT; one finds surprisingly good agreement between theory and  experiment, with no adjustable parameters (see Fig. \ref{resmaxima}). Similar data are found also from experiments on Mott organics and other conventional Mott systems, where successful comparison with DMFT theory has already been established \cite{limelette03prl,radonjic10}. 

Further comparison between the experiment and theory is obtained by plotting $T_{max}$ vs. $m^*$, where both quantities can be independently obtained both from DMFT theory and from 2DEG experiments. Here we express $T_{max}$ in units of the Fermi temperature and $m^*$ in the units of the known band mass, in order to perform a comparison with no adjustable parameters. Again, one finds excellent agreement between DMFT and experiments, giving
\begin{equation}
T_{max} / T_F \approx 0.7 (m / m^* ),
\end{equation}
where even the numerical prefactor close to 0.7 is obtained in both cases.

\begin{figure}[h]
\begin{centering}
\includegraphics[width=0.48\columnwidth]{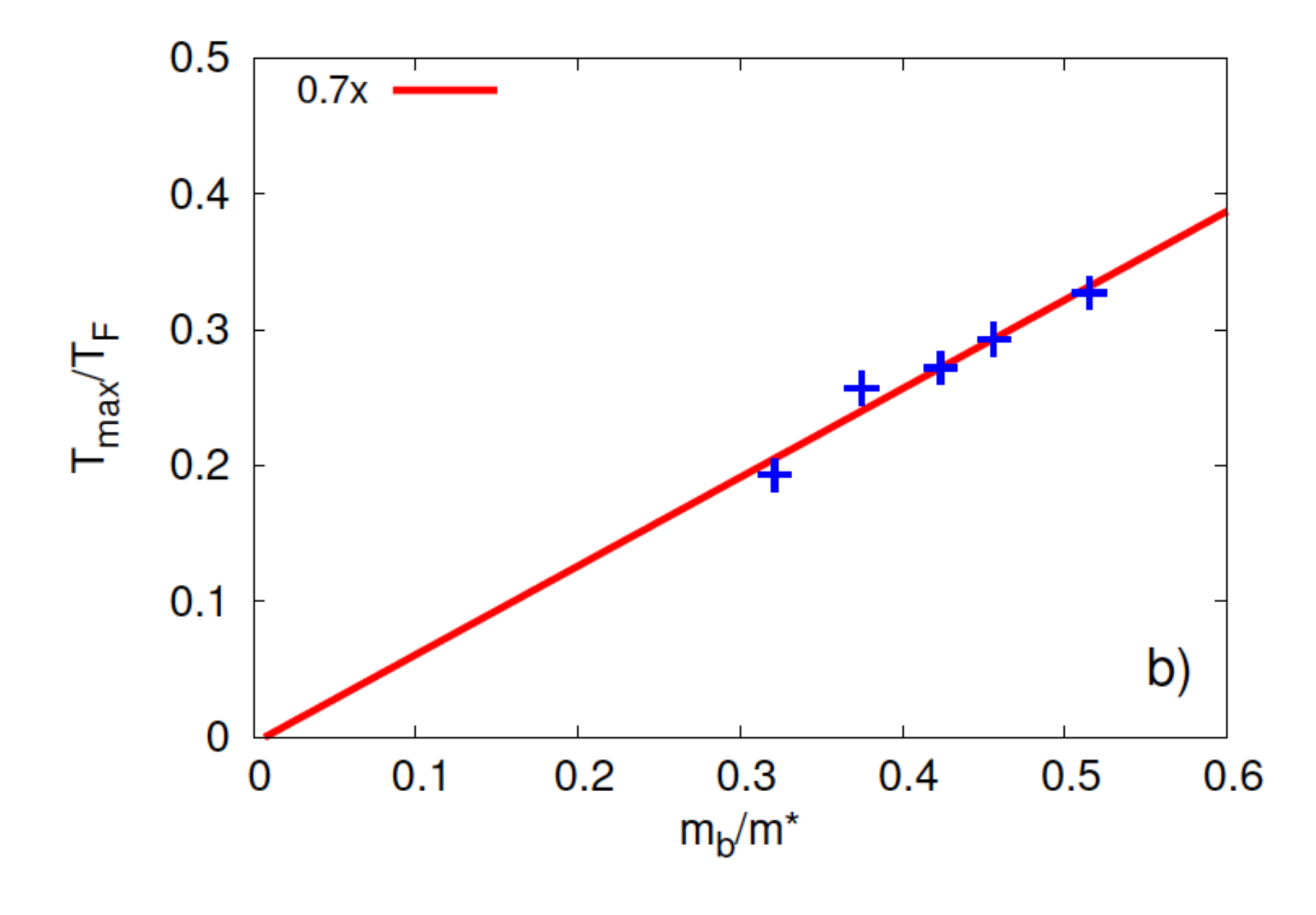}
\includegraphics[width=0.46\columnwidth]{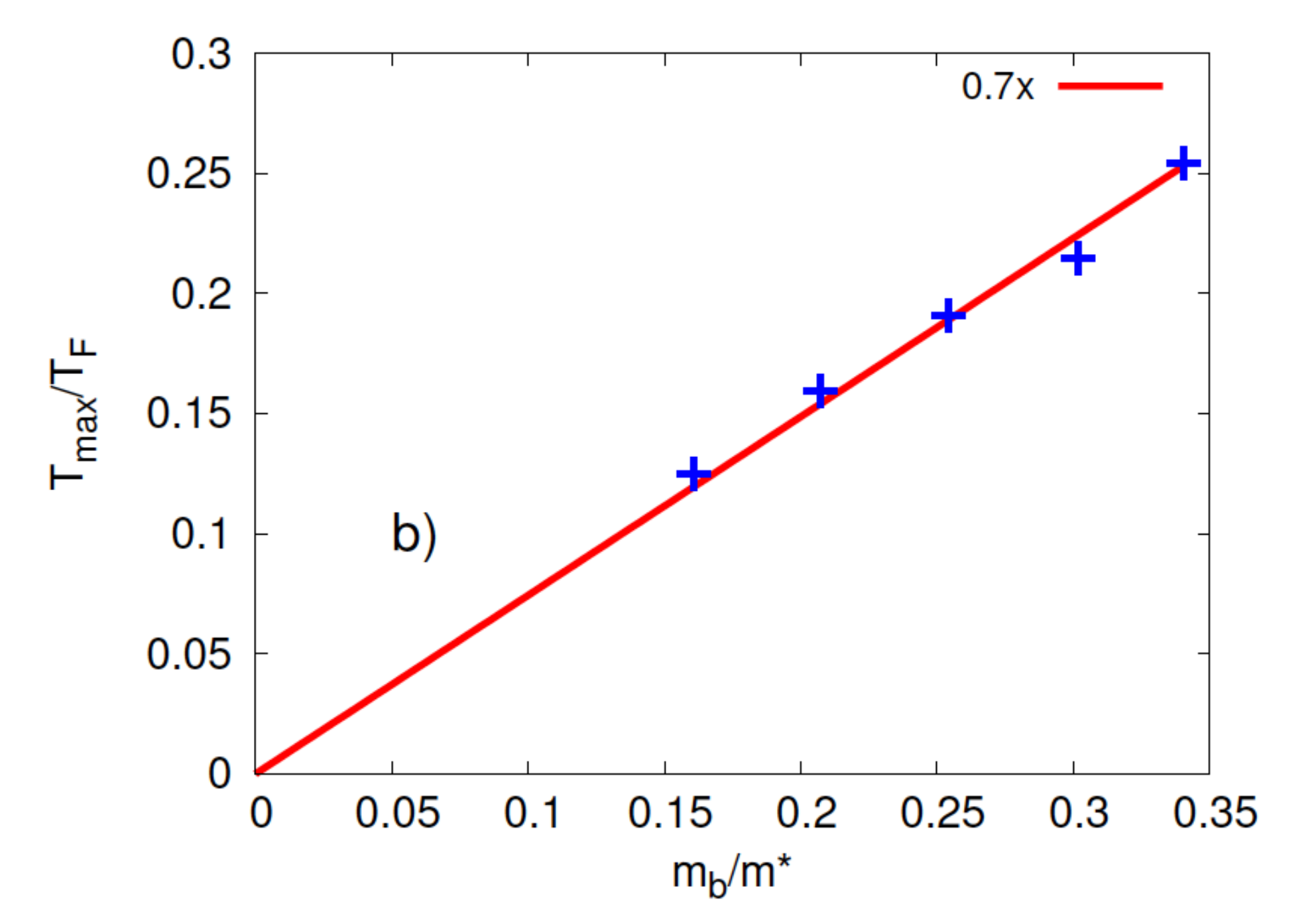}
\par\end{centering}
\caption{ Dependence of $T_{max}$ on $m^*$ from (a) DMFT theory; (b) experiments on 2DEG close to 2D-MIT. (from \cite{radonjic12prb})}
\centering{}\label{tvsm}
\end{figure}

Clear physical understanding of the Mott transition within DMFT, together with spectacular agreement between DMFT theory and 2DEG experiments, paints a convincing and transparent picture of the mechanism for the resistivity maxima, as follows. 

\begin{itemize}
\item Heavy quasiparticles that exist near the Mott transition are characterized by a small coherence temperature $T^* \sim 1/m^*$. 
\vspace{6pt}

\item Given that both theory and experiments find $T_{max} \sim 1/ m^*$, we can directly identify the resistivity maxima with the thermal destruction of heavy quasiparticles. 
\vspace{6pt}

\item The above physical picture is not only well established within DMFT theory, but is also well documented in all strongly correlated electronic systems such as various TMO materials, organic Mott systems, and even heavy fermion compounds.
\vspace{6pt}

\item The complete  thermal destruction of quasiparticles around $T \sim T_{max}$ is a physical picture completely different than the  situation described by the disordered Fermi liquid scenario \cite{punnoose02}, which does not even provide a good data collapse. In this scenario, the relevant quasiparticle regime extends both below and above $T_{max}$, and the maxima result from the competition of two different elastic (but temperature dependent) scattering mechanisms \cite{Zala}. 
\vspace{6pt}
\item We have seen that, both in DMFT theory and in all known Mott systems, the metal-insulator coexistence region does exist close to the MIT, but it remains confined to very low temperatures, typically two orders of magnitude smaller than the Fermi energy or the Coulomb repulsion $U$. 
\vspace{6pt}

\item The resistivity maxima, in contrast, are found at much higher temperatures, typically as high as a 10-20\% of the Fermi temperature. This feature is clearly found both in DMFT, in all conventional Mott systems, but also in 2DEG materials close to 2D-MIT. 
\vspace{6pt}

\item In contrast to DMFT theory and the known behavior of many conventional Mott systems, the phase separation scenario proposed by Kivelson and Spivak \cite{spivak05} can hold only within the metal-insulator coexistence region. It therefore appears very unconvincing as a possible mechanism behind the resistivity maxima found in 2DEG systems. 

\end{itemize}

\subsection{Quantum criticality and scaling}

What is the physical nature of the Mott transition? The physical picture of  Brinkmann and Rice (BR) \cite{brinkmann70prb} makes it plausible that the Mott transition should be viewed as a quantum critical point, since the characteristic energy scale of the correlated Fermi liquid $T^* \sim T_F /m^*$ continuously decreases as the transition is approached. On the other hand, the Mott transition discussed here describes the opening of a correlation-induced spectral
gap in absence of magnetic ordering, i.e. within the paramagnetic
phase. Such a phase transition is not associated with spontaneous symmetry breaking associated with any static order parameter. Why should the phase transition then have any second-order (continuous) character at all?

\begin{figure}[h]
\begin{centering}
\includegraphics[width=0.6\columnwidth]{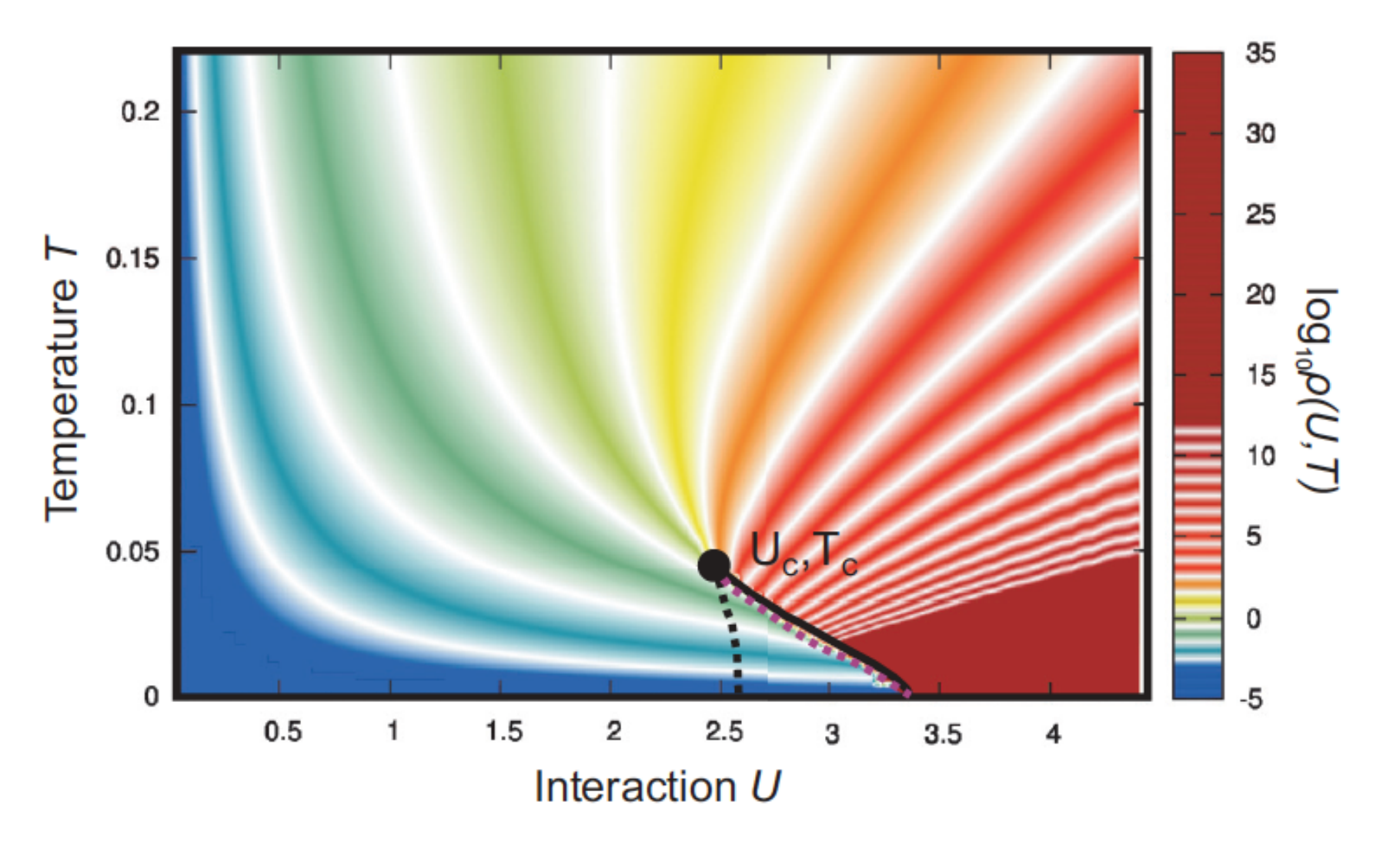}
\par\end{centering}
\caption{Color-coded plot of resistivity across the DMFT phase diagram of a half-filled Hubbard model (from \cite{vucicevic13prb}).}
\centering{}\label{color code}
\end{figure}

More insight into this fundamental question followed the development of DMFT methods \cite{dmft96}, which extended the BR theory to include incoherent (inelastic) processes at finite temperature. According to this formulation, the Mott transition does assume first-order character with the associated metal-insulator coexistence dome. However, this is found only below the very low critical end-point temperature $T_c \sim 0.02 T_F$, and a smooth crossover behavior arises in the very broad intermediate temperature interval $T_c < T < T_F$. Precisely the same features for the Mott transition phase diagram are found not only in many TMO materials, but also in Mott organics we discussed above.

A visually striking illustration of what happens in the intermediate temperature range is seen by color-coding the resistivity across the DMFT phase diagram \cite{vucicevic13prb}, where each white band indicates another order of magnitude in resistivity. If one ignores the low-temperature  coexistence region, one immediately notices a fan-like shape of the constant resistivity lines, characteristic of what is generally expected \cite{sachdevbook} for quantum criticality. Given the fact that the coexistence dome is a very small energy feature, one should consider the Mott transition as having "weakly first-order" (WFO) character. At WFO points, which are well know in standard critical phenomena \cite{goldenfeldbook}, the transition assumes a first-order character only very close to the critical point; further away, the behavior is precisely that of a conventional second-order phase transition, including all aspects of the scaling phenomenology. 

In the case of MITs, the quantum critical point (QCP) is expected \cite{dobrosavljevic2012conductor} to occur at $T=0$. Therefore, if a QPT assumes WFO character, as we find here, we still expect to observe all features of Quantum Criticality (QC) in a broad intermediate temperature region, above the coexistence dome. Similar situations are not uncommon in general QC phenomena \cite{sachdevbook}, since the immediate vicinity of many QCPs is often "masked" by the dome of an appropriate ordered state induced by critical fluctuations.

\begin{figure}[t]
\begin{centering}
\includegraphics[width=0.7\columnwidth]{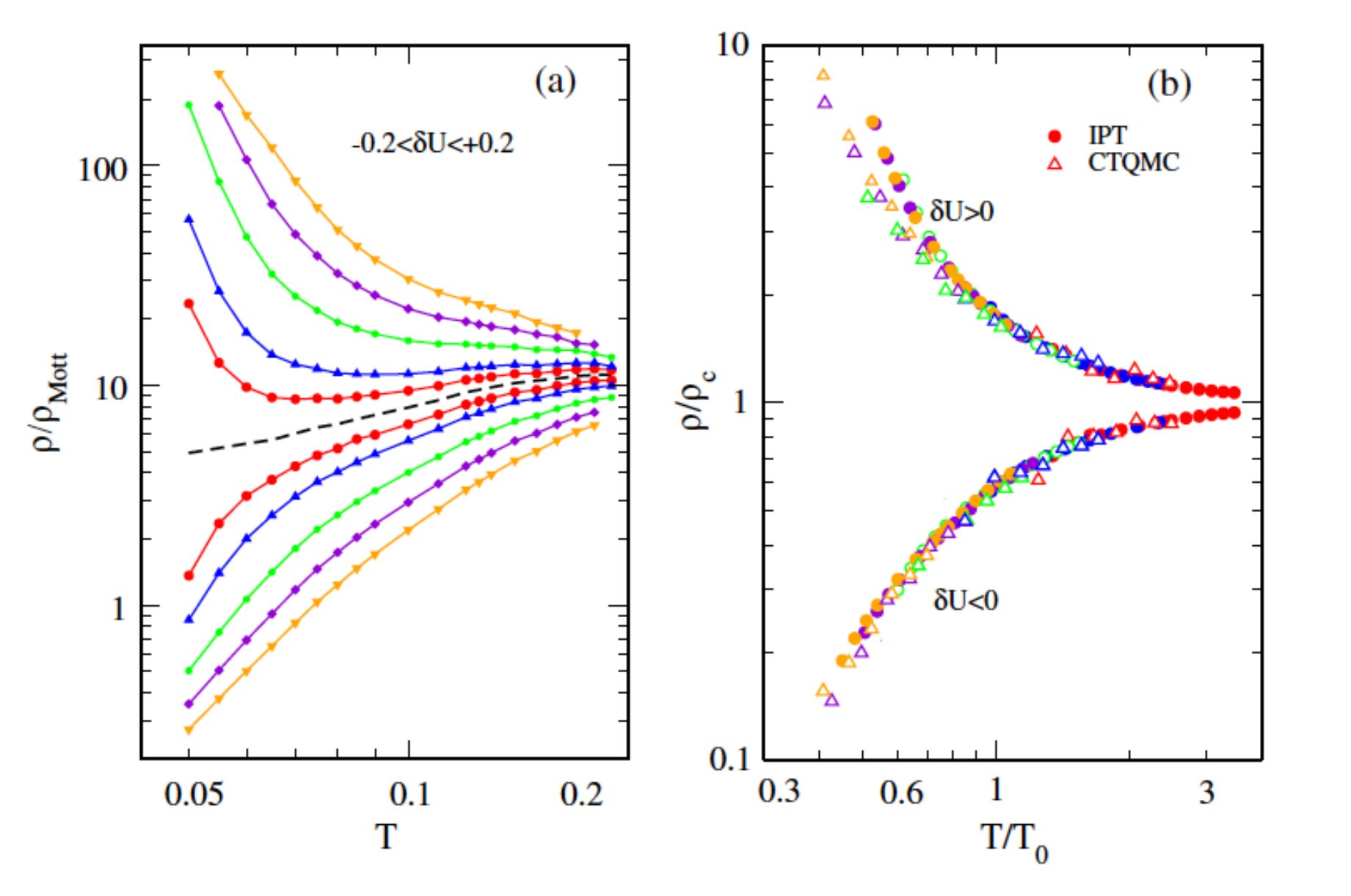}
\par\end{centering}
\caption{Scaling behavior of the resistivity curves in the Quantum Critical region of the Mott transition at half filling, as obtained from DMFT theory (from \cite{terletska11prl}). Note the remarkable "mirror symmetry" of the corresponding scaling function, as experimentally found both in 2DEG systems near 2D-MIT \cite{simonian97}, and also in very recent experiments in Mott organics \cite{kanoda2015nphys}.}
\centering{}\label{QC scaling}
\end{figure}

If these ideas are correct, then an appropriate scaling analysis should be able to reveal the expected QC scaling of the resistivity curves in a broad intermediate temperature range $T_c < T < T_F$. To perform such an analysis, one needs to follow a judiciously chosen trajectory across the phase diagram, corresponding to the center of the QC region. Such a trajectory - the so-called Widom line - was identified a long time ago \cite{widom65}  in the context of conventional (thermal) critical phenomena. Directly applying these standard scaling procedures to the DMFT description of the Mott transition was first carried out a few years ago \cite{terletska11prl}, revealing compelling evidence of QC behavior. Follow-up work  further established that such QC behavior is a robust feature of the Mott point both at half-filling \cite{vucicevic13prb}, and for doping-driven Mott transitions \cite{vucicevic15prl}. The latter result also provided important new insight into the origin of the so-called "Bad Metal" (linear resistivity) behavior in doped Mott insulator, a long-time puzzle \cite{emery95prl} in the field of correlated electrons. 

The discovery of the Mott QC regime within DMFT \cite{terletska11prl} was directly motivated by the pioneering works on 2D-MIT \cite{kravchenko95, popovic97}, which first revealed intriguing features of QC scaling near the MIT. The results obtained from DMFT presented surprising similarity to the experiment, both in identifying previously overlooked scaling behavior in interaction-driven MITs without disorder, and in producing microscopic underpinning for early scaling phenomenology \cite{gang4me} for the "mirror symmetry". Still, this model calculation, based on the single-band Hubbard model at half filling, should really not be viewed an accurate description of dilute 2DEG systems, where the experiments have been performed. 

In contrast, organic Mott systems of the $\kappa$-family should be considered \cite{mckenzie2011review} as a much more faithful realization of the half-filled Hubbard model. This notion stimulated further experimental investigation, following the 2011 theoretical discovery, focusing on the previously overlooked intermediate temperature region $T_c < T < T_F$. The experiments took several years of very careful work on several different materials in this family, but the results published by 2015 provided \cite{kanoda2015nphys} spectacular confirmation of the DMFT theoretical predictions. 

\subsection{Is Wigner crystallization a Mott transition in disguise?}

The original ideas of Mott \cite{mott1949}, who thought about doped
semiconductors, envisioned electrons hopping between well localized
atomic orbitals corresponding to donor ions. In other Mott systems,
such as transition metal oxides, the electrons travel between the
atomic orbitals of the appropriate transition metal ions. In all these
cases, the Coulomb repulsion restricts the occupation of such localized
orbitals, leading to the Mott insulating state, but it does not provide
the essential mechanism for the formation such tightly bound electronic
states. The atomic orbitals in all these examples result from the
(partially screened) ionic potential within the crystal lattice.

The situation is more interesting if one considers an idealized situation
describing an interacting electron gas in absence of any periodic
(or random) lattice potential due to ions. Such a physical situation
is achieved, for example, when dilute carriers are injected in a semiconductor quantum well \cite{AFS1982}, where all the effects of the crystal lattice can be treated within the effective mass approximation \cite{ashcroft}.
This picture is valid if the Fermi wavelength of the electron is much
longer then the lattice spacing, and the quantum mechanical dynamics
of the Bloch electron can be reduced to that of a free itinerant particle
with a band mass $m_{b}$. In such situations, the only potential
energy in the problem corresponds to the Coulomb repulsion $E_{C}$
between the electrons, which is the dominant energy scale in low carrier density 
systems. At the lowest densities, $E_{C}\gg E_{F}$, and the electrons
form a Wigner crystal lattice \cite{wigner34pr} to minimize the Coulomb
repulsion.

\begin{figure}[h]
\begin{centering}
\includegraphics[width=0.7\columnwidth]{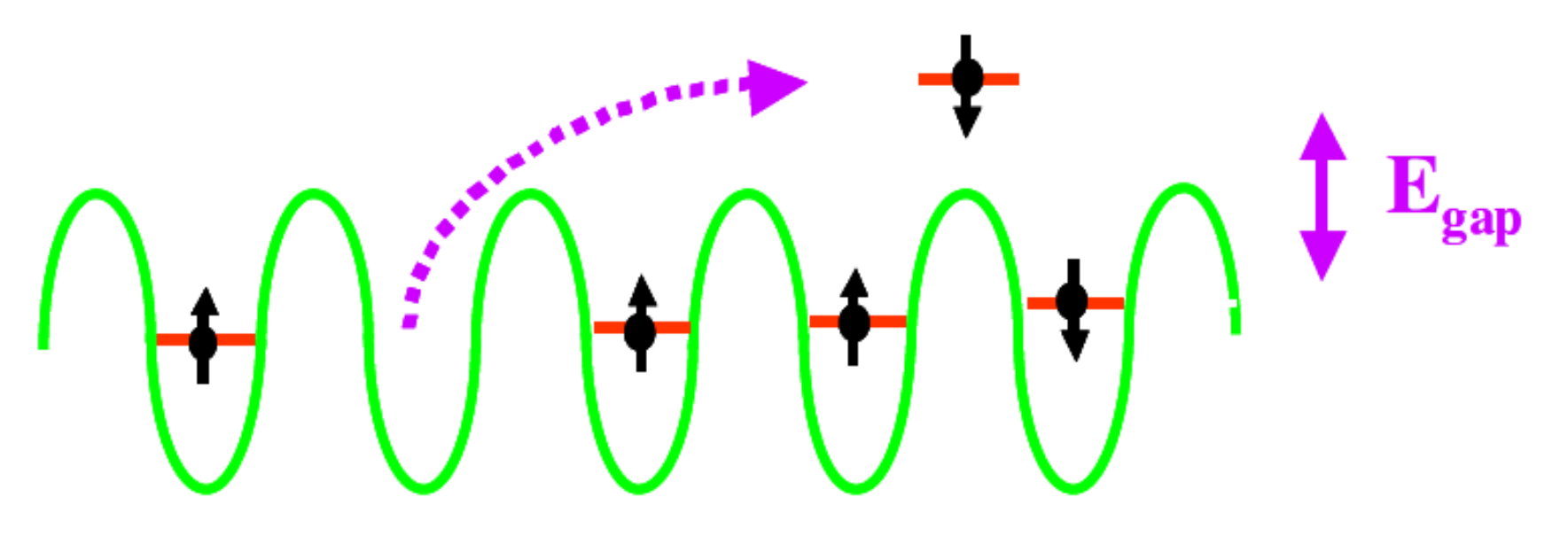}
\par\end{centering}
\caption{In a Wigner crystal, each electron is confined to a potential well
produced by Coulomb repulsion from neighboring electrons, forming
a spin 1/2 local moment. The lowest energy particle hole excitation
creates a vacancy-interstitial pair \protect\cite{candidi01prb},
which costs an energy $E_{gap}$ comparable to the Coulomb repulsion.}
\centering{}\label{wigner}
\end{figure}

Here, each electron is confined not by an ionic potential, but due
to the formation of a deep potential well produced by repulsion from
other electrons. The same mechanism prevents double occupation of
such localized orbitals, and each electron in the Wigner lattice reduces
to a localizes $S=1/2$ localized magnetic moment. A Wigner crystal
is therefore nothing but a magnetic insulator: a Mott insulator in
disguise. At higher densities, the Fermi energy becomes sufficiently
large to overcome the Coulomb repulsion, and the Wigner lattice melts
\cite{ceperley89prb}. The electrons then form a Fermi liquid. The
quantum melting of a Wigner crystal is therefore a metal-insulator
transition, perhaps in many ways similar to a conventional Mott transition.
What kind of phase transition is this? Despite years of effort, this
important question is still not fully resolved.

What degrees of freedom play the leading role in destabilizing the
Wigner crystal as it melts? Even in absence of an accepted and detailed
theoretical picture describing this transition, we may immediately
identify two possible classes of elementary excitations which potentially
contribute to melting, as follows.
\begin{enumerate}

\item \textit{Collective charge excitations} (``elastic'' deformations)
of the Wigner crystal. In the quantum limit, these excitations have
a bosonic character, but they persist and play an important role even
in the semi-classical ($k_{B}T\gg E_{F}$) limit, where they contribute
to the thermal melting of the Wigner lattice \cite{thouless78jphysc}.
They clearly dominate in the quantum Hall regime \cite{chen-2006-2},
where both the spin degrees of freedom and the kinetic energy are
suppressed due to Landau quantization. But for 2D-MIT in zero magnetic field, these degrees of freedom may not be so important.
\vspace{12pt}

\item \textit{Single-particle excitations} leading to vacancy-interstitial
pair formation (Fig. \ref{wigner}). These excitations have a fermionic
character, where the spin degrees of freedom play an important role.
 Recent quantum Monte-Carlo simulations indicate \cite{candidi01prb}
that the effective gap for vacancy-interstitial pair formation seems
to collapse precisely around the quantum melting of the Wigner crystal. If
these excitation dominate, then quantum melting of the Wigner crystal
is a process very similar to the Mott metal-insulator transition,
and may be expected to produce a strongly correlated  Fermi liquid on the metallic side. This physical picture is the central idea of this article. 
\end{enumerate}

\begin{figure}[b]
\begin{centering}
\includegraphics[width=0.48\columnwidth]{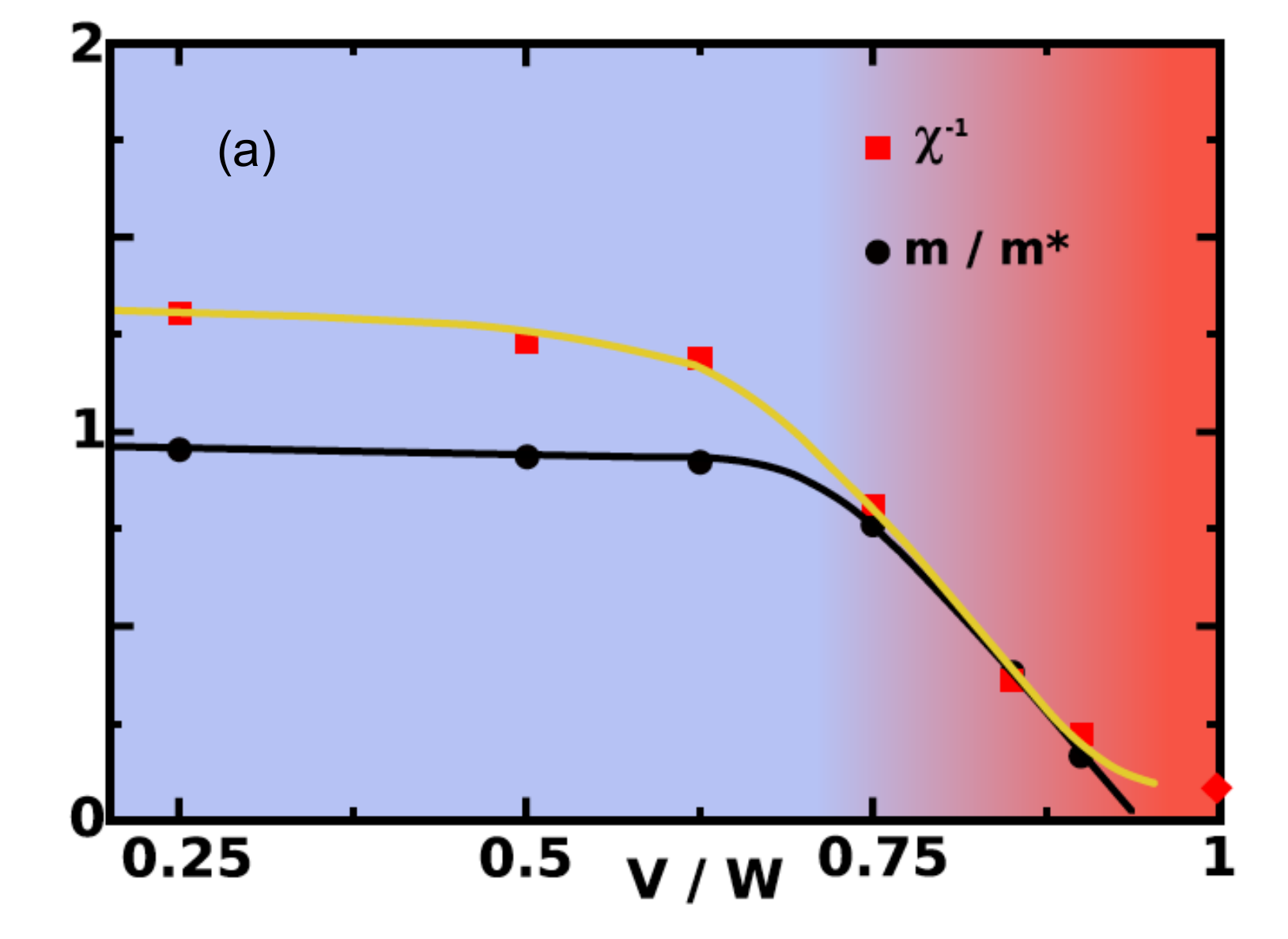}
\includegraphics[width=0.505\columnwidth]{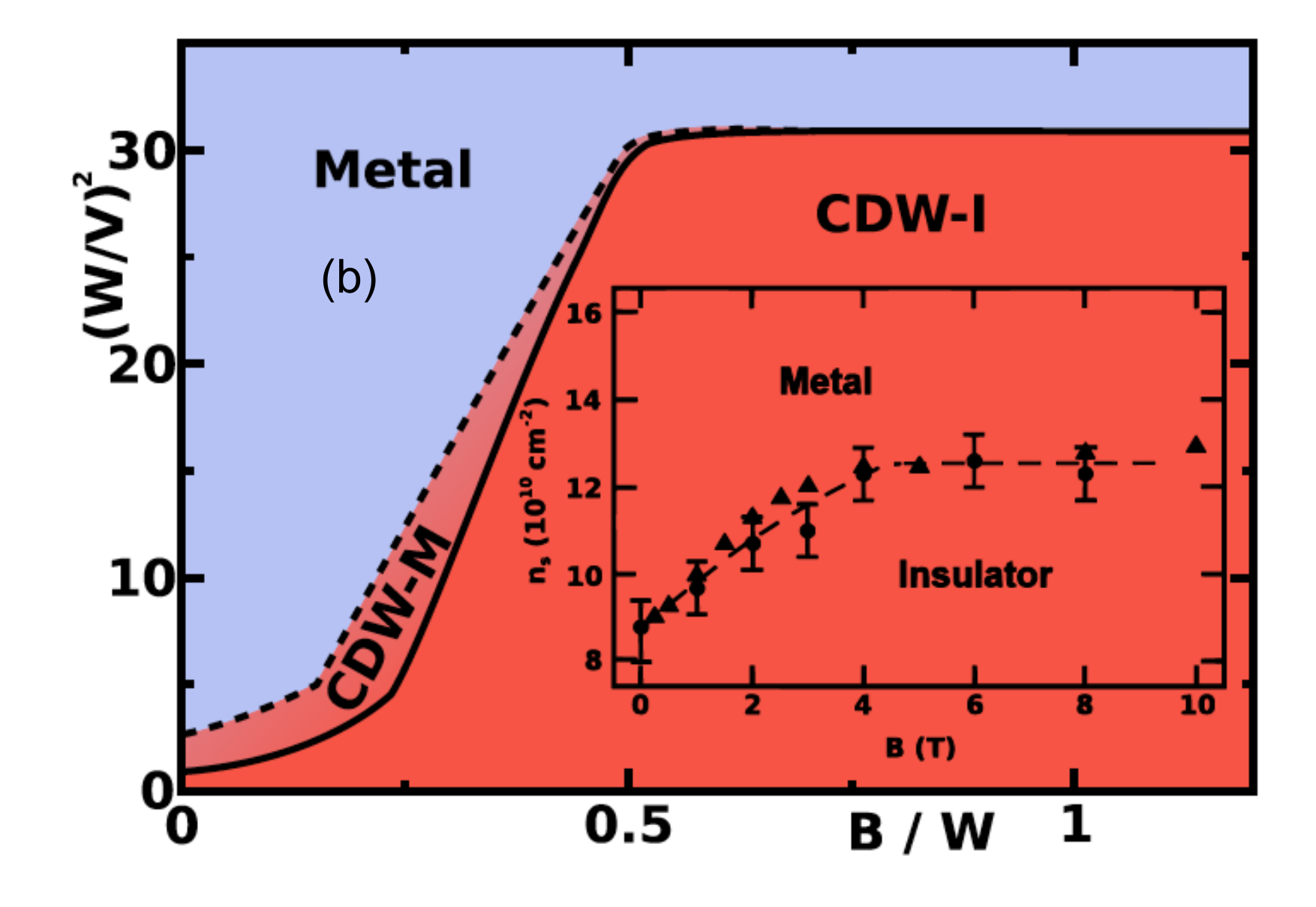}
\par\end{centering}
\caption{Toy model for Wigner crystallization. Results from the DMFT solution of a Hubbard model at quarter filling (from \cite{camjayi08nphys}) show (a)  that both the spin susceptibility $\chi$ and the effective mass $m^*$ are strongly enhanced as the transition is approached. (b) Phase diagram in presence of (parallel) magnetic field. A field-driven transition is found only sufficiently close to the insulating state, in agreement with experiments (the inset shows the experimental phase diagram \cite{shashkin2001,eng2001}).}
\vspace{-12pt}
\centering{}\label{wigner-mott}
\end{figure}

We should mention, however, that the Wigner crystal melting in zero
magnetic field is believed \cite{ceperley89prb} to be a weakly first
order phase transition. Conventional (e.g. liquid-gas or liquid-crystal)
first order transitions are normally associated with a density discontinuity
and global phase separation within the coexistence dome. For charged
systems, however, global phase separation is precluded by charge neutrality 
\protect\cite{gorkov-JETP87}. In this case, one may expect the emergence of
various modulated intermediate phases, leading to bubble or stripe
\protect\cite{spivak05}, or possibly even ``stripe
glass'' \protect\cite{schmalian-prl00, mahmoudian15prl} order. While
convincing evidence for the relevance of such ``nano-scale phase
separation'' has been identified \cite{terletska11aprl} in certain 
systems \cite{jan07prb}, recent work seems to indicate
\cite{waitnal06prb,ceperley09prl} that such effects may be negligibly
small for Wigner crystal melting.

A complementary line of work, where the Mott character of the transition is the focus  \cite{pankov08prb}, has been the subject of recent model calculations based on DMFT approaches. These works considered a toy lattice model for the Wigner-Mott transition by examining an extended Hubbard model at quarter-filling \cite{camjayi08nphys, camjayi10prb}, which has also been discussed \cite{fratini09prb,merino13prl} in the context of the $\theta$-family of Mott organics.  In general,  a spatially uniform phase cannot become a Mott insulator away from half filling, even for arbitrarily large values of the on-site repulsion $U$. However, when the inter-site interaction $V$ is sufficiently large (as compared to the bandwidth $W$), the electronic system undergoes charge ordering, where one sub-lattice becomes close to half-filling, while the other one become nearly empty. Such charge ordering, which is the lattice analogue of Wigner crystallization, results in an Mott insulating state, having localized spin-1/2 magnetic moments on each site of the carrier-rich sublattice. 

The corresponding MIT has a character very similar to the conventional Mott transition, and the approach from the metallic side resembles the familiar BR scenario \cite{brinkmann70prb}. One finds strong enhancement in both the spin susceptibility and the QP effective mass $m^*$, in agreement with experiments. A new feature, which differs from the standard Mott transition at half-filling, is the response to (parallel) magnetic fields (Zeeman coupling). This calculation finds \cite{camjayi08nphys}, in agreement with experiment on 2DEG systems \cite{shashkin2001,eng2001}, that a field-driven MIT arises only sufficiently close to the $B=0$ MIT, while deeper in the metallic phase even full spin polarization cannot completely destroy the metal. 

\section{Conclusions}

In this chapter we presented evidence suggesting that 2D-MIT found in ultra-clean 2DEG devices should be viewed as an interaction-driven MIT with many features in common with conventional Mott systems.  Striking similarities were established between thermodynamic and transport properties  of the respective experiments, but also with the predictions of generic theoretical models describing Mott and Wigner-Mott transitions. The resulting DMFT physical picture seems to offer, for quantum fluids, a perspective comparable to what the very successful Van der Waals theory provided for the classical liquid-gas critical point. It should give the proper starting point for future theoretical studies of more complicated experiments containing nontrivial interplay between strong correlations and disorder, as described in the chapter by D. Popovic. This fascinating research direction remains a challenge for upcoming theoretical work. 

\section{Acknowledgements} 

The authors are indebted to their  collaborators in the field: E. Abrahams, S. Chakravarty, S. Florens, S. Fratini, K. Haule, G. Kotliar, E. Manousakis, E. Miranda, G. Moeller, S. Pankov, D. Popovi\'c, Y. Pramudya, M. Radonji\'c, A. Ralko, M. J. Rozenberg, A. E. Ruckenstein, H. Terletska, and J. Vu\v ci\v cevi\'c, and have benefited from very useful discussions with A. Georges, A. M. Finkel'stein, S. Hartnoll, K. Kanoda, S. Kravchenko, R. H. McKenzie, Z. Y. Meng, A. J. Millis, P. Monceau, A. Punnoose, M. P. Sarachik, J. Saunders, T. Senthil, J. Schmalian, G. Sordi, A. M. Tramblay, M. Vojta, and J. Zaanen. V. D. was supported by the NSF grant  DMR-1410132 and the National High Magnetic Field Laboratory. D. T. acknowledges support from the Serbian Ministry of Education, Science and Technological Development under Project No.~ON171017.


\begin{thebibliography}{91}
\expandafter\ifx\csname natexlab\endcsname\relax\def\natexlab#1{#1}\fi
\expandafter\ifx\csname bibnamefont\endcsname\relax
  \def\bibnamefont#1{#1}\fi
\expandafter\ifx\csname bibfnamefont\endcsname\relax
  \def\bibfnamefont#1{#1}\fi
\expandafter\ifx\csname citenamefont\endcsname\relax
  \def\citenamefont#1{#1}\fi
\expandafter\ifx\csname url\endcsname\relax
  \def\url#1{\texttt{#1}}\fi
\expandafter\ifx\csname urlprefix\endcsname\relax\def\urlprefix{URL }\fi
\providecommand{\bibinfo}[2]{#2}
\providecommand{\eprint}[2][]{\url{#2}}

\bibitem[{\citenamefont{Dobbs}(2002)}]{newton-alchemist}
\bibinfo{author}{\bibfnamefont{B.~J.~T.} \bibnamefont{Dobbs}},
  \emph{\bibinfo{title}{The Janus Faces of Genius: The Role of Alchemy in
  Newton's Thought}} (\bibinfo{publisher}{Cambridge University Press},
  \bibinfo{address}{UK}, \bibinfo{year}{2002}).

\bibitem[{\citenamefont{Ashcroft and Mermin}(1976)}]{ashcroft}
\bibinfo{author}{\bibfnamefont{N.~W.} \bibnamefont{Ashcroft}} \bibnamefont{and}
  \bibinfo{author}{\bibfnamefont{D.}~\bibnamefont{Mermin}},
  \emph{\bibinfo{title}{Solid State Physics}} (\bibinfo{publisher}{Saunders
  College Publishing}, \bibinfo{year}{1976}), \bibinfo{edition}{international}
  ed.

\bibitem[{\citenamefont{Mott}(1990)}]{mott-book90}
\bibinfo{author}{\bibfnamefont{N.~F.} \bibnamefont{Mott}},
  \emph{\bibinfo{title}{Metal-{I}nsulator {T}ransition}}
  (\bibinfo{publisher}{Taylor \& Francis}, \bibinfo{address}{London},
  \bibinfo{year}{1990}).

\bibitem[{\citenamefont{Dobrosavljevi\'c
  et~al.}(2012)\citenamefont{Dobrosavljevi\'c, Trivedi, and
  Valles~Jr.}}]{dobrosavljevic2012conductor}
\bibinfo{author}{\bibfnamefont{V.}~\bibnamefont{Dobrosavljevi\'c}},
  \bibinfo{author}{\bibfnamefont{N.}~\bibnamefont{Trivedi}}, \bibnamefont{and}
  \bibinfo{author}{\bibfnamefont{J.~M.} \bibnamefont{Valles~Jr.}},
  \emph{\bibinfo{title}{Conductor Insulator Quantum Phase Transitions}}
  (\bibinfo{publisher}{Oxford University Press}, \bibinfo{address}{UK},
  \bibinfo{year}{2012}).

\bibitem[{\citenamefont{Slater}(1951)}]{slater51}
\bibinfo{author}{\bibfnamefont{J.~C.} \bibnamefont{Slater}},
  \bibinfo{journal}{Phys. Rev.} \textbf{\bibinfo{volume}{82}},
  \bibinfo{pages}{538} (\bibinfo{year}{1951}).

\bibitem[{\citenamefont{Mott}(1949)}]{mott1949}
\bibinfo{author}{\bibfnamefont{N.~F.} \bibnamefont{Mott}},
  \bibinfo{journal}{Proc. Phys. Soc. (London) A} \textbf{\bibinfo{volume}{62}},
  \bibinfo{pages}{416} (\bibinfo{year}{1949}).

\bibitem[{\citenamefont{Anderson}(1958)}]{anderson1958}
\bibinfo{author}{\bibfnamefont{P.}~\bibnamefont{Anderson}},
  \bibinfo{journal}{Physical Review} \textbf{\bibinfo{volume}{109}},
  \bibinfo{pages}{1492} (\bibinfo{year}{1958}).

\bibitem[{\citenamefont{Lee and Ramakrishnan}(1985)}]{leeramakrishnan}
\bibinfo{author}{\bibfnamefont{P.~A.} \bibnamefont{Lee}} \bibnamefont{and}
  \bibinfo{author}{\bibfnamefont{T.~V.} \bibnamefont{Ramakrishnan}},
  \bibinfo{journal}{Rev. Mod. Phys.} \textbf{\bibinfo{volume}{57}},
  \bibinfo{pages}{287} (\bibinfo{year}{1985}).

\bibitem[{\citenamefont{Finkel'stein}(1983)}]{fink-jetp83}
\bibinfo{author}{\bibfnamefont{A.~M.} \bibnamefont{Finkel'stein}},
  \bibinfo{journal}{Zh. Eksp. Teor. Fiz.} \textbf{\bibinfo{volume}{84}},
  \bibinfo{pages}{168} (\bibinfo{year}{1983}), \bibinfo{note}{[Sov. Phys. JETP
  {\bf 57}, 97 (1983)]}.

\bibitem[{\citenamefont{Finkel'stein}(1984)}]{fink-jetp84}
\bibinfo{author}{\bibfnamefont{A.~M.} \bibnamefont{Finkel'stein}},
  \bibinfo{journal}{Zh. Eksp. Teor. Fiz.} \textbf{\bibinfo{volume}{86}},
  \bibinfo{pages}{367} (\bibinfo{year}{1984}), \bibinfo{note}{[Sov. Phys. JETP
  {\bf 59}, 212 (1983)]}.

\bibitem[{\citenamefont{Miranda and Dobrosavljevic}(2005)}]{RoP2005review}
\bibinfo{author}{\bibfnamefont{E.}~\bibnamefont{Miranda}} \bibnamefont{and}
  \bibinfo{author}{\bibfnamefont{V.}~\bibnamefont{Dobrosavljevic}},
  \bibinfo{journal}{Reports on Progress in Physics}
  \textbf{\bibinfo{volume}{68}}, \bibinfo{pages}{2337} (\bibinfo{year}{2005}).

\bibitem[{\citenamefont{Fowler et~al.}(1966)\citenamefont{Fowler, Fang, Howard,
  and Stiles}}]{fowler1966}
\bibinfo{author}{\bibfnamefont{A.~B.} \bibnamefont{Fowler}},
  \bibinfo{author}{\bibfnamefont{F.~F.} \bibnamefont{Fang}},
  \bibinfo{author}{\bibfnamefont{W.~E.} \bibnamefont{Howard}},
  \bibnamefont{and} \bibinfo{author}{\bibfnamefont{P.~J.}
  \bibnamefont{Stiles}}, \bibinfo{journal}{Phys. Rev. Lett.}
  \textbf{\bibinfo{volume}{16}}, \bibinfo{pages}{901} (\bibinfo{year}{1966}).

\bibitem[{\citenamefont{Ando et~al.}(1982)\citenamefont{Ando, Fowler, and
  Stern}}]{AFS1982}
\bibinfo{author}{\bibfnamefont{T.}~\bibnamefont{Ando}},
  \bibinfo{author}{\bibfnamefont{A.~B.} \bibnamefont{Fowler}},
  \bibnamefont{and} \bibinfo{author}{\bibfnamefont{F.}~\bibnamefont{Stern}},
  \bibinfo{journal}{Rev. Mod. Phys.} \textbf{\bibinfo{volume}{54}},
  \bibinfo{pages}{437} (\bibinfo{year}{1982}).

\bibitem[{\citenamefont{Kravchenko et~al.}(1995)\citenamefont{Kravchenko,
  Mason, Bowker, Furneaux, Pudalov, and D'Iorio}}]{kravchenko95}
\bibinfo{author}{\bibfnamefont{S.~V.} \bibnamefont{Kravchenko}},
  \bibinfo{author}{\bibfnamefont{W.~E.} \bibnamefont{Mason}},
  \bibinfo{author}{\bibfnamefont{G.~E.} \bibnamefont{Bowker}},
  \bibinfo{author}{\bibfnamefont{J.~E.} \bibnamefont{Furneaux}},
  \bibinfo{author}{\bibfnamefont{V.~M.} \bibnamefont{Pudalov}},
  \bibnamefont{and} \bibinfo{author}{\bibfnamefont{M.}~\bibnamefont{D'Iorio}},
  \bibinfo{journal}{Phys. Rev. B} \textbf{\bibinfo{volume}{51}},
  \bibinfo{pages}{7038} (\bibinfo{year}{1995}).

\bibitem[{\citenamefont{Abrahams et~al.}(1979)\citenamefont{Abrahams, Anderson,
  Licciardello, and Ramakrishnan}}]{gang4}
\bibinfo{author}{\bibfnamefont{E.}~\bibnamefont{Abrahams}},
  \bibinfo{author}{\bibfnamefont{P.~W.} \bibnamefont{Anderson}},
  \bibinfo{author}{\bibfnamefont{D.~C.} \bibnamefont{Licciardello}},
  \bibnamefont{and} \bibinfo{author}{\bibfnamefont{T.~V.}
  \bibnamefont{Ramakrishnan}}, \bibinfo{journal}{Phys. Rev. Lett.}
  \textbf{\bibinfo{volume}{42}}, \bibinfo{pages}{673} (\bibinfo{year}{1979}).

\bibitem[{\citenamefont{Altshuler et~al.}(2001)\citenamefont{Altshuler, Maslov,
  and Pudalov}}]{maslov2001}
\bibinfo{author}{\bibfnamefont{B.~L.} \bibnamefont{Altshuler}},
  \bibinfo{author}{\bibfnamefont{D.~L.} \bibnamefont{Maslov}},
  \bibnamefont{and} \bibinfo{author}{\bibfnamefont{V.~M.}
  \bibnamefont{Pudalov}}, \bibinfo{journal}{Physica (Amsterdam)}
  \textbf{\bibinfo{volume}{9E}}, \bibinfo{pages}{209} (\bibinfo{year}{2001}).

\bibitem[{\citenamefont{Popovi\'{c} et~al.}(1997)\citenamefont{Popovi\'{c},
  Fowler, and Washburn}}]{popovic97}
\bibinfo{author}{\bibfnamefont{D.}~\bibnamefont{Popovi\'{c}}},
  \bibinfo{author}{\bibfnamefont{A.~B.} \bibnamefont{Fowler}},
  \bibnamefont{and} \bibinfo{author}{\bibfnamefont{S.}~\bibnamefont{Washburn}},
  \bibinfo{journal}{Phys. Rev. Lett.} \textbf{\bibinfo{volume}{79}},
  \bibinfo{pages}{1543} (\bibinfo{year}{1997}).

\bibitem[{\citenamefont{Abrahams et~al.}(2001)\citenamefont{Abrahams,
  Kravchenko, and Sarachik}}]{abrahams01}
\bibinfo{author}{\bibfnamefont{E.}~\bibnamefont{Abrahams}},
  \bibinfo{author}{\bibfnamefont{S.~V.} \bibnamefont{Kravchenko}},
  \bibnamefont{and} \bibinfo{author}{\bibfnamefont{M.~P.}
  \bibnamefont{Sarachik}}, \bibinfo{journal}{Rev. Mod. Phys.}
  \textbf{\bibinfo{volume}{73}}, \bibinfo{pages}{251} (\bibinfo{year}{2001}).

\bibitem[{\citenamefont{Zala et~al.}(2001)\citenamefont{Zala, Narozhny, and
  Aleiner}}]{Zala}
\bibinfo{author}{\bibfnamefont{G.}~\bibnamefont{Zala}},
  \bibinfo{author}{\bibfnamefont{B.~N.} \bibnamefont{Narozhny}},
  \bibnamefont{and} \bibinfo{author}{\bibfnamefont{I.~L.}
  \bibnamefont{Aleiner}}, \bibinfo{journal}{Phys. Rev. B}
  \textbf{\bibinfo{volume}{64}}, \bibinfo{pages}{214204}
  (\bibinfo{year}{2001}).

\bibitem[{\citenamefont{Andrade et~al.}(2009)\citenamefont{Andrade, Miranda,
  and Dobrosavljevic}}]{andrade09prl}
\bibinfo{author}{\bibfnamefont{E.~C.} \bibnamefont{Andrade}},
  \bibinfo{author}{\bibfnamefont{E.}~\bibnamefont{Miranda}}, \bibnamefont{and}
  \bibinfo{author}{\bibfnamefont{V.}~\bibnamefont{Dobrosavljevic}},
  \bibinfo{journal}{Phys. Rev. Lett.} \textbf{\bibinfo{volume}{102}},
  \bibinfo{pages}{206403} (\bibinfo{year}{2009}).

\bibitem[{\citenamefont{Punnoose and Finkel'stein}(2002)}]{punnoose02}
\bibinfo{author}{\bibfnamefont{A.}~\bibnamefont{Punnoose}} \bibnamefont{and}
  \bibinfo{author}{\bibfnamefont{A.~M.} \bibnamefont{Finkel'stein}},
  \bibinfo{journal}{Phys. Rev. Lett.} \textbf{\bibinfo{volume}{88}},
  \bibinfo{pages}{016802} (\bibinfo{year}{2002}).

\bibitem[{\citenamefont{Stewart}(1984)}]{stewartrev1}
\bibinfo{author}{\bibfnamefont{G.~R.} \bibnamefont{Stewart}},
  \bibinfo{journal}{Rev. Mod. Phys.} \textbf{\bibinfo{volume}{56}},
  \bibinfo{pages}{755} (\bibinfo{year}{1984}).

\bibitem[{\citenamefont{Goodenough}(1963)}]{goddenough63book}
\bibinfo{author}{\bibfnamefont{J.~B.} \bibnamefont{Goodenough}},
  \emph{\bibinfo{title}{Magnetism and the {C}hemical {B}ond}}
  (\bibinfo{publisher}{John Wiley \& Sons}, \bibinfo{address}{New York -
  London}, \bibinfo{year}{1963}).

\bibitem[{\citenamefont{Georges et~al.}(1996)\citenamefont{Georges, Kotliar,
  Krauth, and Rozenberg}}]{dmft96}
\bibinfo{author}{\bibfnamefont{A.}~\bibnamefont{Georges}},
  \bibinfo{author}{\bibfnamefont{G.}~\bibnamefont{Kotliar}},
  \bibinfo{author}{\bibfnamefont{W.}~\bibnamefont{Krauth}}, \bibnamefont{and}
  \bibinfo{author}{\bibfnamefont{M.~J.} \bibnamefont{Rozenberg}},
  \bibinfo{journal}{Rev. Mod. Phys.} \textbf{\bibinfo{volume}{68}},
  \bibinfo{pages}{13} (\bibinfo{year}{1996}).

\bibitem[{\citenamefont{Goldenfeld}(1992)}]{goldenfeldbook}
\bibinfo{author}{\bibfnamefont{N.}~\bibnamefont{Goldenfeld}},
  \emph{\bibinfo{title}{Lectures on phase transitions and the renormalization
  group}} (\bibinfo{publisher}{Addison-Wesley}, \bibinfo{address}{Reading},
  \bibinfo{year}{1992}).

\bibitem[{\citenamefont{Sachdev}(2011)}]{sachdevbook}
\bibinfo{author}{\bibfnamefont{S.}~\bibnamefont{Sachdev}},
  \emph{\bibinfo{title}{Quantum phase transitions, 2nd Edition}}
  (\bibinfo{publisher}{Cambridge University Press}, \bibinfo{address}{UK},
  \bibinfo{year}{2011}).

\bibitem[{\citenamefont{Simonian
  et~al.}(1997{\natexlab{a}})\citenamefont{Simonian, Kravchenko, and
  Sarachik}}]{simonian97}
\bibinfo{author}{\bibfnamefont{D.}~\bibnamefont{Simonian}},
  \bibinfo{author}{\bibfnamefont{S.~V.} \bibnamefont{Kravchenko}},
  \bibnamefont{and} \bibinfo{author}{\bibfnamefont{M.~P.}
  \bibnamefont{Sarachik}}, \bibinfo{journal}{Phys. Rev. B}
  \textbf{\bibinfo{volume}{55}}, \bibinfo{pages}{R13421}
  (\bibinfo{year}{1997}{\natexlab{a}}).

\bibitem[{\citenamefont{Dobrosavljevi\'{c}
  et~al.}(1997)\citenamefont{Dobrosavljevi\'{c}, Abrahams, Miranda, and
  Chakravarty}}]{gang4me}
\bibinfo{author}{\bibfnamefont{V.}~\bibnamefont{Dobrosavljevi\'{c}}},
  \bibinfo{author}{\bibfnamefont{E.}~\bibnamefont{Abrahams}},
  \bibinfo{author}{\bibfnamefont{E.}~\bibnamefont{Miranda}}, \bibnamefont{and}
  \bibinfo{author}{\bibfnamefont{S.}~\bibnamefont{Chakravarty}},
  \bibinfo{journal}{Phys. Rev. Lett.} \textbf{\bibinfo{volume}{79}},
  \bibinfo{pages}{455} (\bibinfo{year}{1997}).

\bibitem[{\citenamefont{Shklovskii and Efros}(1984)}]{efros-book}
\bibinfo{author}{\bibfnamefont{B.~I.} \bibnamefont{Shklovskii}}
  \bibnamefont{and} \bibinfo{author}{\bibfnamefont{A.~L.} \bibnamefont{Efros}},
  \emph{\bibinfo{title}{Electronic Properties of Doped Semiconductors}}
  (\bibinfo{publisher}{Springer-Verlag}, \bibinfo{year}{1984}).

\bibitem[{\citenamefont{Furukawa et~al.}(2015)\citenamefont{Furukawa, Miyagawa,
  Taniguchi, Kato, and Kanoda}}]{kanoda2015nphys}
\bibinfo{author}{\bibfnamefont{T.}~\bibnamefont{Furukawa}},
  \bibinfo{author}{\bibfnamefont{K.}~\bibnamefont{Miyagawa}},
  \bibinfo{author}{\bibfnamefont{H.}~\bibnamefont{Taniguchi}},
  \bibinfo{author}{\bibfnamefont{R.}~\bibnamefont{Kato}}, \bibnamefont{and}
  \bibinfo{author}{\bibfnamefont{K.}~\bibnamefont{Kanoda}},
  \bibinfo{journal}{Nat. Phys.} \textbf{\bibinfo{volume}{11}},
  \bibinfo{pages}{221} (\bibinfo{year}{2015}).

\bibitem[{\citenamefont{Pudalov et~al.}(1998)\citenamefont{Pudalov, Brunthaler,
  Prinz, and Bauer}}]{pudalov98}
\bibinfo{author}{\bibfnamefont{V.}~\bibnamefont{Pudalov}},
  \bibinfo{author}{\bibfnamefont{G.}~\bibnamefont{Brunthaler}},
  \bibinfo{author}{\bibfnamefont{A.}~\bibnamefont{Prinz}}, \bibnamefont{and}
  \bibinfo{author}{\bibfnamefont{G.}~\bibnamefont{Bauer}},
  \bibinfo{journal}{Physica E: Low-dimensional Systems and Nanostructures}
  \textbf{\bibinfo{volume}{3}}, \bibinfo{pages}{79} (\bibinfo{year}{1998}).

\bibitem[{\citenamefont{Radonji\ifmmode~\acute{c}\else \'{c}\fi{}
  et~al.}(2012)\citenamefont{Radonji\ifmmode~\acute{c}\else \'{c}\fi{},
  Tanaskovi\ifmmode~\acute{c}\else \'{c}\fi{},
  Dobrosavljevi\ifmmode~\acute{c}\else \'{c}\fi{}, Haule, and
  Kotliar}}]{radonjic12prb}
\bibinfo{author}{\bibfnamefont{M.~M.}
  \bibnamefont{Radonji\ifmmode~\acute{c}\else \'{c}\fi{}}},
  \bibinfo{author}{\bibfnamefont{D.}~\bibnamefont{Tanaskovi\ifmmode~\acute{c}\else
  \'{c}\fi{}}},
  \bibinfo{author}{\bibfnamefont{V.}~\bibnamefont{Dobrosavljevi\ifmmode~\acute{c}\else
  \'{c}\fi{}}}, \bibinfo{author}{\bibfnamefont{K.}~\bibnamefont{Haule}},
  \bibnamefont{and} \bibinfo{author}{\bibfnamefont{G.}~\bibnamefont{Kotliar}},
  \bibinfo{journal}{Phys. Rev. B} \textbf{\bibinfo{volume}{85}},
  \bibinfo{pages}{085133} (\bibinfo{year}{2012}).

\bibitem[{\citenamefont{Simonian
  et~al.}(1997{\natexlab{b}})\citenamefont{Simonian, Kravchenko, Sarachik, and
  Pudalov}}]{simonian1997}
\bibinfo{author}{\bibfnamefont{D.}~\bibnamefont{Simonian}},
  \bibinfo{author}{\bibfnamefont{S.~V.} \bibnamefont{Kravchenko}},
  \bibinfo{author}{\bibfnamefont{M.~P.} \bibnamefont{Sarachik}},
  \bibnamefont{and} \bibinfo{author}{\bibfnamefont{V.~M.}
  \bibnamefont{Pudalov}}, \bibinfo{journal}{Phys. Rev. Lett.}
  \textbf{\bibinfo{volume}{79}}, \bibinfo{pages}{2304}
  (\bibinfo{year}{1997}{\natexlab{b}}).

\bibitem[{\citenamefont{Shashkin et~al.}(2001)\citenamefont{Shashkin,
  Kravchenko, Dolgopolov, and Klapwijk}}]{shashkin2001}
\bibinfo{author}{\bibfnamefont{A.~A.} \bibnamefont{Shashkin}},
  \bibinfo{author}{\bibfnamefont{S.~V.} \bibnamefont{Kravchenko}},
  \bibinfo{author}{\bibfnamefont{V.~T.} \bibnamefont{Dolgopolov}},
  \bibnamefont{and} \bibinfo{author}{\bibfnamefont{T.~M.}
  \bibnamefont{Klapwijk}}, \bibinfo{journal}{Phys. Rev. Lett.}
  \textbf{\bibinfo{volume}{87}}, \bibinfo{pages}{086801}
  (\bibinfo{year}{2001}).

\bibitem[{\citenamefont{Jamei et~al.}(2005)\citenamefont{Jamei, Kivelson, and
  Spivak}}]{spivak05}
\bibinfo{author}{\bibfnamefont{R.}~\bibnamefont{Jamei}},
  \bibinfo{author}{\bibfnamefont{S.}~\bibnamefont{Kivelson}}, \bibnamefont{and}
  \bibinfo{author}{\bibfnamefont{B.}~\bibnamefont{Spivak}},
  \bibinfo{journal}{Phys. Rev. Lett.} \textbf{\bibinfo{volume}{94}},
  \bibinfo{pages}{056805} (\bibinfo{year}{2005}).

\bibitem[{\citenamefont{Eng et~al.}(2002)\citenamefont{Eng, Feng,
  Popovi\ifmmode~\acute{c}\else \'{c}\fi{}, and Washburn}}]{eng2001}
\bibinfo{author}{\bibfnamefont{K.}~\bibnamefont{Eng}},
  \bibinfo{author}{\bibfnamefont{X.~G.} \bibnamefont{Feng}},
  \bibinfo{author}{\bibfnamefont{D.}~\bibnamefont{Popovi\ifmmode~\acute{c}\else
  \'{c}\fi{}}}, \bibnamefont{and}
  \bibinfo{author}{\bibfnamefont{S.}~\bibnamefont{Washburn}},
  \bibinfo{journal}{Phys. Rev. Lett.} \textbf{\bibinfo{volume}{88}},
  \bibinfo{pages}{136402} (\bibinfo{year}{2002}).

\bibitem[{\citenamefont{Sarachik}(1995)}]{sarachik95}
\bibinfo{author}{\bibfnamefont{M.~P.} \bibnamefont{Sarachik}}, in
  \emph{\bibinfo{booktitle}{Metal-Insulator Transitions Revisited}}, edited by
  \bibinfo{editor}{\bibfnamefont{P.}~\bibnamefont{Edwards}} \bibnamefont{and}
  \bibinfo{editor}{\bibfnamefont{C.~N.~R.} \bibnamefont{Rao}}
  (\bibinfo{publisher}{Taylor and Francis}, \bibinfo{year}{1995}).

\bibitem[{\citenamefont{Kravchenko and Sarachik}(2004)}]{kravchenko2004}
\bibinfo{author}{\bibfnamefont{S.~V.} \bibnamefont{Kravchenko}}
  \bibnamefont{and} \bibinfo{author}{\bibfnamefont{M.~P.}
  \bibnamefont{Sarachik}}, \bibinfo{journal}{Reports on Progress in Physics}
  \textbf{\bibinfo{volume}{67}}, \bibinfo{pages}{1} (\bibinfo{year}{2004}).

\bibitem[{\citenamefont{Prus et~al.}(2003)\citenamefont{Prus, Yaish, Reznikov,
  Sivan, and Pudalov}}]{reznikov2003}
\bibinfo{author}{\bibfnamefont{O.}~\bibnamefont{Prus}},
  \bibinfo{author}{\bibfnamefont{Y.}~\bibnamefont{Yaish}},
  \bibinfo{author}{\bibfnamefont{M.}~\bibnamefont{Reznikov}},
  \bibinfo{author}{\bibfnamefont{U.}~\bibnamefont{Sivan}}, \bibnamefont{and}
  \bibinfo{author}{\bibfnamefont{V.}~\bibnamefont{Pudalov}},
  \bibinfo{journal}{Phys. Rev. B} \textbf{\bibinfo{volume}{67}},
  \bibinfo{pages}{205407} (\bibinfo{year}{2003}).

\bibitem[{\citenamefont{Shashkin et~al.}(2002)\citenamefont{Shashkin,
  Kravchenko, Dolgopolov, and Klapwijk}}]{shashkin2002}
\bibinfo{author}{\bibfnamefont{A.~A.} \bibnamefont{Shashkin}},
  \bibinfo{author}{\bibfnamefont{S.~V.} \bibnamefont{Kravchenko}},
  \bibinfo{author}{\bibfnamefont{V.~T.} \bibnamefont{Dolgopolov}},
  \bibnamefont{and} \bibinfo{author}{\bibfnamefont{T.~M.}
  \bibnamefont{Klapwijk}}, \bibinfo{journal}{Phys. Rev. B}
  \textbf{\bibinfo{volume}{66}}, \bibinfo{pages}{073303}
  (\bibinfo{year}{2002}).

\bibitem[{\citenamefont{Mokashi et~al.}(2012)\citenamefont{Mokashi, Li, Wen,
  Kravchenko, Shashkin, Dolgopolov, and Sarachik}}]{kravchenko2012}
\bibinfo{author}{\bibfnamefont{A.}~\bibnamefont{Mokashi}},
  \bibinfo{author}{\bibfnamefont{S.}~\bibnamefont{Li}},
  \bibinfo{author}{\bibfnamefont{B.}~\bibnamefont{Wen}},
  \bibinfo{author}{\bibfnamefont{S.~V.} \bibnamefont{Kravchenko}},
  \bibinfo{author}{\bibfnamefont{A.~A.} \bibnamefont{Shashkin}},
  \bibinfo{author}{\bibfnamefont{V.~T.} \bibnamefont{Dolgopolov}},
  \bibnamefont{and} \bibinfo{author}{\bibfnamefont{M.~P.}
  \bibnamefont{Sarachik}}, \bibinfo{journal}{Phys. Rev. Lett.}
  \textbf{\bibinfo{volume}{109}}, \bibinfo{pages}{096405}
  (\bibinfo{year}{2012}).

\bibitem[{\citenamefont{Slater}(1934)}]{slater34rmp}
\bibinfo{author}{\bibfnamefont{J.~C.} \bibnamefont{Slater}},
  \bibinfo{journal}{Rev. Mod. Phys.} \textbf{\bibinfo{volume}{6}},
  \bibinfo{pages}{209} (\bibinfo{year}{1934}).

\bibitem[{\citenamefont{Vollhardt}(1984)}]{vollhardt84rmp}
\bibinfo{author}{\bibfnamefont{D.}~\bibnamefont{Vollhardt}},
  \bibinfo{journal}{Rev. Mod. Phys.} \textbf{\bibinfo{volume}{56}},
  \bibinfo{pages}{99} (\bibinfo{year}{1984}).

\bibitem[{\citenamefont{Pines and Nozi{\`e}res}(1965)}]{pinesnozieres}
\bibinfo{author}{\bibfnamefont{D.}~\bibnamefont{Pines}} \bibnamefont{and}
  \bibinfo{author}{\bibfnamefont{P.}~\bibnamefont{Nozi{\`e}res}},
  \emph{\bibinfo{title}{The {T}heory of {Q}uantum {L}iquids}}
  (\bibinfo{publisher}{Benjamin}, \bibinfo{address}{New York},
  \bibinfo{year}{1965}).

\bibitem[{\citenamefont{Landau}(1957)}]{landauFL1}
\bibinfo{author}{\bibfnamefont{L.~D.} \bibnamefont{Landau}},
  \bibinfo{journal}{Sov. Phys. JETP} \textbf{\bibinfo{volume}{3}},
  \bibinfo{pages}{920} (\bibinfo{year}{1957}).

\bibitem[{\citenamefont{Brinkman and Rice}(1970)}]{brinkmann70prb}
\bibinfo{author}{\bibfnamefont{W.~F.} \bibnamefont{Brinkman}} \bibnamefont{and}
  \bibinfo{author}{\bibfnamefont{T.}~\bibnamefont{Rice}},
  \bibinfo{journal}{Phys. Rev. B} \textbf{\bibinfo{volume}{2}},
  \bibinfo{pages}{4302} (\bibinfo{year}{1970}).

\bibitem[{\citenamefont{Casey et~al.}(2003)\citenamefont{Casey, Patel, Ny\'eki,
  Cowan, and Saunders}}]{casey03prl}
\bibinfo{author}{\bibfnamefont{A.}~\bibnamefont{Casey}},
  \bibinfo{author}{\bibfnamefont{H.}~\bibnamefont{Patel}},
  \bibinfo{author}{\bibfnamefont{J.}~\bibnamefont{Ny\'eki}},
  \bibinfo{author}{\bibfnamefont{B.~P.} \bibnamefont{Cowan}}, \bibnamefont{and}
  \bibinfo{author}{\bibfnamefont{J.}~\bibnamefont{Saunders}},
  \bibinfo{journal}{Phys. Rev. Lett} \textbf{\bibinfo{volume}{90}},
  \bibinfo{pages}{115301} (\bibinfo{year}{2003}).

\bibitem[{\citenamefont{Misguich et~al.}(1998)\citenamefont{Misguich, Bernu,
  Lhuillier, and Waldtmann}}]{misguich98prl}
\bibinfo{author}{\bibfnamefont{G.}~\bibnamefont{Misguich}},
  \bibinfo{author}{\bibfnamefont{B.}~\bibnamefont{Bernu}},
  \bibinfo{author}{\bibfnamefont{C.}~\bibnamefont{Lhuillier}},
  \bibnamefont{and}
  \bibinfo{author}{\bibfnamefont{C.}~\bibnamefont{Waldtmann}},
  \bibinfo{journal}{Phys. Rev. Lett.} \textbf{\bibinfo{volume}{81}},
  \bibinfo{pages}{1098} (\bibinfo{year}{1998}).

\bibitem[{\citenamefont{Powell and McKenzie}(2011)}]{mckenzie2011review}
\bibinfo{author}{\bibfnamefont{B.~J.} \bibnamefont{Powell}} \bibnamefont{and}
  \bibinfo{author}{\bibfnamefont{R.~H.} \bibnamefont{McKenzie}},
  \bibinfo{journal}{Reports on Progress in Physics}
  \textbf{\bibinfo{volume}{74}}, \bibinfo{pages}{056501}
  (\bibinfo{year}{2011}).

\bibitem[{\citenamefont{Els\"asser et~al.}(2012)\citenamefont{Els\"asser, Wu,
  Dressel, and Schlueter}}]{elsasser2012}
\bibinfo{author}{\bibfnamefont{S.}~\bibnamefont{Els\"asser}},
  \bibinfo{author}{\bibfnamefont{D.}~\bibnamefont{Wu}},
  \bibinfo{author}{\bibfnamefont{M.}~\bibnamefont{Dressel}}, \bibnamefont{and}
  \bibinfo{author}{\bibfnamefont{J.~A.} \bibnamefont{Schlueter}},
  \bibinfo{journal}{Phys. Rev. B} \textbf{\bibinfo{volume}{86}},
  \bibinfo{pages}{155150} (\bibinfo{year}{2012}).

\bibitem[{\citenamefont{Kurosaki et~al.}(2005)\citenamefont{Kurosaki, Shimizu,
  Miyagawa, Kanoda, and Saito}}]{kanoda2005prl}
\bibinfo{author}{\bibfnamefont{Y.}~\bibnamefont{Kurosaki}},
  \bibinfo{author}{\bibfnamefont{Y.}~\bibnamefont{Shimizu}},
  \bibinfo{author}{\bibfnamefont{K.}~\bibnamefont{Miyagawa}},
  \bibinfo{author}{\bibfnamefont{K.}~\bibnamefont{Kanoda}}, \bibnamefont{and}
  \bibinfo{author}{\bibfnamefont{G.}~\bibnamefont{Saito}},
  \bibinfo{journal}{Phys. Rev. Lett.} \textbf{\bibinfo{volume}{95}},
  \bibinfo{pages}{177001} (\bibinfo{year}{2005}).

\bibitem[{\citenamefont{Limelette et~al.}(2003)\citenamefont{Limelette,
  Wzietek, Florens, Georges, Costi, Pasquier, Jerome, Meziere, and
  Batail}}]{limelette03prl}
\bibinfo{author}{\bibfnamefont{P.}~\bibnamefont{Limelette}},
  \bibinfo{author}{\bibfnamefont{P.}~\bibnamefont{Wzietek}},
  \bibinfo{author}{\bibfnamefont{S.}~\bibnamefont{Florens}},
  \bibinfo{author}{\bibfnamefont{A.}~\bibnamefont{Georges}},
  \bibinfo{author}{\bibfnamefont{T.~A.} \bibnamefont{Costi}},
  \bibinfo{author}{\bibfnamefont{C.}~\bibnamefont{Pasquier}},
  \bibinfo{author}{\bibfnamefont{D.}~\bibnamefont{Jerome}},
  \bibinfo{author}{\bibfnamefont{C.}~\bibnamefont{Meziere}}, \bibnamefont{and}
  \bibinfo{author}{\bibfnamefont{P.}~\bibnamefont{Batail}},
  \bibinfo{journal}{Phys. Rev. Lett.} \textbf{\bibinfo{volume}{91}},
  \bibinfo{pages}{016401} (\bibinfo{year}{2003}).

\bibitem[{\citenamefont{Terletska et~al.}(2011)\citenamefont{Terletska,
  Vu\ifmmode \check{c}\else \v{c}\fi{}i\ifmmode \check{c}\else
  \v{c}\fi{}evi\ifmmode~\acute{c}\else \'{c}\fi{},
  Tanaskovi\ifmmode~\acute{c}\else \'{c}\fi{}, and
  Dobrosavljevi\ifmmode~\acute{c}\else \'{c}\fi{}}}]{terletska11prl}
\bibinfo{author}{\bibfnamefont{H.}~\bibnamefont{Terletska}},
  \bibinfo{author}{\bibfnamefont{J.}~\bibnamefont{Vu\ifmmode \check{c}\else
  \v{c}\fi{}i\ifmmode \check{c}\else \v{c}\fi{}evi\ifmmode~\acute{c}\else
  \'{c}\fi{}}},
  \bibinfo{author}{\bibfnamefont{D.}~\bibnamefont{Tanaskovi\ifmmode~\acute{c}\else
  \'{c}\fi{}}}, \bibnamefont{and}
  \bibinfo{author}{\bibfnamefont{V.}~\bibnamefont{Dobrosavljevi\ifmmode~\acute{c}\else
  \'{c}\fi{}}}, \bibinfo{journal}{Phys. Rev. Lett.}
  \textbf{\bibinfo{volume}{107}}, \bibinfo{pages}{026401}
  (\bibinfo{year}{2011}).

\bibitem[{\citenamefont{Hubbard}(1963)}]{hubbard3}
\bibinfo{author}{\bibfnamefont{J.}~\bibnamefont{Hubbard}},
  \bibinfo{journal}{Proc. R. Soc. (London) A} \textbf{\bibinfo{volume}{276}},
  \bibinfo{pages}{238} (\bibinfo{year}{1963}).

\bibitem[{\citenamefont{Kotliar et~al.}(2006)\citenamefont{Kotliar, Savrasov,
  Haule, Oudovenko, Parcollet, and Marianetti}}]{LDA-DMFT06rmp}
\bibinfo{author}{\bibfnamefont{G.}~\bibnamefont{Kotliar}},
  \bibinfo{author}{\bibfnamefont{S.~Y.} \bibnamefont{Savrasov}},
  \bibinfo{author}{\bibfnamefont{K.}~\bibnamefont{Haule}},
  \bibinfo{author}{\bibfnamefont{V.~S.} \bibnamefont{Oudovenko}},
  \bibinfo{author}{\bibfnamefont{O.}~\bibnamefont{Parcollet}},
  \bibnamefont{and} \bibinfo{author}{\bibfnamefont{C.~A.}
  \bibnamefont{Marianetti}}, \bibinfo{journal}{Rev. Mod. Phys.}
  \textbf{\bibinfo{volume}{78}}, \bibinfo{pages}{865} (\bibinfo{year}{2006}).

\bibitem[{\citenamefont{Tanaskovi\ifmmode~\acute{c}\else \'{c}\fi{}
  et~al.}(2011)\citenamefont{Tanaskovi\ifmmode~\acute{c}\else \'{c}\fi{},
  Haule, Kotliar, and Dobrosavljevi\ifmmode~\acute{c}\else
  \'{c}\fi{}}}]{tanaskovic11prb}
\bibinfo{author}{\bibfnamefont{D.}~\bibnamefont{Tanaskovi\ifmmode~\acute{c}\else
  \'{c}\fi{}}}, \bibinfo{author}{\bibfnamefont{K.}~\bibnamefont{Haule}},
  \bibinfo{author}{\bibfnamefont{G.}~\bibnamefont{Kotliar}}, \bibnamefont{and}
  \bibinfo{author}{\bibfnamefont{V.}~\bibnamefont{Dobrosavljevi\ifmmode~\acute{c}\else
  \'{c}\fi{}}}, \bibinfo{journal}{Phys. Rev. B} \textbf{\bibinfo{volume}{84}},
  \bibinfo{pages}{115105} (\bibinfo{year}{2011}).

\bibitem[{\citenamefont{Vu\ifmmode \check{c}\else \v{c}\fi{}i\ifmmode
  \check{c}\else \v{c}\fi{}evi\ifmmode~\acute{c}\else \'{c}\fi{}
  et~al.}(2013)\citenamefont{Vu\ifmmode \check{c}\else \v{c}\fi{}i\ifmmode
  \check{c}\else \v{c}\fi{}evi\ifmmode~\acute{c}\else \'{c}\fi{}, Terletska,
  Tanaskovi\ifmmode~\acute{c}\else \'{c}\fi{}, and
  Dobrosavljevi\ifmmode~\acute{c}\else \'{c}\fi{}}}]{vucicevic13prb}
\bibinfo{author}{\bibfnamefont{J.}~\bibnamefont{Vu\ifmmode \check{c}\else
  \v{c}\fi{}i\ifmmode \check{c}\else \v{c}\fi{}evi\ifmmode~\acute{c}\else
  \'{c}\fi{}}}, \bibinfo{author}{\bibfnamefont{H.}~\bibnamefont{Terletska}},
  \bibinfo{author}{\bibfnamefont{D.}~\bibnamefont{Tanaskovi\ifmmode~\acute{c}\else
  \'{c}\fi{}}}, \bibnamefont{and}
  \bibinfo{author}{\bibfnamefont{V.}~\bibnamefont{Dobrosavljevi\ifmmode~\acute{c}\else
  \'{c}\fi{}}}, \bibinfo{journal}{Phys. Rev. B} \textbf{\bibinfo{volume}{88}},
  \bibinfo{pages}{075143} (\bibinfo{year}{2013}).

\bibitem[{\citenamefont{Vu\ifmmode \check{c}\else \v{c}\fi{}i\ifmmode
  \check{c}\else \v{c}\fi{}evi\ifmmode~\acute{c}\else \'{c}\fi{}
  et~al.}(2015)\citenamefont{Vu\ifmmode \check{c}\else \v{c}\fi{}i\ifmmode
  \check{c}\else \v{c}\fi{}evi\ifmmode~\acute{c}\else \'{c}\fi{},
  Tanaskovi\ifmmode~\acute{c}\else \'{c}\fi{}, Rozenberg, and
  Dobrosavljevi\ifmmode~\acute{c}\else \'{c}\fi{}}}]{vucicevic15prl}
\bibinfo{author}{\bibfnamefont{J.}~\bibnamefont{Vu\ifmmode \check{c}\else
  \v{c}\fi{}i\ifmmode \check{c}\else \v{c}\fi{}evi\ifmmode~\acute{c}\else
  \'{c}\fi{}}},
  \bibinfo{author}{\bibfnamefont{D.}~\bibnamefont{Tanaskovi\ifmmode~\acute{c}\else
  \'{c}\fi{}}}, \bibinfo{author}{\bibfnamefont{M.~J.} \bibnamefont{Rozenberg}},
  \bibnamefont{and}
  \bibinfo{author}{\bibfnamefont{V.}~\bibnamefont{Dobrosavljevi\ifmmode~\acute{c}\else
  \'{c}\fi{}}}, \bibinfo{journal}{Phys. Rev. Lett.}
  \textbf{\bibinfo{volume}{114}}, \bibinfo{pages}{246402}
  (\bibinfo{year}{2015}).

\bibitem[{\citenamefont{Anderson}(1959)}]{anderson59superexchange}
\bibinfo{author}{\bibfnamefont{P.~W.} \bibnamefont{Anderson}},
  \bibinfo{journal}{Phys. Rev.} \textbf{\bibinfo{volume}{115}},
  \bibinfo{pages}{2} (\bibinfo{year}{1959}).

\bibitem[{\citenamefont{Kotliar and Ruckenstein}(1986)}]{kotliarruckenstein}
\bibinfo{author}{\bibfnamefont{G.}~\bibnamefont{Kotliar}} \bibnamefont{and}
  \bibinfo{author}{\bibfnamefont{A.~E.} \bibnamefont{Ruckenstein}},
  \bibinfo{journal}{Phys. Rev. Lett.} \textbf{\bibinfo{volume}{57}},
  \bibinfo{pages}{1362} (\bibinfo{year}{1986}).

\bibitem[{\citenamefont{Flouquet et~al.}(1982)\citenamefont{Flouquet,
  Lasjaunias, Peyrard, and Ribault}}]{flouquet1982}
\bibinfo{author}{\bibfnamefont{J.}~\bibnamefont{Flouquet}},
  \bibinfo{author}{\bibfnamefont{J.~C.} \bibnamefont{Lasjaunias}},
  \bibinfo{author}{\bibfnamefont{J.}~\bibnamefont{Peyrard}}, \bibnamefont{and}
  \bibinfo{author}{\bibfnamefont{M.}~\bibnamefont{Ribault}},
  \bibinfo{journal}{J. Appl. Phys.} \textbf{\bibinfo{volume}{53}},
  \bibinfo{pages}{2127} (\bibinfo{year}{1982}).

\bibitem[{\citenamefont{Flouquet}(2005)}]{Flouquet2005}
\bibinfo{author}{\bibfnamefont{J.}~\bibnamefont{Flouquet}}
  (\bibinfo{publisher}{Elsevier}, \bibinfo{year}{2005}),
  vol.~\bibinfo{volume}{15} of \emph{\bibinfo{series}{Progress in Low
  Temperature Physics}}, pp. \bibinfo{pages}{139 -- 281}.

\bibitem[{\citenamefont{Moeller et~al.}(1999)\citenamefont{Moeller,
  Dobrosavljevi\'{c}, and Ruckenstein}}]{moeller99prb}
\bibinfo{author}{\bibfnamefont{G.}~\bibnamefont{Moeller}},
  \bibinfo{author}{\bibfnamefont{V.}~\bibnamefont{Dobrosavljevi\'{c}}},
  \bibnamefont{and} \bibinfo{author}{\bibfnamefont{A.~E.}
  \bibnamefont{Ruckenstein}}, \bibinfo{journal}{Phys. Rev. B}
  \textbf{\bibinfo{volume}{59}}, \bibinfo{pages}{6846} (\bibinfo{year}{1999}).

\bibitem[{\citenamefont{Park et~al.}(2008)\citenamefont{Park, Haule, and
  Kotliar}}]{park08prl}
\bibinfo{author}{\bibfnamefont{H.}~\bibnamefont{Park}},
  \bibinfo{author}{\bibfnamefont{K.}~\bibnamefont{Haule}}, \bibnamefont{and}
  \bibinfo{author}{\bibfnamefont{G.}~\bibnamefont{Kotliar}},
  \bibinfo{journal}{Phys. Rev. Lett.} \textbf{\bibinfo{volume}{101}},
  \bibinfo{pages}{186403} (\bibinfo{year}{2008}).

\bibitem[{\citenamefont{C\^andido et~al.}(2004)\citenamefont{C\^andido, Bernu,
  and Ceperley}}]{candido-2004-70}
\bibinfo{author}{\bibfnamefont{L.}~\bibnamefont{C\^andido}},
  \bibinfo{author}{\bibfnamefont{B.}~\bibnamefont{Bernu}}, \bibnamefont{and}
  \bibinfo{author}{\bibfnamefont{D.~M.} \bibnamefont{Ceperley}},
  \bibinfo{journal}{Physical Review B} \textbf{\bibinfo{volume}{70}},
  \bibinfo{pages}{094413} (\bibinfo{year}{2004}).

\bibitem[{\citenamefont{Hewson}(1993)}]{hewson}
\bibinfo{author}{\bibfnamefont{A.~C.} \bibnamefont{Hewson}},
  \emph{\bibinfo{title}{The {K}ondo {P}roblem to {H}eavy {F}ermions}}
  (\bibinfo{publisher}{Cambrige University Press},
  \bibinfo{address}{Cambridge}, \bibinfo{year}{1993}).

\bibitem[{\citenamefont{Millis}(1993)}]{millis}
\bibinfo{author}{\bibfnamefont{A.~J.} \bibnamefont{Millis}},
  \bibinfo{journal}{Phys. Rev. B} \textbf{\bibinfo{volume}{48}},
  \bibinfo{pages}{7183} (\bibinfo{year}{1993}).

\bibitem[{\citenamefont{Ting et~al.}(1975)\citenamefont{Ting, Lee, and
  Quinn}}]{quinn75prl}
\bibinfo{author}{\bibfnamefont{C.~S.} \bibnamefont{Ting}},
  \bibinfo{author}{\bibfnamefont{T.~K.} \bibnamefont{Lee}}, \bibnamefont{and}
  \bibinfo{author}{\bibfnamefont{J.~J.} \bibnamefont{Quinn}},
  \bibinfo{journal}{Phys. Rev. Lett.} \textbf{\bibinfo{volume}{34}},
  \bibinfo{pages}{870} (\bibinfo{year}{1975}).

\bibitem[{\citenamefont{Zhang and {Das Sarma}}(2005)}]{dassarma05prb}
\bibinfo{author}{\bibfnamefont{Y.}~\bibnamefont{Zhang}} \bibnamefont{and}
  \bibinfo{author}{\bibfnamefont{S.}~\bibnamefont{{Das Sarma}}},
  \bibinfo{journal}{Phys. Rev. B} \textbf{\bibinfo{volume}{71}},
  \bibinfo{pages}{045322} (\bibinfo{year}{2005}).

\bibitem[{\citenamefont{Jacko et~al.}(2009)\citenamefont{Jacko, Fjaerestad, and
  Powell}}]{powell09nphys}
\bibinfo{author}{\bibfnamefont{A.~C.} \bibnamefont{Jacko}},
  \bibinfo{author}{\bibfnamefont{J.~O.} \bibnamefont{Fjaerestad}},
  \bibnamefont{and} \bibinfo{author}{\bibfnamefont{B.~J.}
  \bibnamefont{Powell}}, \bibinfo{journal}{Nat. Phys.}
  \textbf{\bibinfo{volume}{5}}, \bibinfo{pages}{422} (\bibinfo{year}{2009}).

\bibitem[{\citenamefont{Gull et~al.}(2011)\citenamefont{Gull, Millis,
  Lichtenstein, Rubtsov, Troyer, and Werner}}]{ctqmc2011rmp}
\bibinfo{author}{\bibfnamefont{E.}~\bibnamefont{Gull}},
  \bibinfo{author}{\bibfnamefont{A.~J.} \bibnamefont{Millis}},
  \bibinfo{author}{\bibfnamefont{A.~I.} \bibnamefont{Lichtenstein}},
  \bibinfo{author}{\bibfnamefont{A.~N.} \bibnamefont{Rubtsov}},
  \bibinfo{author}{\bibfnamefont{M.}~\bibnamefont{Troyer}}, \bibnamefont{and}
  \bibinfo{author}{\bibfnamefont{P.}~\bibnamefont{Werner}},
  \bibinfo{journal}{Rev. Mod. Phys.} \textbf{\bibinfo{volume}{83}},
  \bibinfo{pages}{349} (\bibinfo{year}{2011}).

\bibitem[{\citenamefont{Radonjic et~al.}(2010)\citenamefont{Radonjic,
  Tanaskovic, Dobrosavljevic, and Haule}}]{radonjic10}
\bibinfo{author}{\bibfnamefont{M.~M.} \bibnamefont{Radonjic}},
  \bibinfo{author}{\bibfnamefont{D.}~\bibnamefont{Tanaskovic}},
  \bibinfo{author}{\bibfnamefont{V.}~\bibnamefont{Dobrosavljevic}},
  \bibnamefont{and} \bibinfo{author}{\bibfnamefont{K.}~\bibnamefont{Haule}},
  \bibinfo{journal}{Phys. Rev. B} \textbf{\bibinfo{volume}{81}},
  \bibinfo{pages}{075118} (\bibinfo{year}{2010}).

\bibitem[{\citenamefont{Widom}(1965)}]{widom65}
\bibinfo{author}{\bibfnamefont{B.}~\bibnamefont{Widom}}, \bibinfo{journal}{J.
  Chem. Phys.} \textbf{\bibinfo{volume}{43}}, \bibinfo{pages}{3898}
  (\bibinfo{year}{1965}).

\bibitem[{\citenamefont{Emery and Kivelson}(1995)}]{emery95prl}
\bibinfo{author}{\bibfnamefont{V.~J.} \bibnamefont{Emery}} \bibnamefont{and}
  \bibinfo{author}{\bibfnamefont{S.~A.} \bibnamefont{Kivelson}},
  \bibinfo{journal}{Phys. Rev. Lett.} \textbf{\bibinfo{volume}{74}},
  \bibinfo{pages}{3253} (\bibinfo{year}{1995}).

\bibitem[{\citenamefont{Wigner}(1934)}]{wigner34pr}
\bibinfo{author}{\bibfnamefont{E.}~\bibnamefont{Wigner}},
  \bibinfo{journal}{Phys. Rev.} \textbf{\bibinfo{volume}{46}},
  \bibinfo{pages}{1002} (\bibinfo{year}{1934}).

\bibitem[{\citenamefont{C\^andido et~al.}(2001)\citenamefont{C\^andido,
  Phillips, and Ceperley}}]{candidi01prb}
\bibinfo{author}{\bibfnamefont{L.}~\bibnamefont{C\^andido}},
  \bibinfo{author}{\bibfnamefont{P.}~\bibnamefont{Phillips}}, \bibnamefont{and}
  \bibinfo{author}{\bibfnamefont{D.~M.} \bibnamefont{Ceperley}},
  \bibinfo{journal}{Phys. Rev. Lett.} \textbf{\bibinfo{volume}{86}},
  \bibinfo{pages}{492} (\bibinfo{year}{2001}).

\bibitem[{\citenamefont{Tanatar and Ceperley}(1989)}]{ceperley89prb}
\bibinfo{author}{\bibfnamefont{B.}~\bibnamefont{Tanatar}} \bibnamefont{and}
  \bibinfo{author}{\bibfnamefont{D.~M.} \bibnamefont{Ceperley}},
  \bibinfo{journal}{Phys. Rev. B} \textbf{\bibinfo{volume}{39}},
  \bibinfo{pages}{5005} (\bibinfo{year}{1989}).

\bibitem[{\citenamefont{Thouless}(1978)}]{thouless78jphysc}
\bibinfo{author}{\bibfnamefont{D.~J.} \bibnamefont{Thouless}},
  \bibinfo{journal}{Journal of Physics C: Solid State Physics}
  \textbf{\bibinfo{volume}{11}}, \bibinfo{pages}{L189} (\bibinfo{year}{1978}).

\bibitem[{\citenamefont{Chen et~al.}(2006)\citenamefont{Chen, Sambandamurthy,
  Wang, Lewis, Engel, Tsui, Ye, Pfeiffer, and West}}]{chen-2006-2}
\bibinfo{author}{\bibfnamefont{Y.~P.} \bibnamefont{Chen}},
  \bibinfo{author}{\bibfnamefont{G.}~\bibnamefont{Sambandamurthy}},
  \bibinfo{author}{\bibfnamefont{Z.~H.} \bibnamefont{Wang}},
  \bibinfo{author}{\bibfnamefont{R.~M.} \bibnamefont{Lewis}},
  \bibinfo{author}{\bibfnamefont{L.~W.} \bibnamefont{Engel}},
  \bibinfo{author}{\bibfnamefont{D.~C.} \bibnamefont{Tsui}},
  \bibinfo{author}{\bibfnamefont{P.~D.} \bibnamefont{Ye}},
  \bibinfo{author}{\bibfnamefont{L.~N.} \bibnamefont{Pfeiffer}},
  \bibnamefont{and} \bibinfo{author}{\bibfnamefont{K.~W.} \bibnamefont{West}},
  \bibinfo{journal}{Nature Physics} \textbf{\bibinfo{volume}{2}},
  \bibinfo{pages}{452} (\bibinfo{year}{2006}).

\bibitem[{\citenamefont{Camjayi et~al.}(2008)\citenamefont{Camjayi, Haule,
  Dobrosavljevic, and Kotliar}}]{camjayi08nphys}
\bibinfo{author}{\bibfnamefont{A.}~\bibnamefont{Camjayi}},
  \bibinfo{author}{\bibfnamefont{K.}~\bibnamefont{Haule}},
  \bibinfo{author}{\bibfnamefont{V.}~\bibnamefont{Dobrosavljevic}},
  \bibnamefont{and} \bibinfo{author}{\bibfnamefont{G.}~\bibnamefont{Kotliar}},
  \bibinfo{journal}{Nature Physics} \textbf{\bibinfo{volume}{4}},
  \bibinfo{pages}{932} (\bibinfo{year}{2008}).

\bibitem[{\citenamefont{Gor'kov and Sokol}(1987)}]{gorkov-JETP87}
\bibinfo{author}{\bibfnamefont{L.~P.} \bibnamefont{Gor'kov}} \bibnamefont{and}
  \bibinfo{author}{\bibfnamefont{A.~V.} \bibnamefont{Sokol}},
  \bibinfo{journal}{JETP Lett.} \textbf{\bibinfo{volume}{46}},
  \bibinfo{pages}{420} (\bibinfo{year}{1987}).

\bibitem[{\citenamefont{Schmalian and Wolynes}(2000)}]{schmalian-prl00}
\bibinfo{author}{\bibfnamefont{J.}~\bibnamefont{Schmalian}} \bibnamefont{and}
  \bibinfo{author}{\bibfnamefont{P.~G.} \bibnamefont{Wolynes}},
  \bibinfo{journal}{Phys. Rev. Lett.} \textbf{\bibinfo{volume}{85}},
  \bibinfo{pages}{836} (\bibinfo{year}{2000}).

\bibitem[{\citenamefont{Mahmoudian et~al.}(2015)\citenamefont{Mahmoudian,
  Rademaker, Ralko, Fratini, and Dobrosavljevi\ifmmode~\acute{c}\else
  \'{c}\fi{}}}]{mahmoudian15prl}
\bibinfo{author}{\bibfnamefont{S.}~\bibnamefont{Mahmoudian}},
  \bibinfo{author}{\bibfnamefont{L.}~\bibnamefont{Rademaker}},
  \bibinfo{author}{\bibfnamefont{A.}~\bibnamefont{Ralko}},
  \bibinfo{author}{\bibfnamefont{S.}~\bibnamefont{Fratini}}, \bibnamefont{and}
  \bibinfo{author}{\bibfnamefont{V.}~\bibnamefont{Dobrosavljevi\ifmmode~\acute{c}\else
  \'{c}\fi{}}}, \bibinfo{journal}{Phys. Rev. Lett.}
  \textbf{\bibinfo{volume}{115}}, \bibinfo{pages}{025701}
  (\bibinfo{year}{2015}).

\bibitem[{\citenamefont{Terletska and Dobrosavljevi\ifmmode~\acute{c}\else
  \'{c}\fi{}}(2011)}]{terletska11aprl}
\bibinfo{author}{\bibfnamefont{H.}~\bibnamefont{Terletska}} \bibnamefont{and}
  \bibinfo{author}{\bibfnamefont{V.}~\bibnamefont{Dobrosavljevi\ifmmode~\acute{c}\else
  \'{c}\fi{}}}, \bibinfo{journal}{Phys. Rev. Lett.}
  \textbf{\bibinfo{volume}{106}}, \bibinfo{pages}{186402}
  (\bibinfo{year}{2011}).

\bibitem[{\citenamefont{Jaroszy\'{n}ski
  et~al.}(2007)\citenamefont{Jaroszy\'{n}ski, Andrearczyk, Karczewski,
  Wr\'{o}bel, Wojtowicz, Popovi\'{c}, and Dietl}}]{jan07prb}
\bibinfo{author}{\bibfnamefont{J.}~\bibnamefont{Jaroszy\'{n}ski}},
  \bibinfo{author}{\bibfnamefont{T.}~\bibnamefont{Andrearczyk}},
  \bibinfo{author}{\bibfnamefont{G.}~\bibnamefont{Karczewski}},
  \bibinfo{author}{\bibfnamefont{J.}~\bibnamefont{Wr\'{o}bel}},
  \bibinfo{author}{\bibfnamefont{T.}~\bibnamefont{Wojtowicz}},
  \bibinfo{author}{\bibfnamefont{D.}~\bibnamefont{Popovi\'{c}}},
  \bibnamefont{and} \bibinfo{author}{\bibfnamefont{T.}~\bibnamefont{Dietl}},
  \bibinfo{journal}{Phys. Rev. B} \textbf{\bibinfo{volume}{76}},
  \bibinfo{pages}{045322} (\bibinfo{year}{2007}).

\bibitem[{\citenamefont{Waintal}(2006)}]{waitnal06prb}
\bibinfo{author}{\bibfnamefont{X.}~\bibnamefont{Waintal}},
  \bibinfo{journal}{Phys. Rev. B} \textbf{\bibinfo{volume}{73}},
  \bibinfo{pages}{075417} (\bibinfo{year}{2006}).

\bibitem[{\citenamefont{Clark et~al.}(2009)\citenamefont{Clark, Casula, and
  Ceperley}}]{ceperley09prl}
\bibinfo{author}{\bibfnamefont{B.~K.} \bibnamefont{Clark}},
  \bibinfo{author}{\bibfnamefont{M.}~\bibnamefont{Casula}}, \bibnamefont{and}
  \bibinfo{author}{\bibfnamefont{D.~M.} \bibnamefont{Ceperley}},
  \bibinfo{journal}{Phys. Rev. Lett.} \textbf{\bibinfo{volume}{103}},
  \bibinfo{pages}{055701} (\bibinfo{year}{2009}).

\bibitem[{\citenamefont{Pankov and Dobrosavljevic}(2008)}]{pankov08prb}
\bibinfo{author}{\bibfnamefont{S.}~\bibnamefont{Pankov}} \bibnamefont{and}
  \bibinfo{author}{\bibfnamefont{V.}~\bibnamefont{Dobrosavljevic}},
  \bibinfo{journal}{Physica B} \textbf{\bibinfo{volume}{403}},
  \bibinfo{pages}{1440} (\bibinfo{year}{2008}).

\bibitem[{\citenamefont{Amaricci et~al.}(2010)\citenamefont{Amaricci, Camjayi,
  Haule, Kotliar, Tanaskovi\ifmmode~\acute{c}\else \'{c}\fi{}, and
  Dobrosavljevi\ifmmode~\acute{c}\else \'{c}\fi{}}}]{camjayi10prb}
\bibinfo{author}{\bibfnamefont{A.}~\bibnamefont{Amaricci}},
  \bibinfo{author}{\bibfnamefont{A.}~\bibnamefont{Camjayi}},
  \bibinfo{author}{\bibfnamefont{K.}~\bibnamefont{Haule}},
  \bibinfo{author}{\bibfnamefont{G.}~\bibnamefont{Kotliar}},
  \bibinfo{author}{\bibfnamefont{D.}~\bibnamefont{Tanaskovi\ifmmode~\acute{c}\else
  \'{c}\fi{}}}, \bibnamefont{and}
  \bibinfo{author}{\bibfnamefont{V.}~\bibnamefont{Dobrosavljevi\ifmmode~\acute{c}\else
  \'{c}\fi{}}}, \bibinfo{journal}{Phys. Rev. B} \textbf{\bibinfo{volume}{82}},
  \bibinfo{pages}{155102} (\bibinfo{year}{2010}).

\bibitem[{\citenamefont{Fratini and Merino}(2009)}]{fratini09prb}
\bibinfo{author}{\bibfnamefont{S.}~\bibnamefont{Fratini}} \bibnamefont{and}
  \bibinfo{author}{\bibfnamefont{J.}~\bibnamefont{Merino}},
  \bibinfo{journal}{Phys. Rev. B} \textbf{\bibinfo{volume}{80}},
  \bibinfo{pages}{165110} (\bibinfo{year}{2009}).

\bibitem[{\citenamefont{Merino et~al.}(2013)\citenamefont{Merino, Ralko, and
  Fratini}}]{merino13prl}
\bibinfo{author}{\bibfnamefont{J.}~\bibnamefont{Merino}},
  \bibinfo{author}{\bibfnamefont{A.}~\bibnamefont{Ralko}}, \bibnamefont{and}
  \bibinfo{author}{\bibfnamefont{S.}~\bibnamefont{Fratini}},
  \bibinfo{journal}{Phys. Rev. Lett.} \textbf{\bibinfo{volume}{111}},
  \bibinfo{pages}{126403} (\bibinfo{year}{2013}).

\end{thebibliography}
\end{document}